\documentclass[12pt]{article}
\usepackage{graphicx}
\usepackage{amsmath}
\graphicspath{{./images2/}}
\usepackage{afterpage}
\usepackage{float}
\restylefloat{figure}
\restylefloat{table}

\title{A microscopic model of evolution of recombination}
\author{\centerline{\normalsize Franco Bagnoli$^{1,2,3}$\thanks{Electronic address: franco.bagnoli@unifi.it} ~~and Carlo Guardiani$^{3}$\thanks{Electronic address: carlo@dma.unifi.it}} \\
\centerline{\textit{\normalsize $^1$ Dipartimento di Energetica, Universit\`a di Firenze,Via S.~Marta 3, I-50139 Firenze, Italy}}\\
\centerline{\textit{\normalsize $^2$ INFN, sez. Firenze}}\\
\centerline{\textit{\normalsize $^3$ Centro Interdipartimentale per lo Studio delle Dinamiche Complesse, }} \\
\centerline{\textit{\normalsize Universit\`a di Firenze, Via Sansone 1, I-50019, Sesto Fiorentino, Italy}}
}

\date{\today}

\begin{document}

\maketitle

\begin{abstract}

We study  the evolution of recombination using a microscopic model developed within the frame of the theory of quantitative traits. Two components of fitness are considered: a static one that describes adaptation to environmental factors not related to the population itself, and a dynamic one that accounts for interactions between organisms \emph{e.g.} competition. We focus on the dynamics of colonization of an empty niche. As competition is a function of the population, selection pressure rapidly changes in time. The simulations show that both in the case of flat and steep static fitness landscapes, recombination provides a high velocity of movement in the phenotypic space thus allowing recombinants to colonize the highest fitness regions earlier than non recombinants that are often driven to extinction. The stabilizing effects of competition and assortativity are also discussed. Finally, the analysis of phase diagrams shows that competition is the key factor for the evolution of recombination, while assortativity plays a significant role only in small populations.

\end{abstract}

\section{The problem}

One of the most challenging problems of evolutionary biology is that of the evolution of sexual reproduction, usually referred to as \emph{The paradox of sex}. 

Sexual reproduction, in fact, is seemingly a very inefficient reproduction strategy, being associated to several costs, yet it is widespread in all taxa of the natural world. Actually~\cite{Butlin}, the only ancient, species-rich, fully asexual taxa are the Rotifer class \emph{Bdelloidea} with 360 described species and an estimated age of 80-100 million years, and the ostracod family \emph{Darwinulidae} with 28 described species and a minimum time without sex of 100 million years. Recent studies on maximum-likelihood gene trees for \textbf{hsp82} proteins in rotifers~\cite{Welch}, however, seem to argue that \emph{Bdelloidea} appeared as the result of several mutations in sexual ancestors, and asexual reproduction was then maintained only thanks to the stability of their environment. Studies on the distribution of sexual and asexual reproduction~\cite{Bell}, on the other hand, show that most asexual animals and plants have close sexual relatives, and fully asexual taxa usually contain few species. This implies that asexual lineages are short lived and they are doomed to extinction before they are able to diversify. Finally, it should be noted that very important taxa do exist, such as \emph{Mammals} and \emph{Gymnosperms}, in which asexual reproduction has never been observed.

The evolutionary success of sexual reproduction is paradoxical if we consider the several costs related to sex. First of all, there are costs related to attracting a partner, as the considerable resources invested by plants in floral display and nectar rewards. Moreover, the secondary sexual characters necessary to attract mates, as the bright colours of the males of many species of birds, make them more vulnerable to predators. Another disadvantage of sex is that it might break up favorable combinations of genes (that enabled the parents to survive to reproductive age) without any guaranty to produce better ones. A final, well-known problem is the \emph{twofold price of sex}. In fact, if we assume that only one parent in sexual species is in charge of feeding and protecting the offsprings, then the fertility of sexual females and asexual individuals is just the same (say $m$ offsprings). As a result, in a population if $N$ individuals, an asexual species will produce $mN$ offsprings, while only $mN/2$ will be produced by a sexual species. Sexual reproduction therefore appears to be a very ineffective reproduction strategy leading to a short-term evolutionary disadvantage.

The first solution to the paradox of sex was proposed in 1889 by August Weismann who argued that the advantage of sex lies in the production of individual variability on which natural selection can act~\cite{Burt}. Later studies, however, showed that Weismann's statement is only partly correct. Syngamy, \emph{i.e.} the process of cell fusion, arose as a strategy to survive in periods of shortage of food, while recombination does not always lead to an increase in variability, which, in any case, can only be regarded as a long-term advantage. In Sections 2  we will discuss the evolution of syngamy and recombination based on the latest advances in the field.

\section{Origin and advantages of sex}

According to several recent studies the evolution of sex took place in two steps:

\begin{enumerate}

\item Evolution of syngamy as a mean to increase survival probability in conditions of starvation.

\item Evolution of recombination as to increase genetic variability and to create new favorable arrangements of genes.

\end{enumerate}

According to Cavalier-Smith~\cite{CavSmith-2001,CavSmith-2002a,CavSmith-1998} the common ancestor of Eukaryotes and Archaebacteria was similar to the current Gram positive bacteria being surrounded by a cell membrane and a peptidoglican wall. The necessity to adapt to a hot acid environment led the common ancestor to lose the peptidoglican wall and develop the ability to synthesize membrane glycoproteins~\cite{CavSmith-1987a, CavSmith-2002a}. The eukaryotes that derived from the common ancestor were therefore characterized by a soft cell and a well-developed cytoskeleton and endomembrane system that made it easy to evolve phagotrophy. A slight modification of the mechanism of phagotrophy then led to syngamy~\cite{MaynardSmith,Michod,Redfield,CavSmith-2002b}. Cell fusion, in fact, doubling the cell food storage greatly increases the survival probability. The survival probability is further increased  if cysts are formed after syngamy because diploidity enables the repair of double strand damages of DNA. Meiosis and recombination then evolved  as a mechanism to restore the ploidity characteristic of the species during excystment; haploid cells then reproduced through mitosis~\cite{CavSmith-1995,Kondrashov}. It should be noted that the evolution of meiosis was a comparatively simple evolutionary step as many recombination enzymes already existed being involved in other processes like replication, transcription and repair~\cite{CavSmith-Heredity}. Even if syngamy originally evolved for trophic reasons, on the long run, sex fixed in populations due to the several advantages of gene mixing.

Recombination first of all increases the probability of fixation of beneficial mutations and decreases the probability of fixation of harmful mutations. In a non recombinant population, in fact, a beneficial mutation can fix only  if it appears in the \emph{progenitor tail} \emph{i.e.} the highest fitness tail of the background fitness frequency distribution, whereas in a recombinant population the association of a mutation with the genetic background is only transient, so that background selection does not affect the fixation of a beneficial mutation~\cite{Rice-Ref-33,Rice-Ref-34,Rice-Ref-36,Rice-Ref-37}. In a similar way, if $s$ is the selective coefficient of a harmful mutation (increment in fitness of the heterozigous caused by the mutation) the mutation can fix in a recombinant population only if $|s| < 1/N$ whereas it fixes in a non recombining population if $|s| < 1/N_p$. As the total population size $N$ is much greater than the progenitor tail population $N_p$, it follows that more harmful mutations will have a chance to be fixed in a non recombining population~\cite{Rice-Ref-32,Rice-Ref-34,Rice-Ref-35,Rice-Ref-36}.

A second theoretical advantage of recombination is that it brings together in the same genome several favorable mutations~\cite{Rice-Ref-3,Rice-Ref-24,Rice-Ref-25}. In non recombinant populations favorable mutations are segregated in different cell lines; only if two or more beneficial mutations occur independently in the same cell, it will be possible to find several favorable mutations in the same cell lineage but this event is very rare. Unfortunately, no experiments so far have been performed to test this hypothesis~\cite{Rice}.

The third theoretical advantage of sex and recombination is that, as already suggested by Weismann, it increases genetic variability, upon which selection can act. However, it must be remarked that recombination increases genetic variability only in the case of negative linkage disequilibrium $D < 0$ (in the case of a quantitative character with two alleles for each gene, it means, roughly speaking, that the frequency of intermediate phenotypes is higher than that of extreme phenotypes)~\cite{SPO-Review}. The other key ingredient for the evolution of recombination is the epistasis $\epsilon$ \emph{i.e.} a measure of fitness interactions between alleles at different loci. Theoretical arguments suggest that recombination can evolve only if $D<0$ and $\epsilon < 0$ (the fitness of intermediate phenotypes is higher than that of the extreme phenotypes)~\cite{SPO-Review,SPO-Ref-37}. Experimental studies however show that this situation is not very common so that another ingredient is apparently missing~\cite{Rice}. Recent studies seem to argue that the missing ingredient is a rapidly changing environment~\cite{SPO-Ref-7,SPO-Ref-37}. A change in the environmental conditions in fact, can reverse the sign of epistasis \emph{i.e.} the highest fitness group can shift from the intermediate to the extreme phenotypes and \emph{vice versa}. It is thus clear that only recombination enables the population to quickly adapt to the changing conditions.

In our work we tested the influence of a changing environment on the evolution of recombination. The change in environmental conditions is not imposed externally (changes of weather in the seasons of the year, or catastrophic changes like ice ages) but is related to the dynamics of the population itself through competition interactions. An increase in the number of competitors in fact, represents a deterioration in the environmental conditions that decreases the fitness of a given phenotype.

The simulations show, that under these conditions recombination provides organisms with a high velocity of movement in the phenotypic space, so that recombinants reach the regions of the phenotypic space with the highest fitness much earlier than non recombinants whose ability of movement is limited by the low mutation rate. We found that the advantage of recombinants increases with genome length which may be represent an explanation of the limited genome size of non recombinant organisms such as prokaryotes~\footnote{The parasexual mechanisms of \emph{conjugation}, \emph{transformation} and \emph{viral transduction} do not combine whole genomes biparentally but typically transfer only a handful of genes unidirectionally so that their contribution to mobility in the phenotype space is very modest.}. 

The simulations also show (in agreement with the findings in our study on sympatric speciation~\cite{Bagnoli-Guardiani}) that competition acts as a stabilizing force allowing the survival of strains in regions of the phenotypic space where the low static fitness level is counteracted by a low competition pressure.

Finally, the study of the phase diagrams in which the final frequency of recombinants is plotted as a function of competition and assortativity, shows that the most important factor for the evolution of recombination is competition, while the role of assortativity is to favor the survival of initially very small recombinant populations.

In Section 3 we describe our model and discuss the simplifications introduced; in Section 4 we report the results of the simulations distinguishing between the case of flat and steep static fitness landscapes and paying special attention to the influence of genome length; in Section 5 we study the evolution of recombination through phase diagrams; finally, in Section 6 we draw the conclusions of our study.

\section{The model }

Our model has been developed within the conceptual framework of the \emph{theory
of quantitative traits}~\cite{Lynch,Falconer} whose basic principles are :

\begin{enumerate}
\item The variability of quantitative traits is due to the simultaneous
      segregation of many genes.
\item The effect of these genes are small, similar and additive.
\item The effect of environmental factors is superimposed to that of genes.
\end{enumerate}

We consider a population of haploid individuals whose genome is represented as a
string of $L$ bits. Each bit represents a \emph{locus} and the Boolean values it
can take are regarded as alternative allelic forms. In particular the  value 0
refers to the \emph{wild-type} allele while the  value 1 to the least deleterious
mutant. The phenotype, in agreement with the theory of quantitative traits is
just the sum of these bits. The mutation is simply implemented by flipping a
randomly chosen bit from 0 to 1 or \emph{vice versa}. This kind of mutations can
only turn a phenotype $x$ into one of its neighbors $x+1$ or $x-1$ and they are
therefore referred to as \emph{short range mutation}.

The initial populations of both recombinants and non recombinants, as will be explained in more detail later on, are binomially distributed and their positions in the phase space can be chosen at will.  

The model is composed of a selection step followed by a reproduction procedure.
  The choice of
the fitness landscape is of paramount importance in selection. In our model we consider a
static and a dynamic component of fitness. The static component describes the
adaptation to environmental factors not related to the population itself
\emph{e.g.} \ abiotic factors such as climate, temperature, etc. The dynamic
component describes how the interactions with other members of the population
(competition, predation, mutualism) affect the fitness and it changes in time as
a function of the population itself. In our very simplified model we considered
only competition. The static component of the fitness is defined as :

\[ H_0(x) = e^{- \frac{1}{\Gamma} \left ( \frac{x}{\Gamma} \right )^{\beta} } \]

One important feature of this function is that it becomes flatter and flatter as
the parameter $\beta$ is increased. When $\beta \rightarrow 0$ we obtain a
sharp peak on the phenotype $x=0$; when $\beta = 1$ the function is a declining
exponential whose steepness depends on the parameter $\Gamma$; and finally when
$\beta \rightarrow \infty$ the fitness landscape is flat in the range
$[0,\Gamma]$.

 The complete expression of the fitness landscape is :

\begin{equation}
H(x) = H_0(x) + \sum_{y}\mathcal{J}_{1}(x|y)P(y) + \sum_{yz}\mathcal{J}_{2}(x|yz)P(y)P(z) + \cdots
\end{equation}

where $H_0(x)$ represents the viability of phenotypes $x$ not depending on the interactions with other individuals while the next terms of the fitness function account for the pair interactions, the three-body interactions, etc. In our model we restricted ourselves to the pair interactions only, so that the fitness function can be written as:

\begin{equation}
H(x) = H_0(x) + \sum_{y}\mathcal{J}(x|y)P(y)
\end{equation}

The matrix $\mathcal{J}$ describes the possible interactions between phenotypes, whereas $H_0$ represents a fixed or slowly changing environment. A strain $x$ with static fitness $H_0(x) > 0$ represents individuals that can survive in isolation (say, after an inoculation in an empty substrate), while a strain with $H_0(x) < 0$ represents predators or parasites that require the presence of some other individuals to survive. The matrix $\mathcal{J}$, that represents the interactions between two strains, provides the inputs necessary for the survival of non autonomous strains. For a classification in terms of usual ecological interrelations, one has to consider together $\mathcal{J}(x,y)$ and $\mathcal{J}(y,x)$. There are thus four possible cases:

\begin{figure}[ht!]
\begin{center}
\begin{tabular}{l l l}
$\mathcal{J}(x|y)<0$ & $\mathcal{J}(y|x)<0$ & competition \\
$\mathcal{J}(x|y)>0$ & $\mathcal{J}(y|x)<0$ & predation or parasitism of species $x$ against species $y$ \\
$\mathcal{J}(x|y)<0$ & $\mathcal{J}(y|x)>0$ & predation or parasitism of species
 $y$ against species $x$ \\
$\mathcal{J}(x|y)>0$ & $\mathcal{J}(y|x)>0$ & cooperation \\
\end{tabular}
\end{center}
\end{figure}

Since the individuals with similar phenotypes  are those sharing the largest quantity of resources , the competition is the stronger  the more similar their phenotypes are. This implies that the interaction matrix $\mathcal{J}$ has negative components on and near the diagonal. In our very simple model we consider only competitive interactions, so that the interaction matrix takes the form:

\[ \mathcal{J}(x|y) = -JK\left( \frac{x-y}{R} \right) \]

with the kernel $K$ given by:

\[ K\left( \frac{x-y}{R} \right) = \exp \left( -\frac{1}{\alpha}\left | \frac{x-y}{R} \right |^{\alpha} \right ) \]
 
The complete expression of fitness in our model is therefore given by:

 \begin{equation}
H(x) = H_0(x) - J\sum_y e^{- \frac{1}{\alpha} \left | \frac{x-y}{R} \right
|^{\alpha}}P(y)
\label{fitness-dinamica}
\end{equation} 

where parameter $J>0$ controls the intensity of competition with respect
to the arbitrary reference value $H_0(0)=1$.
If $ \alpha = 0$ an individual with phenotype $x$ is in competition only with
other organisms with the same phenotype; conversely
in the case $\alpha \rightarrow \infty$ a phenotype $x$
is in competition with all the other phenotypes in the range $[x - R,
x + R]$,  and the boundaries of this competition intervals
blurry when $\alpha$ is decreased.
The expression of the fitness landscape  shows that competition lowers the fitness, and the
competition kernel declines exponentially as the phenotypic distance increases.
It is now easy to work out the fitness function that accounts for the fact that
the number of offsprings must always be a positive quantity:

\[A(x) = e^{H(x)} \]

 The number of survivors of
phenotype $i$ after selection is :

\[ n'(i) = n(i)(1 - m)\frac{A_i}{\bar{A}} \]

where $m$ is the fitness-independent mortality and $\left ( 1 - \frac{A_i}{\bar{A}} \right ) $ is the
fitness-dependent mortality. The introduction of a fitness-independent component of mortality simply accounts
for the loss of individuals that in natural populations occurs even in the presence of a flat fitness
landscape. In this case $A_i = \bar{A}$ \   $\forall{i}$  so that the number of survivors is :

\[ n'(i) = n(i)(1 - m) \]

In the general case, the number of survivors of phenotype $i$ will be higher than this figure if the fitness
is higher than average, and smaller if the fitness is lower than average. In all our simulations we chose the very low value $m = 0.05$ for the fitness independent mortality as this effect is expected to be usually negligeable: even when the static fitness landscape is flat, competition, assortativity and fertility play a much more significant role.

After selection, the population goes through the reproduction process.   
The mechanism of
reproduction of non recombinant individuals is very simple and reminds that of
virus. A copy of the genome is made and a mutation is introduced on a random bit
according to a mutation rate $\mu$.

The reproduction of recombinants is slightly more complex. First, the genome of
the offspring is built by choosing for each \emph{locus} the allele of the first
or second parent with the same probability and then a mutation is introduced
with the same procedure employed in the case of non recombinants. In our model
we therefore assume absence of \emph{linkage} but it must be remembered that it
 is reasonable only in the case of very long genomes subdivided into many 
 independent chromosomes.
 
 The assortativity is introduced through a parameter $\Delta$ which represent
 the maximal phenotypic distance still compatible with reproduction. In other
 words, if the first parent has phenotype $i$ its partners must be chosen in
 the range $[i- \Delta, i+ \Delta]$ ( strictly speaking, the  reproductive range
 is \mbox{$[max \{i - \Delta,0 \}, min \{i + \Delta, L \}]$}- to account for the fact that in
 our model the possible phenotypes are all in the range $[0,L]$).
 
 The population grows according to a logistic law with overlapping generations. The number of offsprings
 produced by recombinants is therefore:
 
 \[ \tilde{M}_F = M'\left (1 - \frac{N'}{K} \right) \]
 
 where $N' = N(1-m)$ and $M' = M(1-m)$ are respectively, the population size and the number of recombinants
 after selection, while $K$ is the carrying capacity of the population. In other words, in our model we assume
 that each of the $M'/2$ pairs of recombinants yields two offsprings with probability $\left ( 1 -
 \frac{N'}{K} \right )$ . In a similar fashion, the number of offsprings produced by non recombinants can be
 calculated as:
 
 \[ \tilde{S}_F = 2S' \left (1 - \frac{N'}{K} \right) \]
 
 where $S'$ is the number of non recombinants after selection. As generations are overlapping, the number of
 recombinants and non recombinants after reproduction is obtained by summing the number of survivors of each
 group with their offsprings:
 
 \[ \tilde{M} = M' + M'\left (1 - \frac{N'}{K} \right) \]

\[ \tilde{S} = S' + 2S'\left (1 - \frac{N'}{K} \right) \]

 As can be easily seen, in our model we assume that each "female" both recombinant and non recombinant
 produces the same number of offsprings: we therefore introduce a \emph{twofold price} of recombination. In
 fact, recombinant females, producing two offsprings, can replace both themselves and their "male" partners,
 while non recombinant "females" can replace themselves and in addition they produce one extra offspring. The
 twofold price of recombination has been introduced to account for the fact that in many species with a real
 sexuality males play an apparently parasitic role. In most cases in fact, the only role of males in
 reproduction is to fertilize the egg-cell while the parental care is completely entrusted to females. As a
 result, there appear to be no logical reason why recombinant females should produce more offsprings than non
 recombinants.

Rigorously, in finite populations one cannot talk of true phase
transitions, and also the concept of invariant distribution is
questionable. On the other hand, the presence of random mutations
should make the system ergodic in the long time limit.
 In order to speed-up simulations, we used a simulated annealing technique: the
 mutation rate $\mu$ depends on time as
 \[
   \mu(t) = \frac{\mu_0 -\mu_{\infty}}{2}
     \left(1-\tanh\left(\frac{t-\tau}{\delta}\right)\right) +
       \mu_{\infty},
       \]
which roughly corresponds to keeping $\mu=\mu_0$  up to a time $\tau-\delta$, then decrease it linearly up to the
desired value $\mu_{\infty}$ is a time interval $2\delta$ and continue
with this value for the rest of simulation.
The limiting case of this procedure is to use $\tau=\delta=1$ and
$\mu_0=1$, which is equivalent to starting from a random genetic
distribution. Simulations show that that the variance and mean  reach their asymptotic value
very quickly.

\section{Simulations}

In our work, the evolution of recombination is supposed to be a dynamical process. In fact, one of the theoretical advantages of sex is that it increases the velocity of movement in the phenotypic space, which is a valuable feature especially in a rapidly changing environment. As a result, the evolutionary dynamics is expected to be dependent on the initial conditions, namely on the positions of the frequency distributions of recombinants and non recombinants in the phenotypic space.

This is why in all the simulations we performed, the initial populations of both recombinants and non recombinants are binomially distributed:

\[ p(x) = \left ( \begin{array}{c}
                   L \\ 
		   x
		  \end{array} \right ) \pi^{x} (1 - \pi)^{L -x} \]

where $\pi$ represents the probability of a single allele being equal to 1. In our simulation $\pi = p$ and $\pi = q$ are respectively the probabilities of a single bit being equal to 1 in the recombinant and non recombinant populations. In the binomial distribution both the mean and the variance depend only on $L$ (the genome length) and $\pi$:

\[ \bar{x} = L\pi \]

\[ var = L\pi(1-\pi) \]

The binomial law thus enables us to move the initial frequency distributions in the phenotypic space by choosing appropriate values for the parameter $\pi$ and to investigate the dependence on initial conditions. As an example, $\pi = 0.5$ means that the initial distribution is centered in the middle of the phenotypic space, whereas $\pi = 0$ and $\pi = 1$ means that the initial distribution is a delta-peak in $x=0$ and $x=1$ respectively.

Another problem we addressed is that of finding a stationary frequency distribution of the population. Our model employs a finite population and is affected by several stochastic elements \emph{e.g.} in the choice of individuals in the reproduction and selection step, so that there is very little chance of finding a distribution that is stationary \emph{strictu sensu}, but there will always be small fluctuations from generation to generation. On the other hand, the use of a \emph{simulated annealing} technique, enables us to find in a reasonably short simulation time a distribution whose fluctuations are very small. In order to monitor the fluctuations, for each simulation we plot the mean and the variance of the frequency distribution. Finally, we also show the plots of frequencies~\footnote{A seemingly odd feature of the frequency plots is that the initial frequency of recombinants is lower and that of non recombinants is higher than the value set in the simulation. This is simply due to the fact that the frequency values are averaged over a number of generations (usually 50 or 10) and in the very first generations the recombinants pay the double price of recombination so that their frequency rapidly drops below the initial values set in the program. Similar remarks apply to the plots of mean and variance.} of recombinants and non recombinants to show how, under appropriate conditions, the recombinants can overcome the handicap due to the twofold price and indeed outperform non recombinants.

\subsection{Flat static fitness landscape}

\label{flat-fit}

\subsubsection{Absence of competition}

\begin{figure}[ht!]

\begin{tabular}{ccc}
\hspace{-2 cm} &
\includegraphics[scale=0.65]{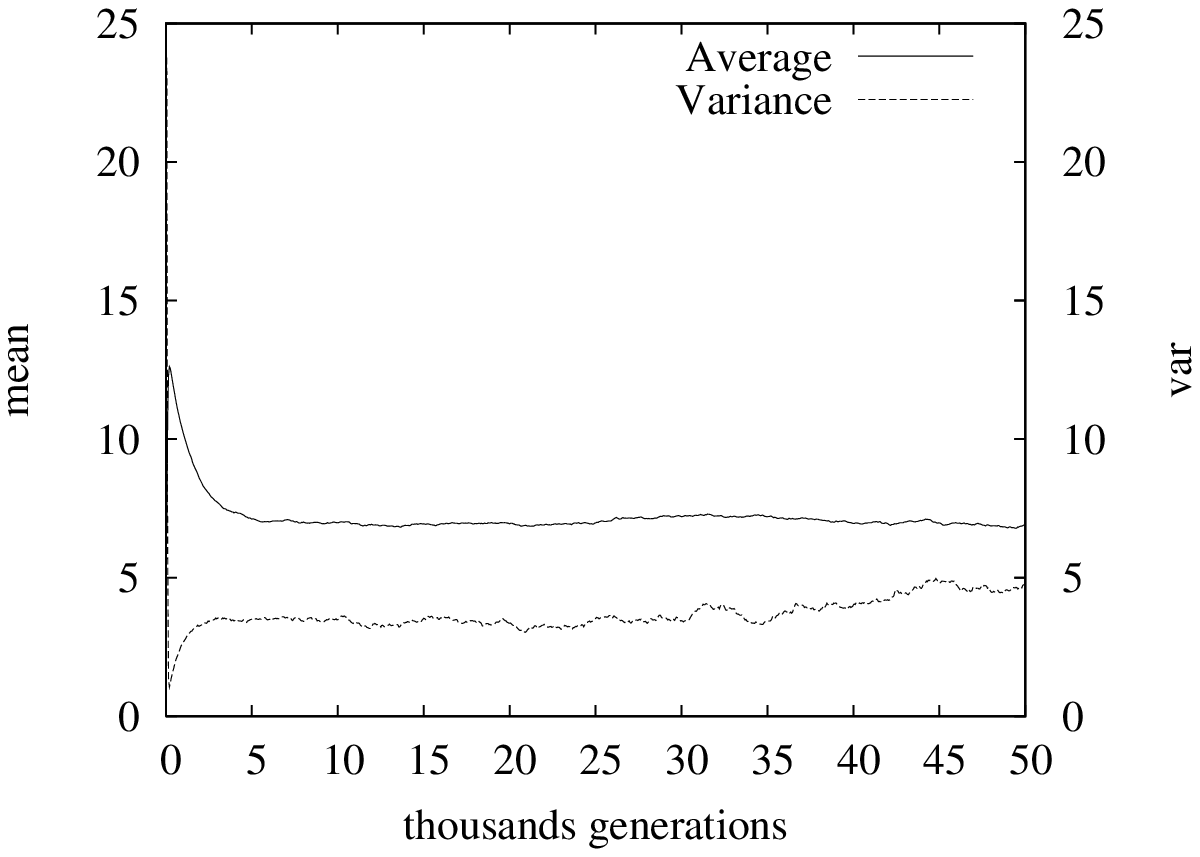} &
\includegraphics[scale=0.65]{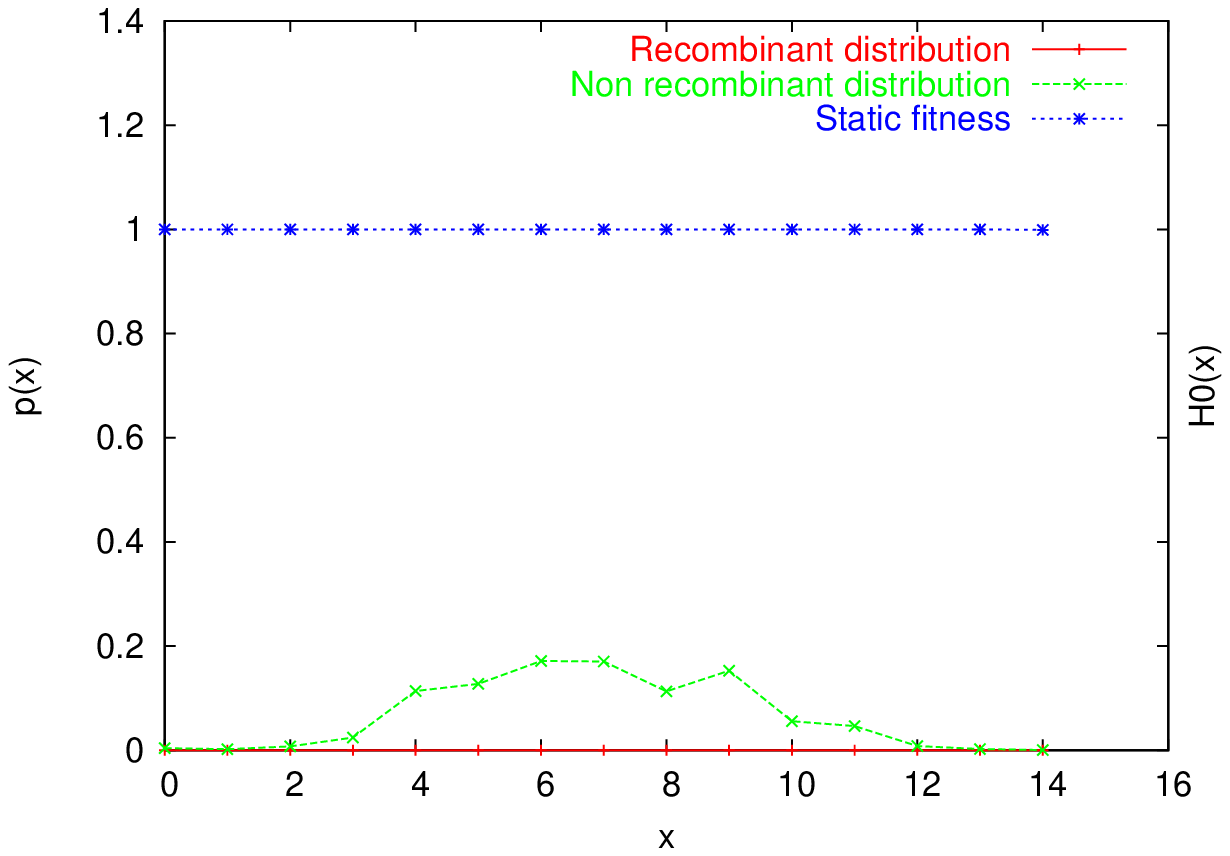} 
\end{tabular}

\caption{ The twofold price of sex: extinction of recombinants in absence of competition ($J=0$) and with flat static fitness ($\beta=100$, $\Gamma=14$). Mating range $\Delta = 14$. Initial frequency of recombinants: 0.9; initial distribution parameters: $p=0$, $q=1$, initial population size $N_0 = 10000$, carrying capacity $K = 100000$. Annealing parameters: $\mu_0 = 10^{-1}$, $ \mu_{\infty} = 10^{-3}$, $\tau = 10000$, $\delta = 100$. Evolution time: 50000 generations. Left panel: plots of mean and variance of the non recombinant distribution; right panel: final distribution in generation 50000. }

\label{fig:1A}

\end{figure}

If competition is absent and the static fitness landscape is flat, all phenotypes have the same fitness and
the evolutionary success depends only on fertility. It is therefore not surprising that, owing to the twofold
price, recombinants are driven to extinction even when their initial frequency is very high. As an example let us consider the case of an initial population with a delta peak of recombinants in $x=0$ (accounting for $90 \%$ of the total population) and a delta peak of non recombinants in $x=14$. Owing to the twofold price, the recombinants become extinct within a few hundreds of generations and only a peak of non recombinants in the neighborhood of $x=14$ survives. As an effect of mutations, the latter distribution moves towards the center of the phenotypic space where more genotypes correspond to each phenotype. After $T = \tau$ the mutation rate decreases and the dynamics of the population becomes dominated by random fluctuations. The effect of fluctuations however can be reduced by choosing a very large population, in particular, we chose an initial size of 10000 individuals that quickly reaches the carrying capacity of 100000 units. In Figure~\ref{fig:1A} we show the plots of mean and variance of the non recombinant population: the average decreases as the distribution moves towards the center of the phenotypic space, while the variance increases because fluctuations lead the population to be distributed over several phenotypes. In the right panel of Figure~\ref{fig:1A} we also show the final distribution featuring a bell shape even if rather irregular due to fluctuations. Finally, it should be noticed that the oscillations in the plots of mean and variance are very small so that the final distribution can be regarded as stationary.

\begin{figure}[ht!]
\begin{center}
\includegraphics[scale=0.65]{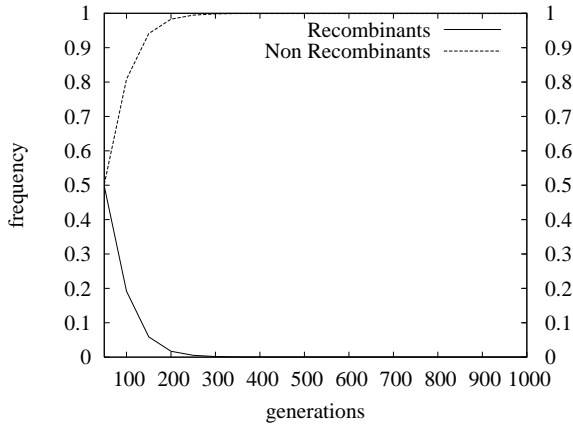} 
\end{center}

\caption{Extinction of recombinants in absence of competition ($J=0$) and with flat static fitness ($\beta=100$, $\Gamma=14$): plot of frequencies of recombinants and non recombinants. Mating range $\Delta = 14$. Initial frequency of recombinants: 0.9; initial distribution parameters: $p=0$, $q=1$, initial population size $N_0 = 10000$, carrying capacity $K = 100000$. Annealing parameters: $\mu_0 = 10^{-1}$, $ \mu_{\infty} = 10^{-3}$, $\tau = 10000$, $\delta = 100$. Evolution time: 50000 generations.}

\label{fig:1B}

\end{figure}

In Figure~\ref{fig:1B} we show the plots of frequencies of recombinants and non recombinants: it can be seen that even if the initial frequency of recombinants was set to $90 \%$, as a consequence of the twofold price, after very few generations the frequencies of recombinants and non recombinants become roughly equal; after that the frequency of recombinants tends to zero and the whole population will be made of non recombinants.

\subsubsection{Weak competition}

\label{weak-comp}

\begin{figure}[ht!]

\begin{tabular}{ccc}
\hspace{-2 cm} & \includegraphics[scale=0.65]{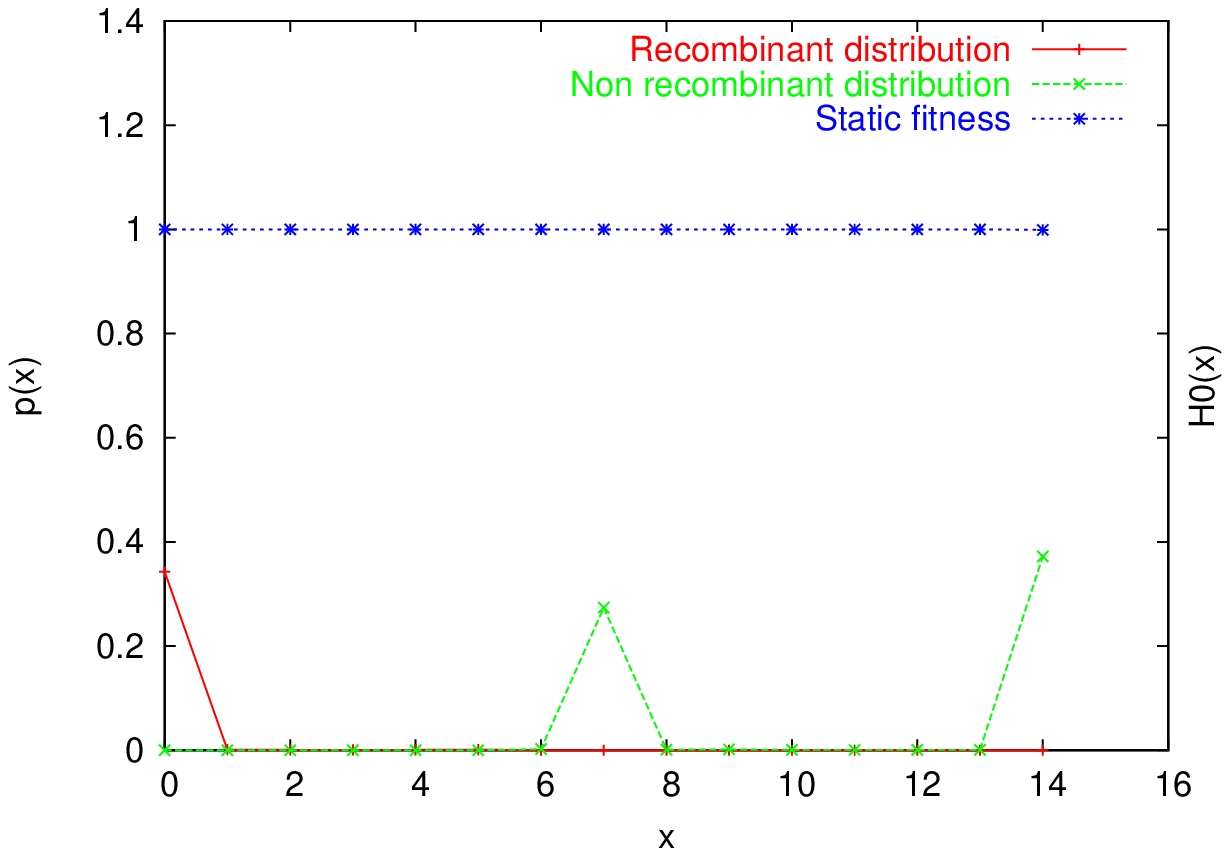} & 
\includegraphics[scale=0.65]{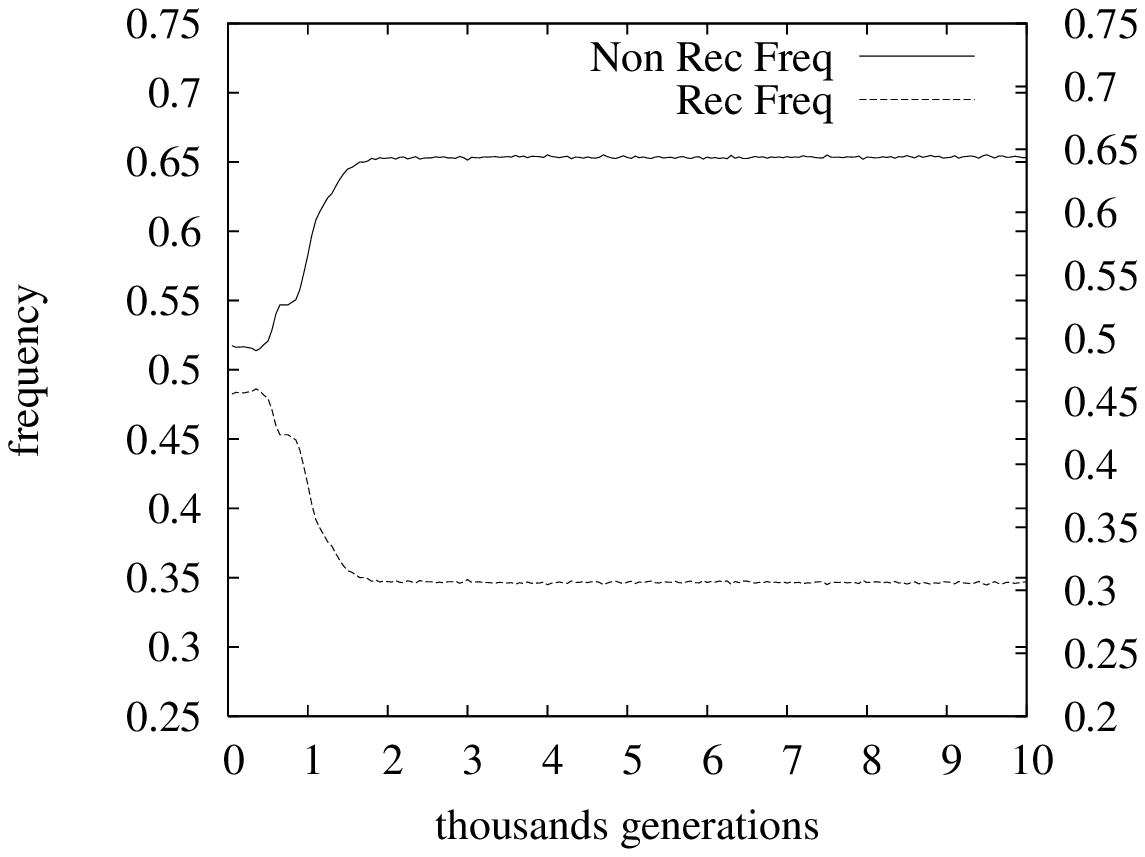} 
\end{tabular}

\caption{Regime of weak  competition ($J=1$, $\alpha = 2$, $R = 4$) and  flat static fitness ($\beta=100$, $\Gamma=14$). Left panel: final distribution; right panel: plot of frequencies of recombinants and non recombinants. Mating range $\Delta = 0$. Initial frequency of recombinants: 0.9; initial distribution parameters: $p=0$, $q=1$, initial population size $N_0 = 1000$, carrying capacity $K = 10000$. Annealing parameters: $\mu_0 =5\times 10^{-4}$, $ \mu_{\infty} = 10^{-6}$, $\tau = 1000$, $\delta = 100$. Total simulation time: 10000 generations.}

\label{fig2:fin-distr}

\end{figure}

Let us start with a simulation in a regime of weak competition ($J=1$, $\alpha = 2$, $R = 4$). The initial distribution is determined by the parameters $p = 0$, $q = 1$, $M = 0.9$ \emph{i.e.} there is a delta-peak of  recombinants in $x=0$ accounting for $90 \%$ of the population and a delta peak of non recombinants in $x=14$ populated by the remaining $10 \%$. Even if the initial frequency of the recombinants is very high, due to the twofold price of recombination, in a few generations their frequency decreases to about $50 \%$; the frequency of non recombinants will then keep on increasing up to $65 \%$ while the  recombinants will only represent the $35 \%$ of the population. The high mutation rate before $T = \tau$ causes the appearance of a new peak of recombinants near $x=0$ and a peak of non recombinants near $x=14$ both moving towards the center of the phenotypic space where the competition pressure is lower. When these two peaks are close enough to each other, because of the higher fertility of non recombinants, the peak of recombinants becomes extinct and the final distribution will typically show a peak of recombinants in $x=0$ and two peaks of non recombinants in $x=7$ and $x=14$. The final distribution and the evolution of the frequencies of recombinants and non recombinants is shown in Figure~\ref{fig2:fin-distr}.

\begin{figure}[ht!]
\begin{center}
\begin{tabular}{c}
\includegraphics[scale=0.7]{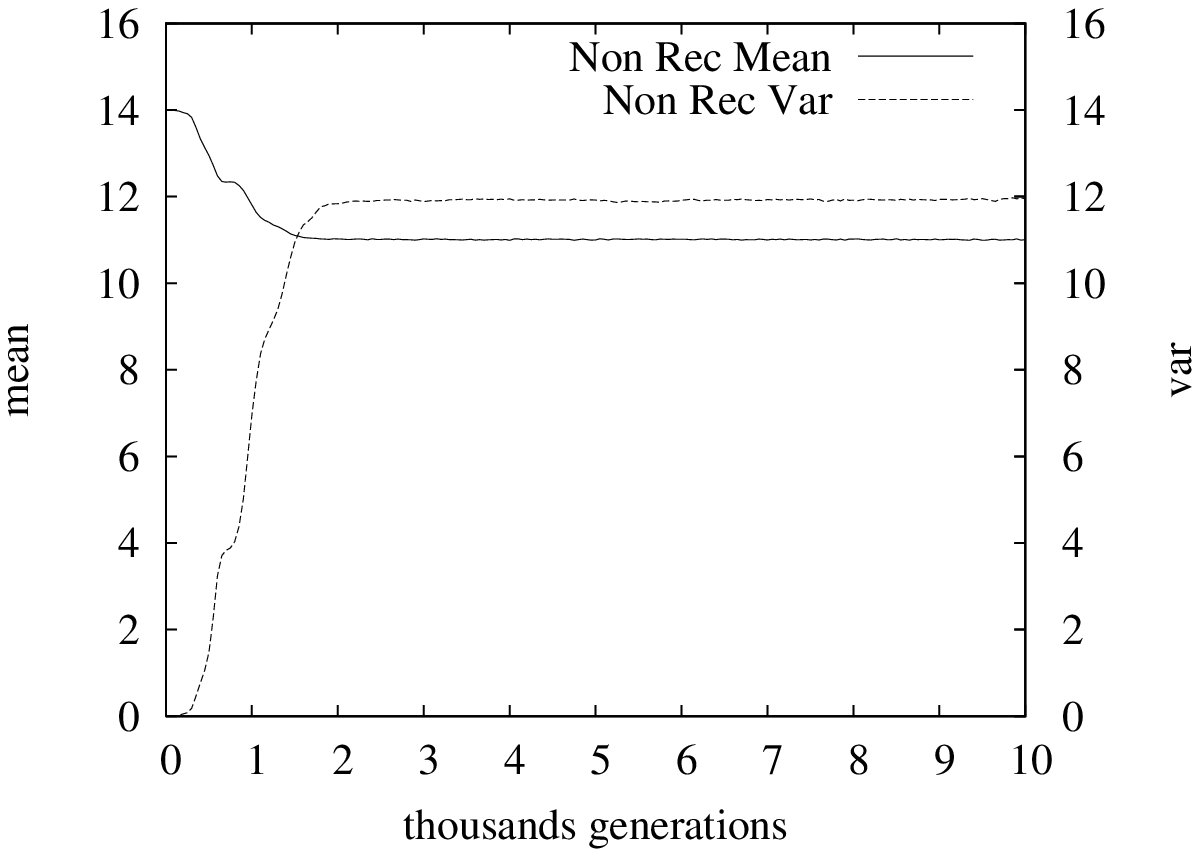} \\
\includegraphics[scale=0.7]{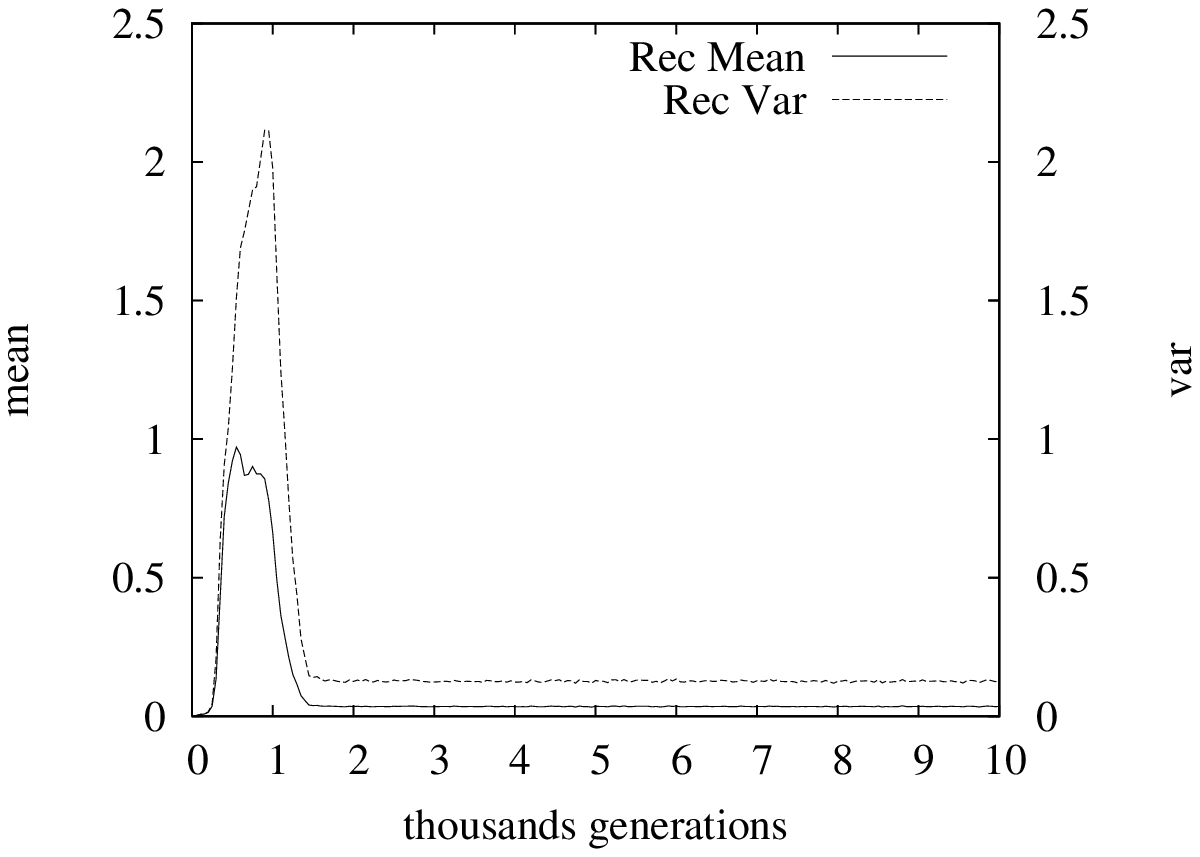} \\
\end{tabular} 
\end{center}

 \caption{Regime of weak  competition ($J=1$, $\alpha = 2$, $R = 4$) and  flat static fitness ($\beta=100$, $\Gamma=14$). Plots of mean and variance; left panel: non recombinants; right panel: recombinants. Mating range $\Delta = 0$. Initial frequency of recombinants: 0.9; initial distribution parameters: $p=0$, $q=1$, initial population size $N_0 = 1000$, carrying capacity $K = 10000$. Annealing parameters: $\mu_0 =5\times 10^{-4}$, $ \mu_{\infty} = 10^{-6}$, $\tau = 1000$, $\delta = 100$. Total simulation time: 10000 generations.}

\label{fig2:var-mean}

\end{figure}

The evolutionary dynamics is well illustrated in the plots of mean and variance of the distributions of recombinants and non recombinants shown in Figure~\ref{fig2:var-mean}. The mean of the non recombinant distribution decreases from $x=14$ to $x=11$ as the second peak moves towards the center of the phenotypic space; the variance accordingly increases from $var = 0$ (typical value of a delta peak) to $var = 12$ as the second peak becomes more and more distant from the first one. As for the recombinant population, the mean and variance are both zero at the beginning (delta peak in $x=0$) and they both increase as a second peak appears and moves towards the center of the space; when this peak disappears, however, variance and mean vanish again.

In order to study the dependence on the initial conditions, we now repeat the simulation setting $p = q = 0.5$ \emph{i.e.} both recombinants and non recombinants are binomially distributed with mean $\bar{x} = 7$. At the beginning of the run, the recombinants react to the competition pressure by widening their frequency distribution that then splits in two peaks in $x=0$ and $x=14$; a single peak of non recombinants in the middle of the phenotypic space is also present. Competition later induces the appearance of colonies of non recombinants near the ends of the phenotypic space. The non recombinants thanks to their higher fertility tend to overwhelm the recombinants. Several different scenarios can be observed according to the ability of recombinants to resist eradication from the ends of the phenotypic space:

\begin{enumerate}

\item The recombinants keep their predominance at both ends of the phenotypic space. Stationary distribution:  peak of recombinants at both ends, peak of non recombinants in the center of the space. Frequency of recombinants: about $70 \%$.

\item The recombinants coexist with non recombinants at one end of the phenotypic space. Stationary distribution: peak of recombinants at one end, peak of non recombinants covering a pair of phenotypes in the middle of the space, peak of recombinants and non recombinants at the other hand of the space. Frequency of recombinants: about $40 \%$.

\item The recombinants are eradicated from one end of the phenotypic space. Stationary distribution: peak of recombinants at one end of the phenotypic space, peak of non recombinants in the center, peak of non recombinants at the other end of the space. Frequency of recombinants: about $35 \%$.

\item The recombinants are eradicated from one end of the space and obliged to coexistence with non recombinants at the other end. Stationary distribution: peak of non recombinants at one end, peak of non recombinants in the center, peaks of recombinants and non recombinants at the other end. Frequency of recombinants: about $20 \%$.

\end{enumerate}

\begin{figure}[ht!]
\begin{center}
\begin{tabular}{ccc}
\hspace{-2 cm} & \includegraphics[scale=0.6]{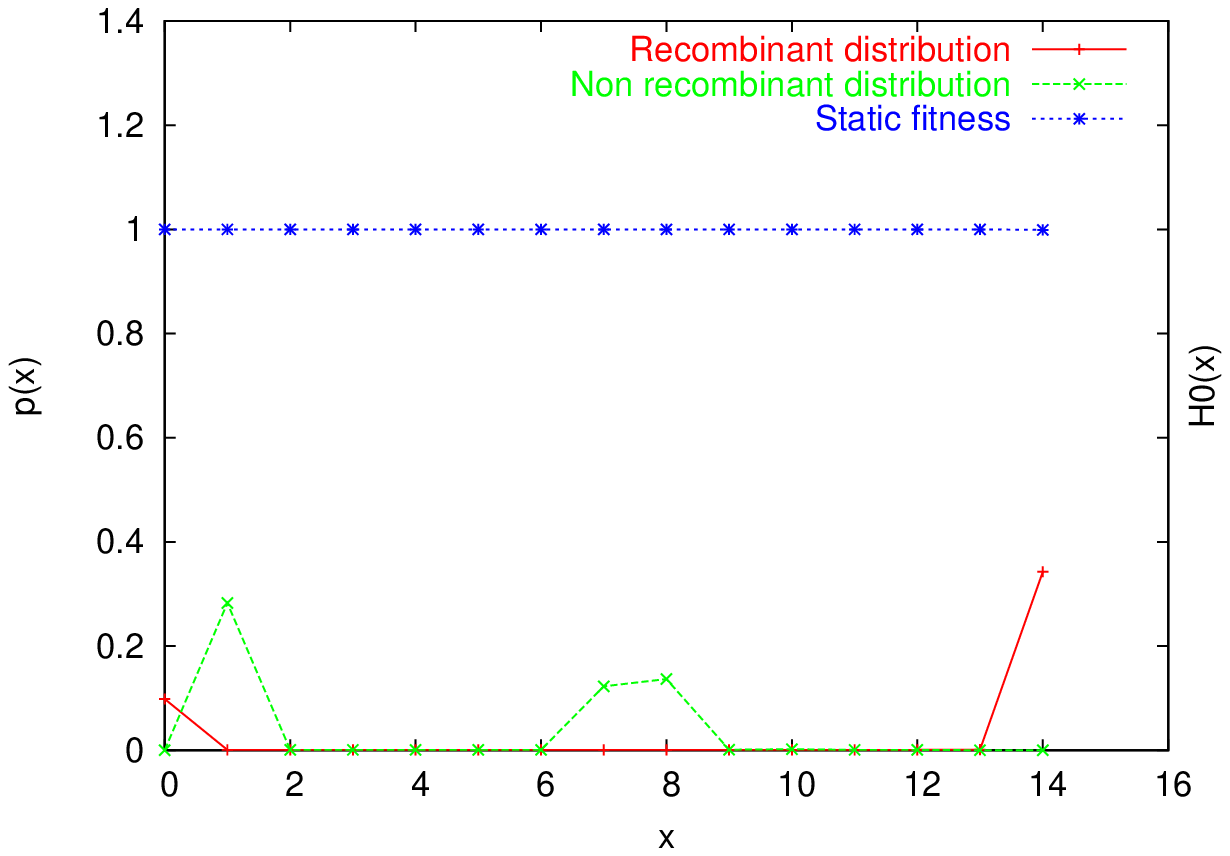} &
\includegraphics[scale=0.6]{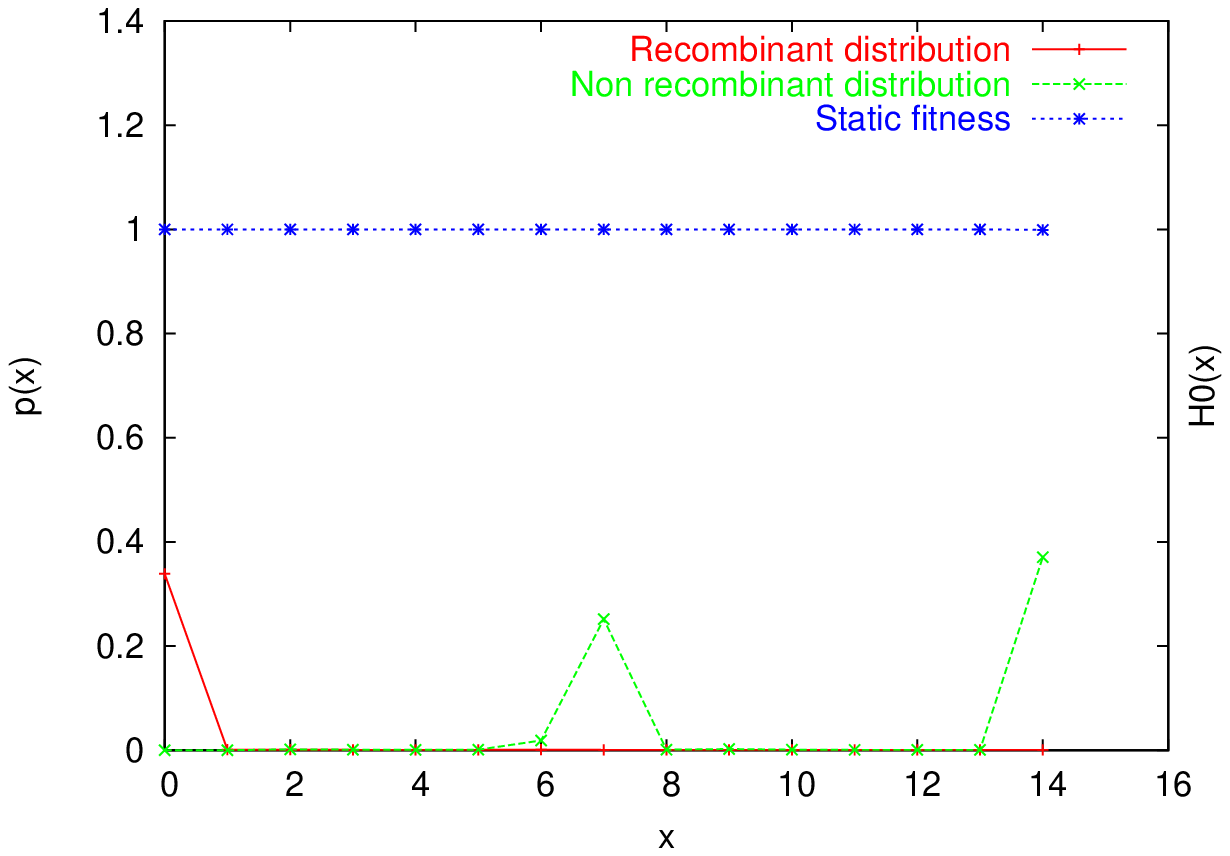} \\
\hspace{-2 cm} & \includegraphics[scale=0.6]{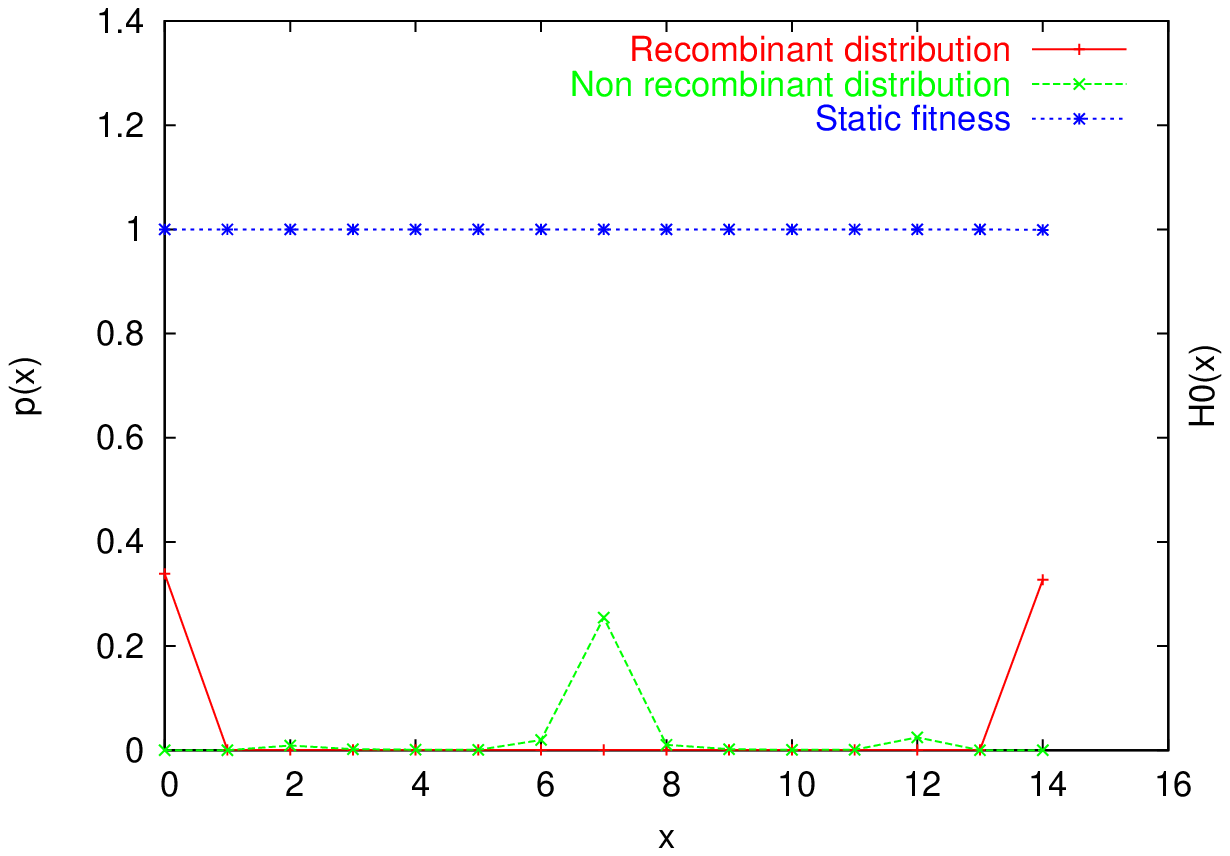} &
\includegraphics[scale=0.6]{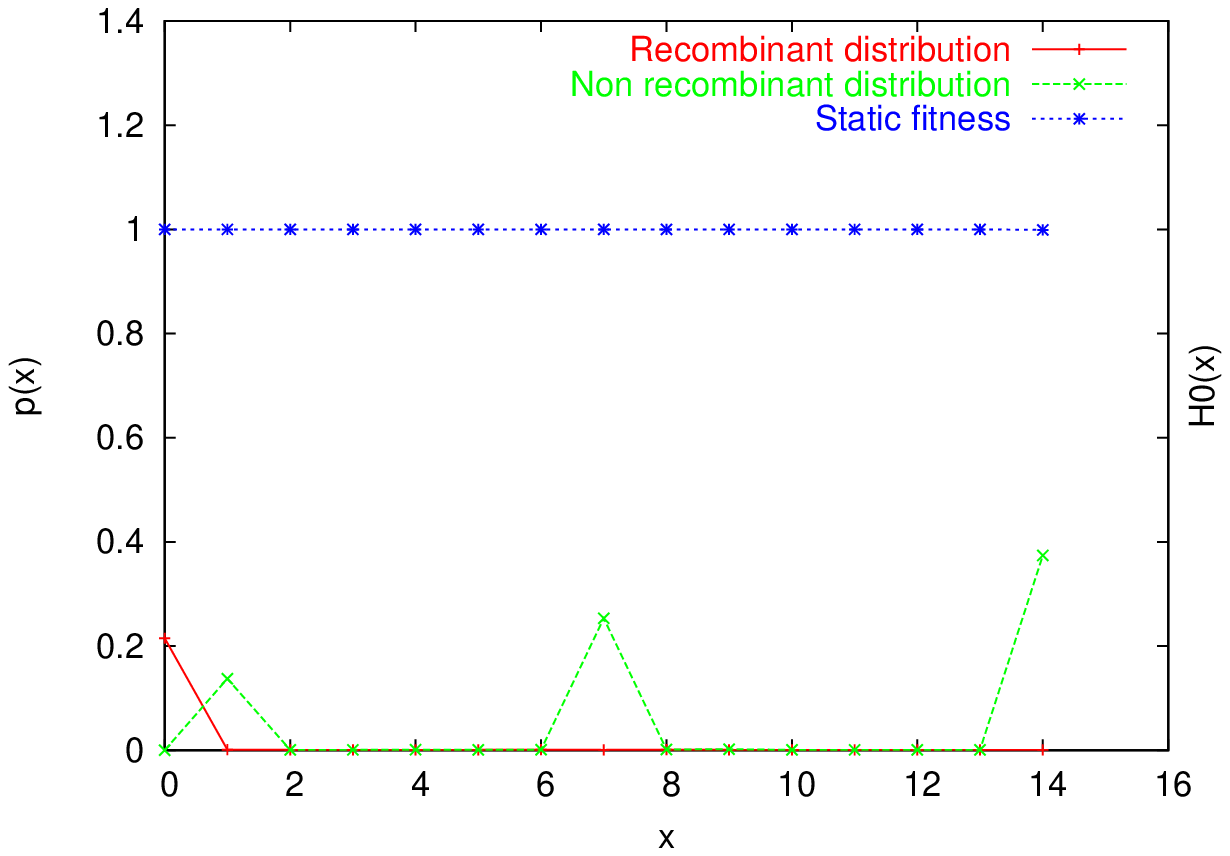} \\

\end{tabular}

 \caption{Examples of stationary distributions in a regime of weak  competition ($J=1$, $\alpha = 2$, $R = 4$) and  flat static fitness ($\beta=100$, $\Gamma=14$). Mating range: $\Delta = 0$. Initial frequency of recombinants: 0.9; initial distribution parameters: $p = q= 0.5$, initial population size $N_0 = 1000$, carrying capacity $K = 10000$. Annealing parameters: $\mu_0 =5\times 10^{-4}$, $ \mu_{\infty} = 10^{-6}$, $\tau = 1000$, $\delta = 100$. Total simulation time: 10000 generations.}

\label{fig3:examples}
\end{center}
\end{figure}

Examples of the possible scenarios are shown in Figure~\ref{fig3:examples}.

\begin{figure}[ht!]
\begin{center}
\begin{tabular}{c}
\includegraphics[scale=0.7]{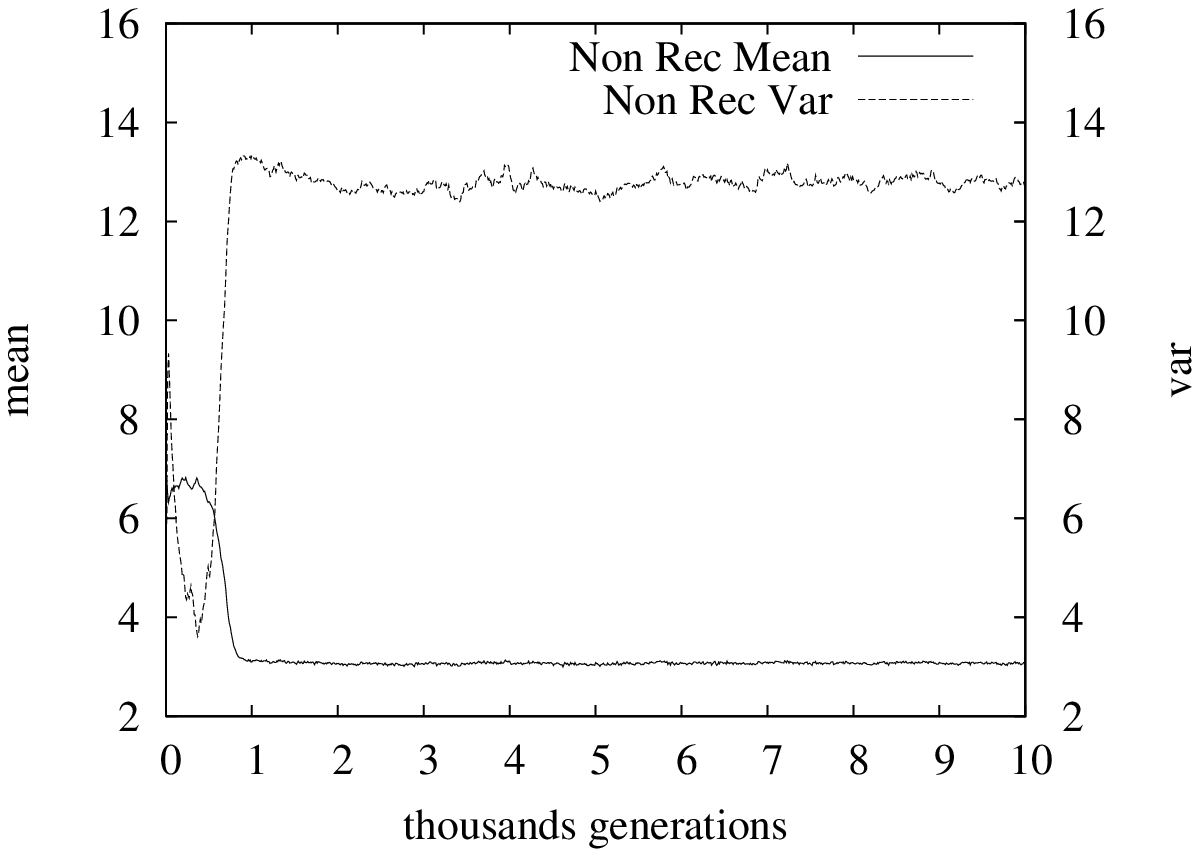} \\
\includegraphics[scale=0.7]{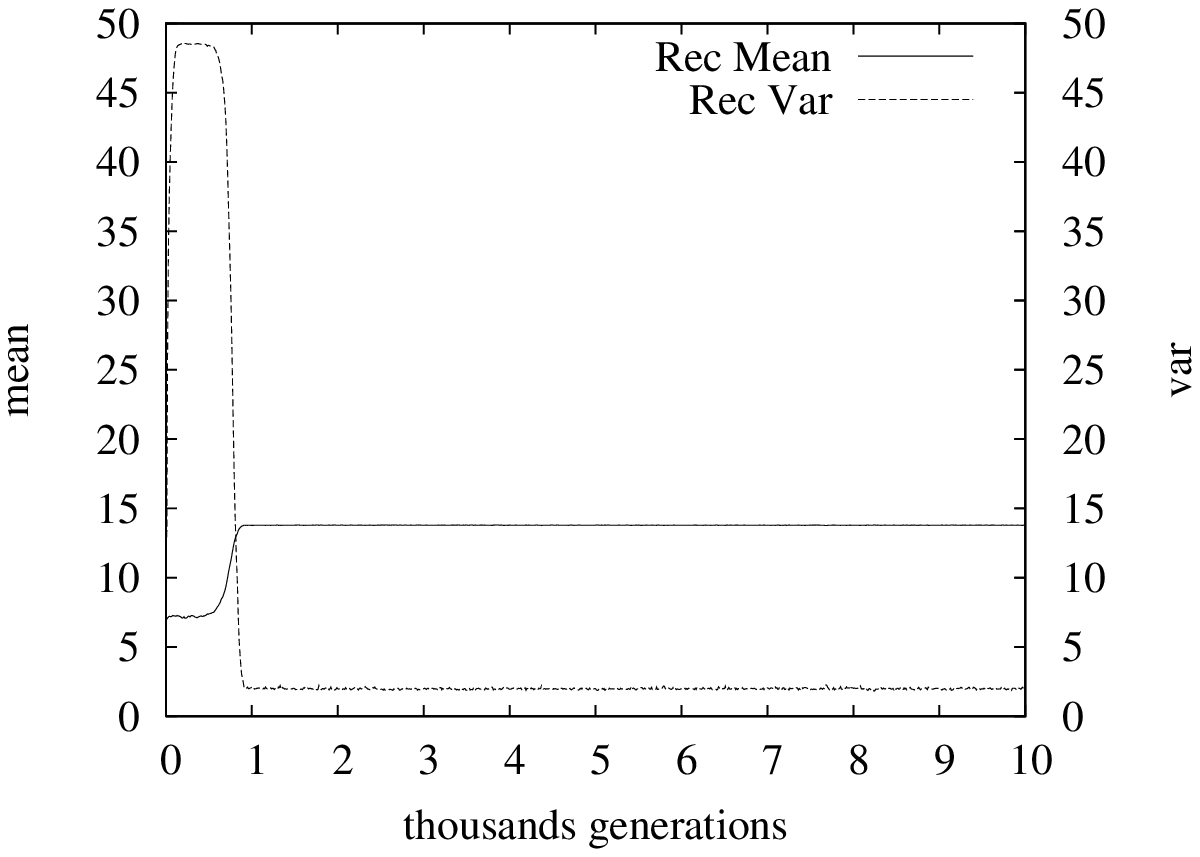} \\
\end{tabular}
\end{center}
 \caption{Regime of weak  competition ($J=1$, $\alpha = 2$, $R = 4$) and  flat static fitness ($\beta=100$, $\Gamma=14$). Plots of mean and variance; left panel: non recombinants; right panel: recombinants. Mating range $\Delta = 0$. Initial frequency of recombinants: 0.9; initial distribution parameters: $p= q =0.5$, initial population size $N_0 = 1000$, carrying capacity $K = 10000$. Annealing parameters: $\mu_0 =5\times 10^{-4}$, $ \mu_{\infty} = 10^{-6}$, $\tau = 1000$, $\delta = 100$. Total simulation time: 10000 generations.}

\label{fig3:var-mean}

\end{figure}

We now discuss the plots of mean and variance in a simulation where the final distribution shows two peaks of non recombinants in $x=0$ and $x=7$ and a peak of recombinants in $x=14$. At the beginning of the run two peaks of recombinants appear at the ends of the phenotypic space, while the non recombinants are concentrated in one peak in the middle of the space that tends to become narrower and narrower. This explains why at the beginning of the run the variance of recombinants increases while that of non recombinants decreases. Later on the recombinants are eradicated from the $x=0$ end of the phenotypic space and they survive in a single delta peak in $x=14$: this explains why their average increases up to 14 while their variance vanishes. The mean of non recombinants conversely decreases to about $x = 4$ (approximately halfway between the peaks at $x=0$ and $x=7$) and their variance increases to $var = 13$. These plots are shown in Figure~\ref{fig3:var-mean}.

The simulations with $p = q = 0.5$ are sensitive to the initial frequency of recombinants. When the initial frequency is $90 \%$, the  recombinants become completely extinct only very seldom and they typically form peaks in one or both ends of the phenotypic space. With an initial frequency $M =0.5$, however, the recombinants become extinct in 4 runs out of 10 and the situation is even worse with $M = 0.2$ with extinction in 9 runs out of 10. When the recombinants completely disappear the final distribution shows three peaks of non recombinants in $x=0$, $x=7$, and $x=14$ so as to minimize competition. In the following sections we will show that the survival of recombinants can be indeed guaranteed by setting a sufficiently high competition intensity or a sufficiently low mutation rate.

\subsubsection{Strong competition}

\begin{figure}[ht!]

\begin{tabular}{ccc}
\hspace{-2 cm} &
\includegraphics[scale=0.65]{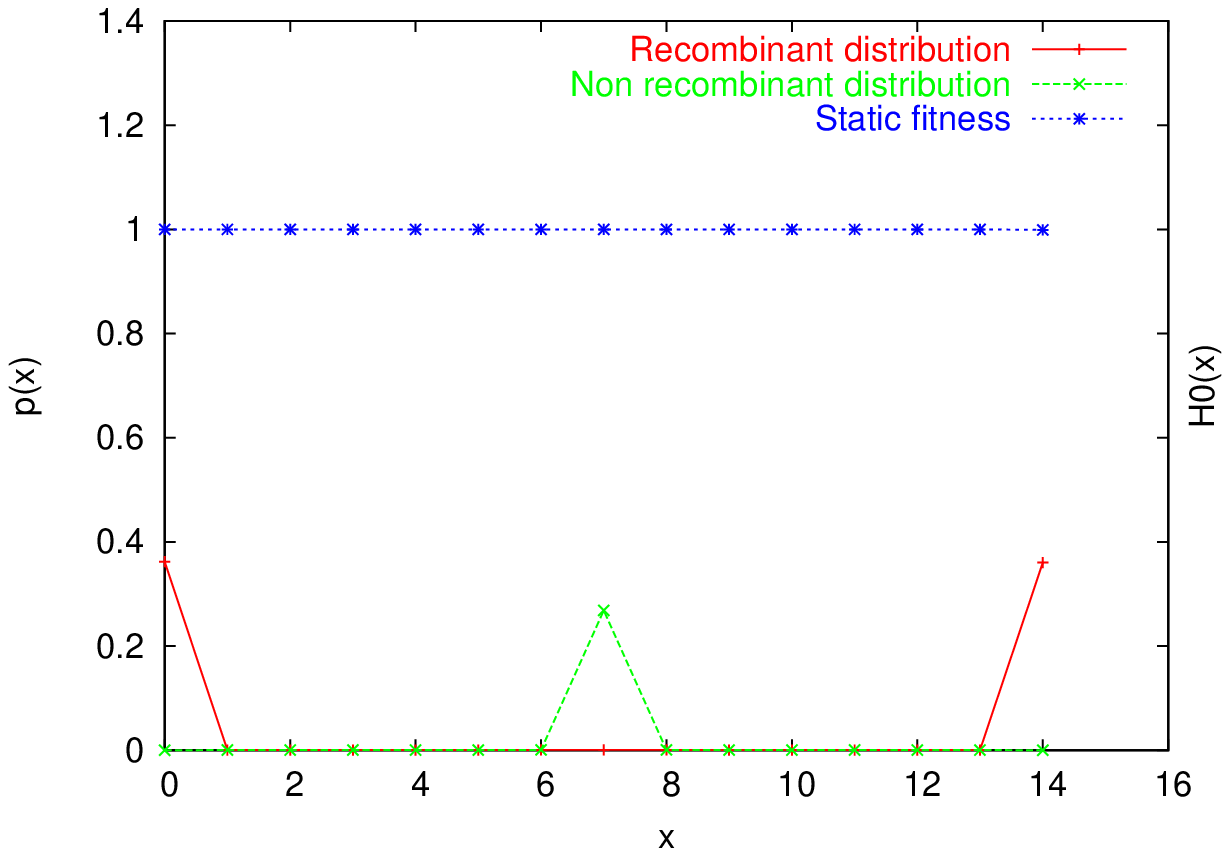} &
\includegraphics[scale=0.65]{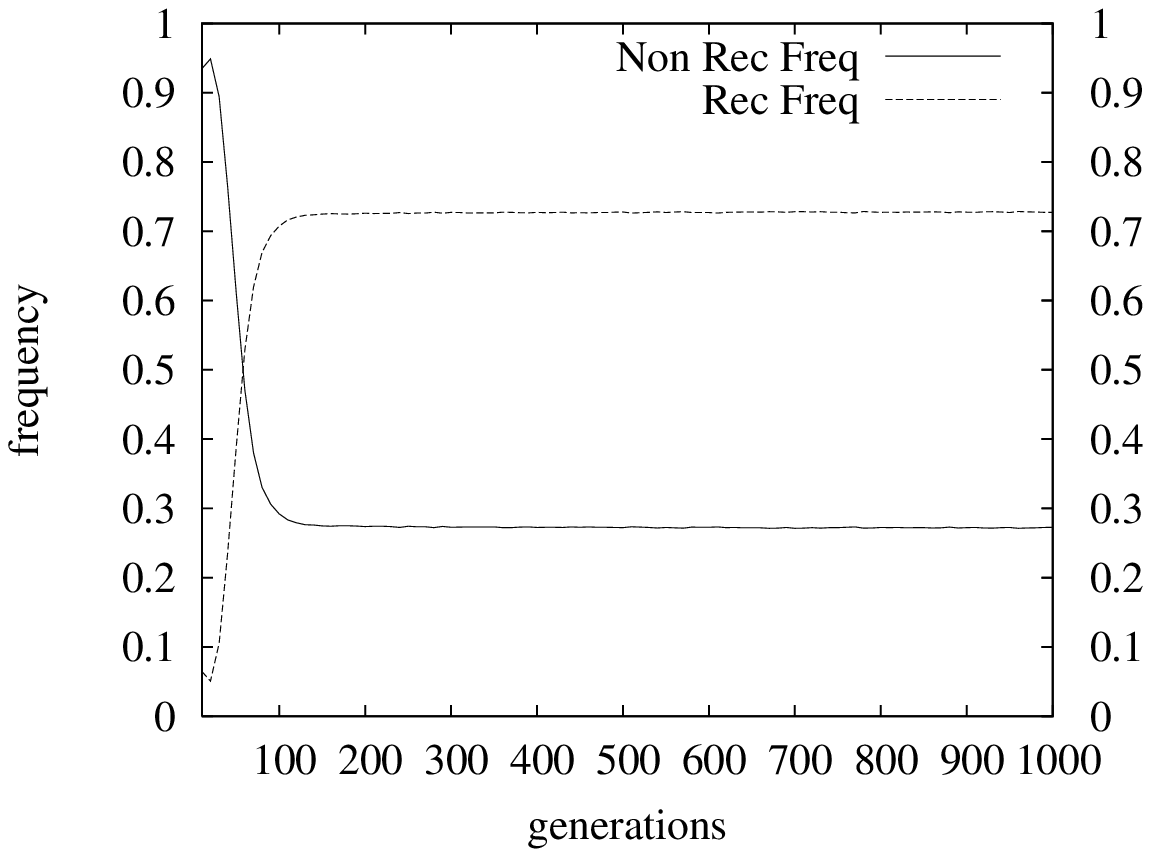}
\end{tabular}

 \caption{Regime of strong  competition ($J=5$, $\alpha = 2$, $R = 4$) and  flat static fitness ($\beta=100$, $\Gamma=14$). The recombinants establish peaks in both ends of the phenotypic space. Left panel: final distribution; right panel: plots of frequency of recombinants and non recombinants. Mating range $\Delta = 0$. Initial frequency of recombinants: 0.2; initial distribution parameters: $p= q =0.5$, initial population size $N_0 = 1000$, carrying capacity $K = 10000$. Annealing parameters: $\mu_0 =5\times 10^{-4}$, $ \mu_{\infty} = 10^{-6}$, $\tau = 1000$, $\delta = 100$. Total evolution time: 10000 generations; for the sake of clarity the plots refer only to the first 1000 generations. }

\label{fig4:final_distr}

\end{figure}

In order to study the effect of competition we repeated the simulation with $p = q = 0.5$, $M = 0.2$ setting $J = 5$. Contrary to the simulation with $J = 1$ where the recombinants become extinct in 9 runs out of 10, when $J = 5$ the recombinants never become extinct and they form peaks in both ends of the phenotypic space with a final frequency of about $72-73 \%$, or at least in one end of the space with with a final frequency of $36 \%$. In this simulation under the pressure of competition the non recombinants form a peak in $x=1$, a peak spanning $x=6$, $x=7$, $x=8$ and a peak in $x=13$. The recombinants, thanks to recombination, quickly widen their distribution and they reach the $x=0$ and $x=14$ ends (where competition is minimal) before the non recombinants. If the recombinants manage to establish colonies of big size in $x=0$ and $x=14$ before non recombinant mutants reach these locations, the peaks of non recombinants in $x=1$ and $x=13$ become extinct and the final distribution will feature two peaks of recombinants in $x=0$ and $x=14$ and a peak of non recombinants in $x=7$. If, conversely, non recombinant mutants reach $x=0$ or $x=14$ before the recombinant peak there has become sufficiently populated, the non recombinants, owing to their larger fertility, become dominant and they drive the recombinants to extinction. The final distribution in this situation shows a peak of recombinants in one end  and two peaks of non recombinants in the other end and in the middle of the phenotypic space. The final distribution and the plots of the frequencies of recombinants and non recombinants are shown in Figure~\ref{fig4:final_distr}.

\begin{figure}[ht!]
\begin{center}
\begin{tabular}{c}
\includegraphics[scale=0.7]{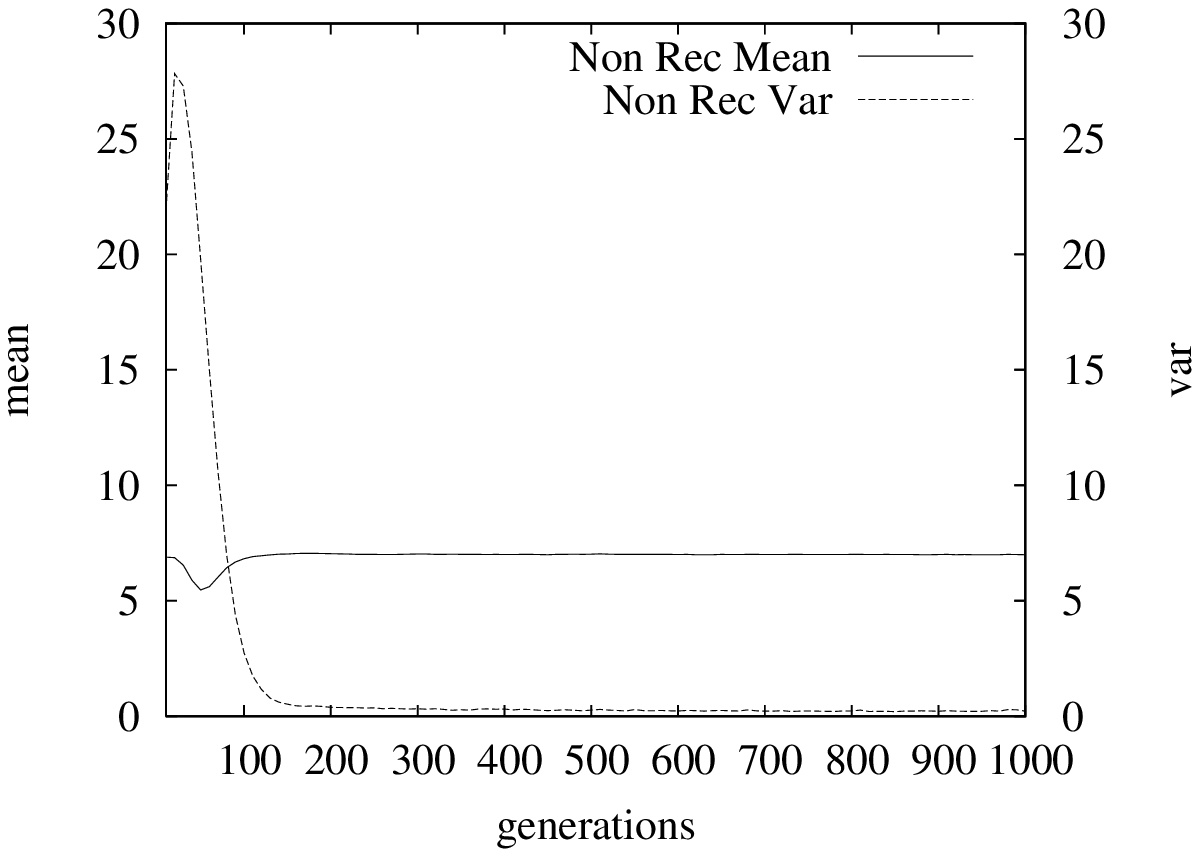} \\
\includegraphics[scale=0.7]{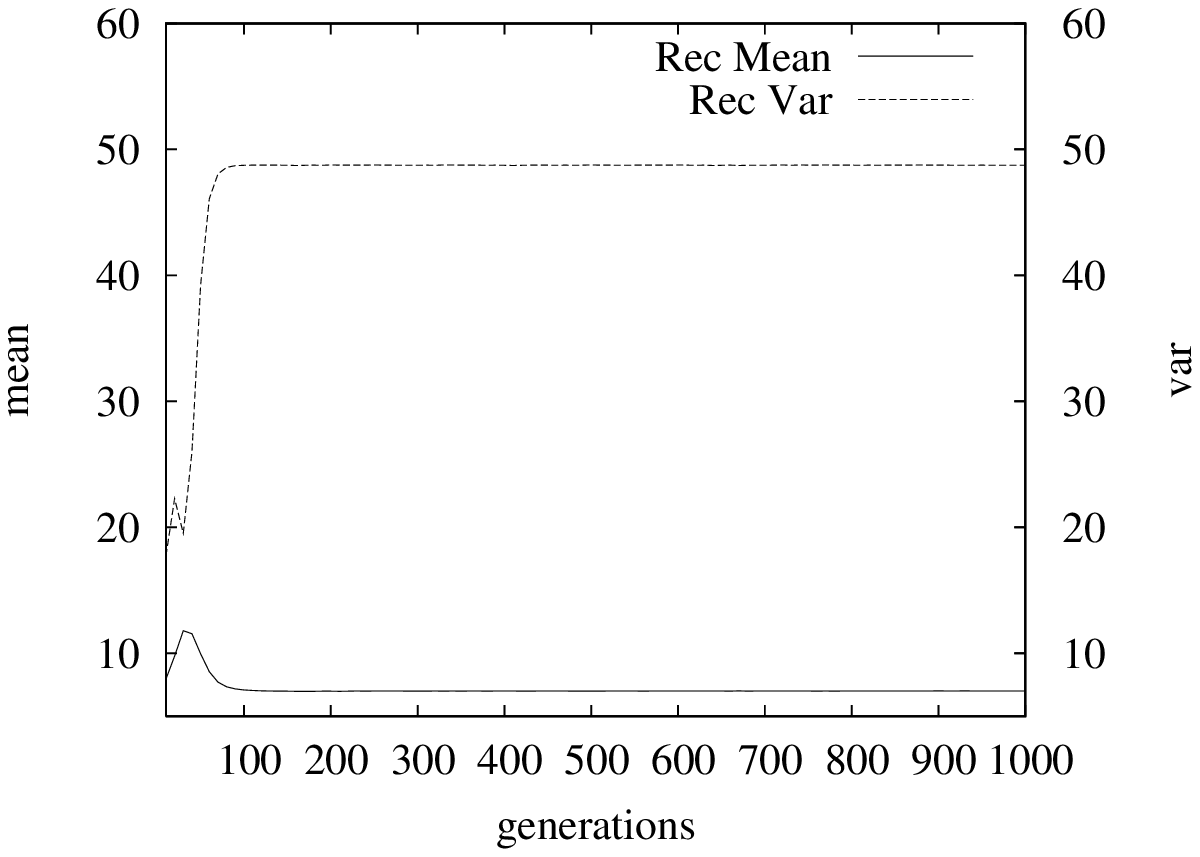} \\
\end{tabular}
\end{center}

 \caption{Regime of strong  competition ($J=5$, $\alpha = 2$, $R = 4$) and  flat static fitness ($\beta=100$, $\Gamma=14$). The recombinants establish peaks in both ends of the phenotypic space. plots of mean and variance. Left panel: non recombinants; right panel: recombinants. Mating range: $\Delta = 0$.  Initial frequency of recombinants: 0.2; initial distribution parameters: $p= q =0.5$, initial population size $N_0 = 1000$, carrying capacity $K = 10000$. Annealing parameters: $\mu_0 =5\times 10^{-4}$, $ \mu_{\infty} = 10^{-6}$, $\tau = 1000$, $\delta = 100$. Total evolution time: 10000 generations; for the sake of clarity the plots refer only to the first 1000 generations. }

\label{fig4:var-mean}

\end{figure}

The evolutionary dynamics can be followed also by plotting variance and mean of recombinant and non recombinant distributions. The plots are shown in Figure~\ref{fig4:var-mean}. The variance of recombinants increases when the distribution splits in two peaks in $x=0$ and $x=14$, while the variance of non recombinants almost vanishes when the peaks in $x=1$ and $x=13$ become extinct and the distribution of non recombinants reduces to a single sharp peak. The mean of the recombinant and non recombinant distributions conversely make small oscillations around $x=7$ because the distributions are deformed in a symmetrical way.

\subsubsection{The role of mutation}

\label{flat-fit-mut}

In the preceding section we have shown that a high intensity of competition induces a fast splitting of the recombinant distribution so that recombinants can colonize the favorable positions $x=0$ and $x=14$  earlier than the non recombinants. In this way the recombinants survive in almost all the runs of the program and attain a stable high level. We will now show that a similar effect can be obtained by keeping the mutation rate at a very low level during the whole simulation. In this more realistic case, in fact, the velocity of movement of non recombinants, that depends mainly on the mutation rate, will be lower than the velocity of recombinants that is related to recombination. As a result, the recombinants reach the ends of the phenotypic space and establish there conspicuous colonies before the arrival of non recombinants. In other words the recombinants counteract the higher fertility of non recombinants by occupying more favorable locations in the phenotypic space.

\begin{figure}[ht!]

\begin{tabular}{ccc}
\hspace{-2 cm} &
\includegraphics[scale=0.65]{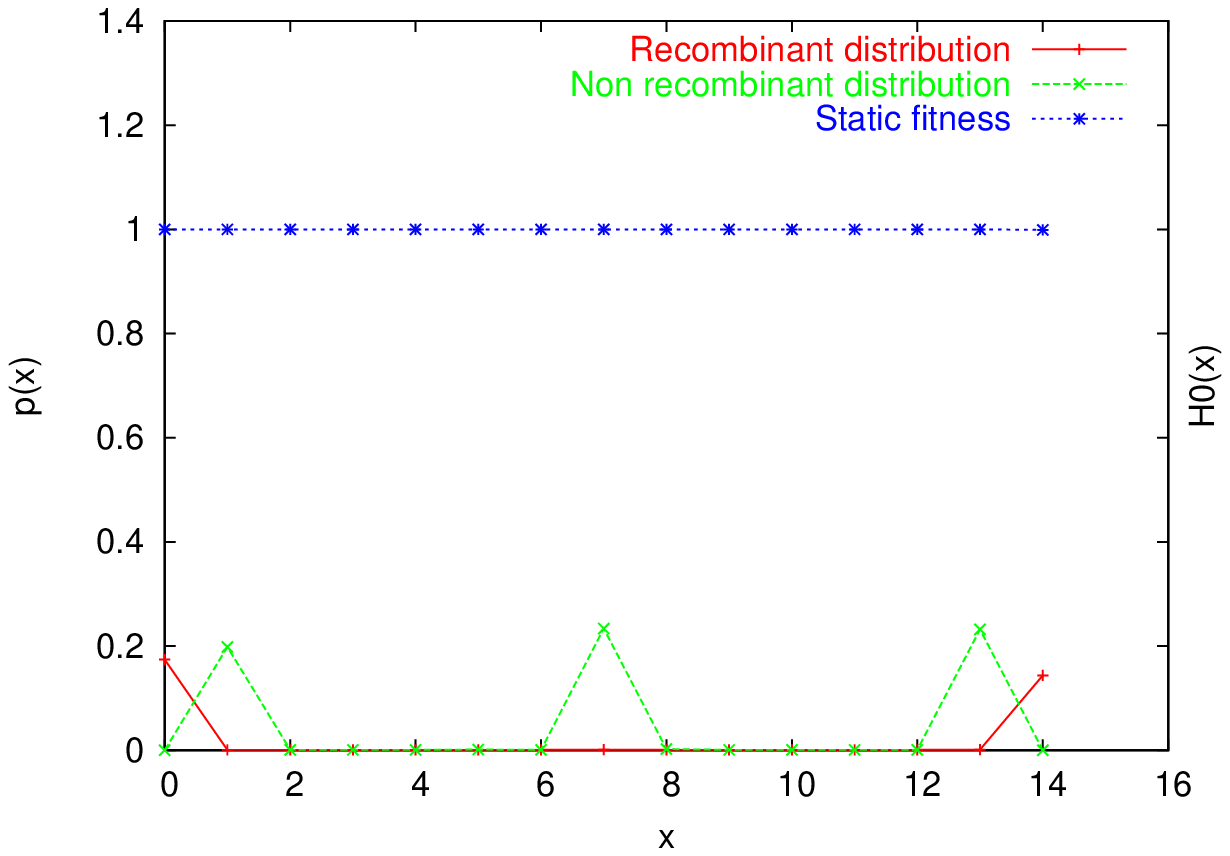} &
\includegraphics[scale=0.65]{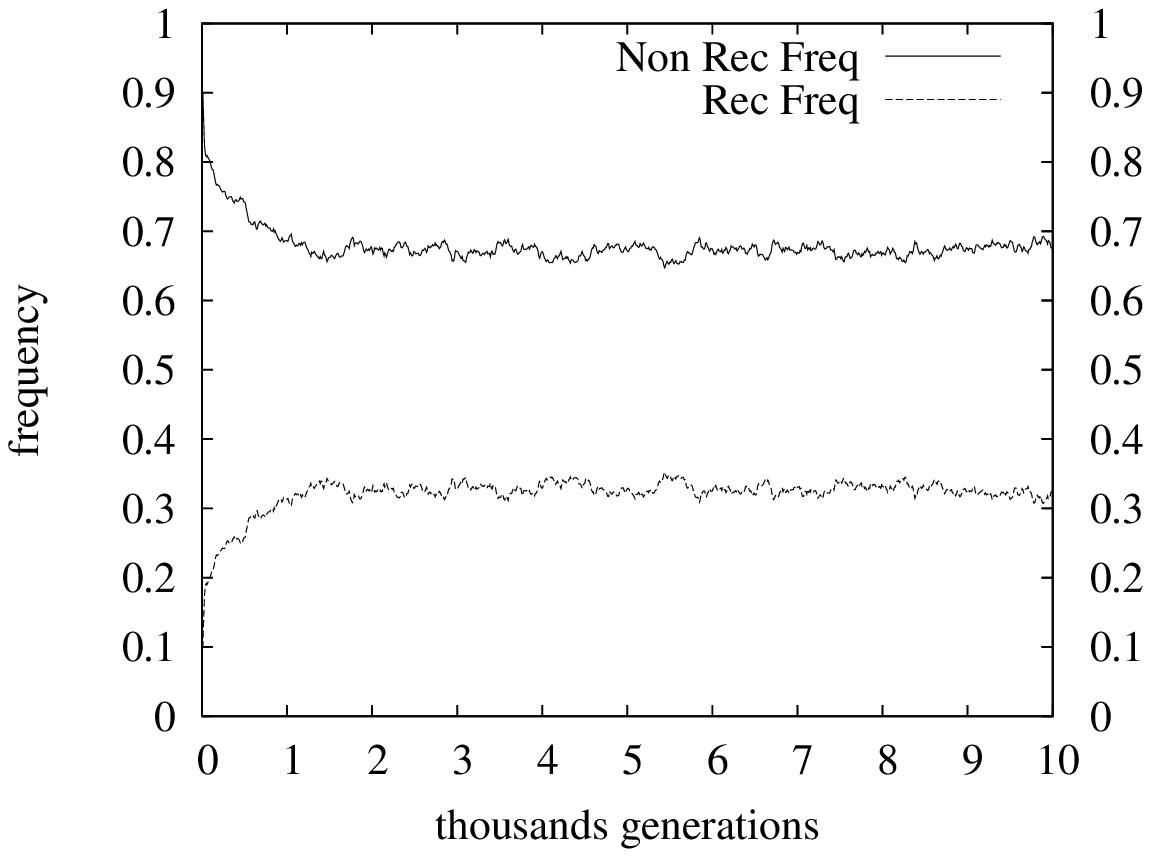}
\end{tabular}

 \caption{Regime of low mutation and small initial frequency of recombinants: stable coexistence of recombinants and non recombinants at both ends of the phenotypic space. Weak  competition ($J=1$, $\alpha = 2$, $R = 4$) and  flat static fitness ($\beta=100$, $\Gamma=14$). Left panel: the final distribution; right panel: frequency plots of recombinants and non recombinants. Mating range: $\Delta = 0$.  Initial frequency of recombinants: 0.2; initial distribution parameters: $p= q =0.5$, initial population size $N_0 = 1000$, carrying capacity $K = 10000$. Annealing parameters: $\mu_0 =10^{-5}$, $ \mu_{\infty} = 10^{-6}$, $\tau = 1000$, $\delta = 100$. Total evolution time: 10000 generations.}

\label{fig5:final_distr}

\end{figure}

Let us start with a simulation with $\mu_0 = 10^{-5}$, $\mu_{\infty} = 10^{-6}$, $J=1$, $p = q = 0.5$ and $M = 0.2$ that can be compared with a similar simulation at the end of Section~\ref{weak-comp} where the initial mutation rate was $5\times 10^{-4}$. In that situation the recombinants became extinct in 9 runs out of 10, whereas in the case of low mutation rate they always survive and attain a final level ranging from $20 \%$ to $33 \%$. At the beginning of the run the non recombinants create three peaks in $x=1$, $x=7$ and $x=13$. Owing to the very low mutation rate however, the non recombinants do not have enough time to reach $x=0$ and $x=14$  before the recombinants. The typical scenario is therefore characterized by a stable coexistence of recombinants and non recombinants at both ends of the phenotypic space: coexistence of recombinants in $x=0$ and non recombinants in $x=1$; a single peak of non recombinants in $x=7$; coexistence of recombinants in $x=14$ and non recombinants in $x=13$. This scenario together with the frequency plots of recombinants and non recombinants is shown in Figure~\ref{fig5:final_distr}.

\begin{figure}[ht!]
\begin{center}
\begin{tabular}{c}
\includegraphics[scale=0.7]{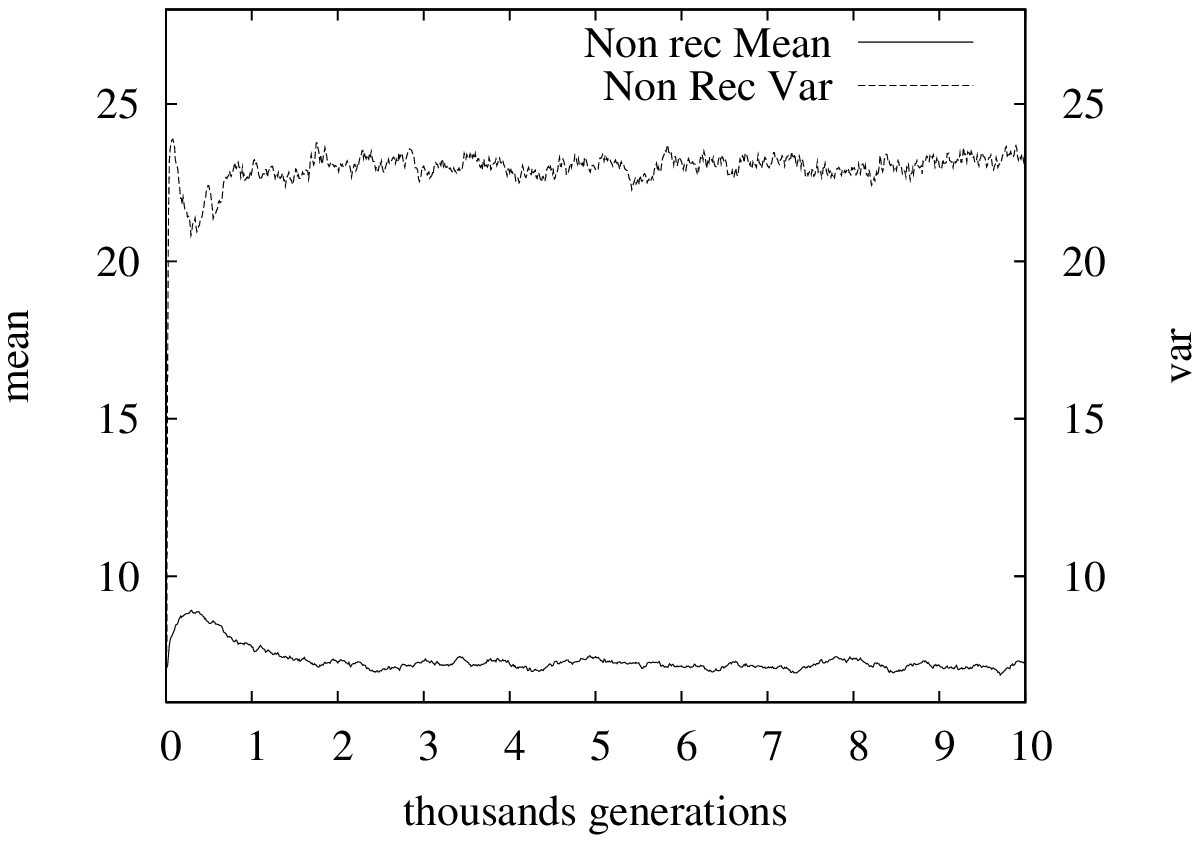} \\
\includegraphics[scale=0.7]{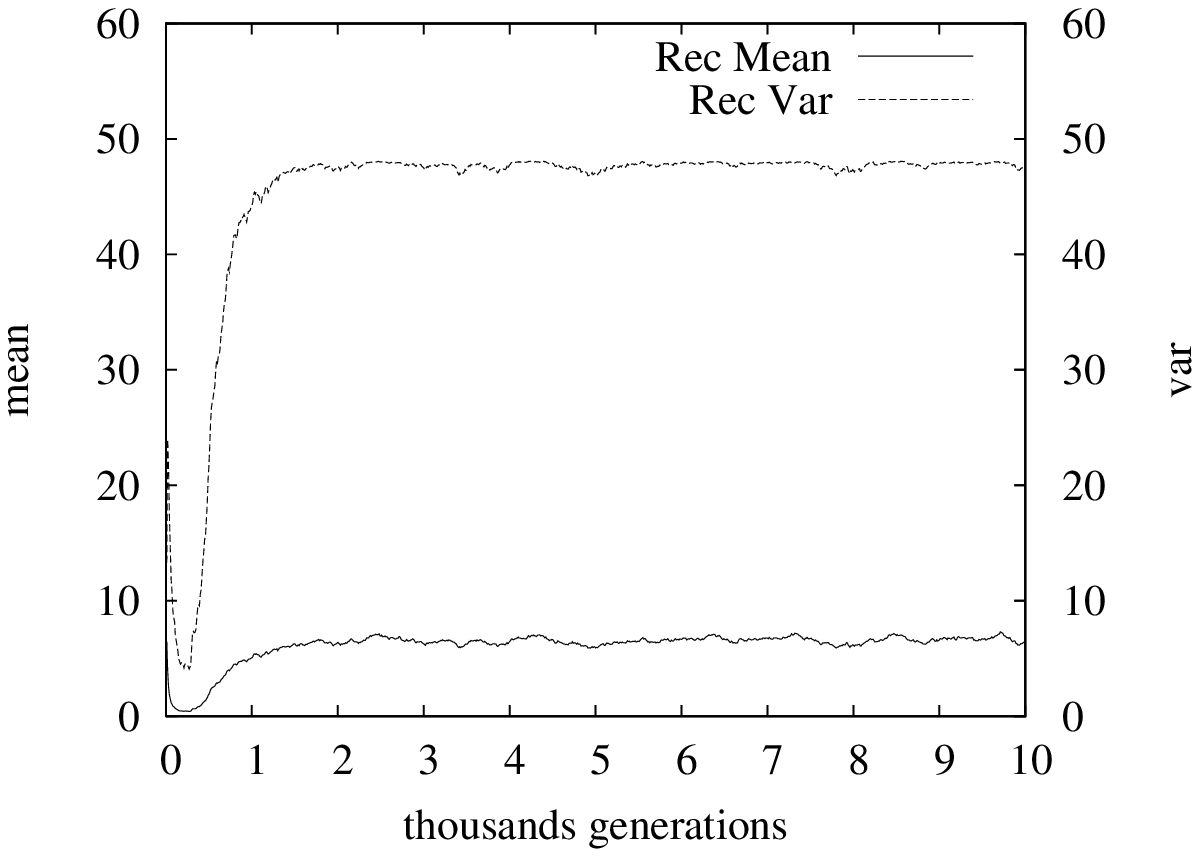} \\
\end{tabular}
\end{center} 
  \caption{Regime of low mutation and small initial frequency of recombinants: stable coexistence of recombinants and non recombinants at both ends of the phenotypic space. Plots of variance and mean of recombinant and non recombinant distributions. Weak  competition ($J=1$, $\alpha = 2$, $R = 4$) and  flat static fitness ($\beta=100$, $\Gamma=14$). Mating range: $\Delta = 0$. Initial frequency of recombinants: 0.2; initial distribution parameters: $p= q =0.5$, initial population size $N_0 = 1000$, carrying capacity $K = 10000$. Annealing parameters: $\mu_0 =10^{-5}$, $ \mu_{\infty} = 10^{-6}$, $\tau = 1000$, $\delta = 100$. Total evolution time: 10000 generations.}

\label{fig5:var-mean}

\end{figure}

The plots of mean and variance of the recombinant and non recombinant distribution provide a good description of the evolutionary dynamics. They are shown in Figure~\ref{fig5:var-mean}. The shape of the mean and variance plots of recombinants depend on the fact that the two peaks do not appear in the same time. In this example the peak in $x=0$ appears first and this implies a decrease both in mean and variance; after the formation of the peak in $x=14$ mean and variance increase again. The variance of non recombinants conversely, increases as the distribution becomes wider and splits in two peaks near the ends of the space; the appearance of the central peak then causes a little decrease in variance.

\begin{figure}[ht!]

\begin{tabular}{ccc}
\hspace{-2 cm} &
\includegraphics[scale=0.65]{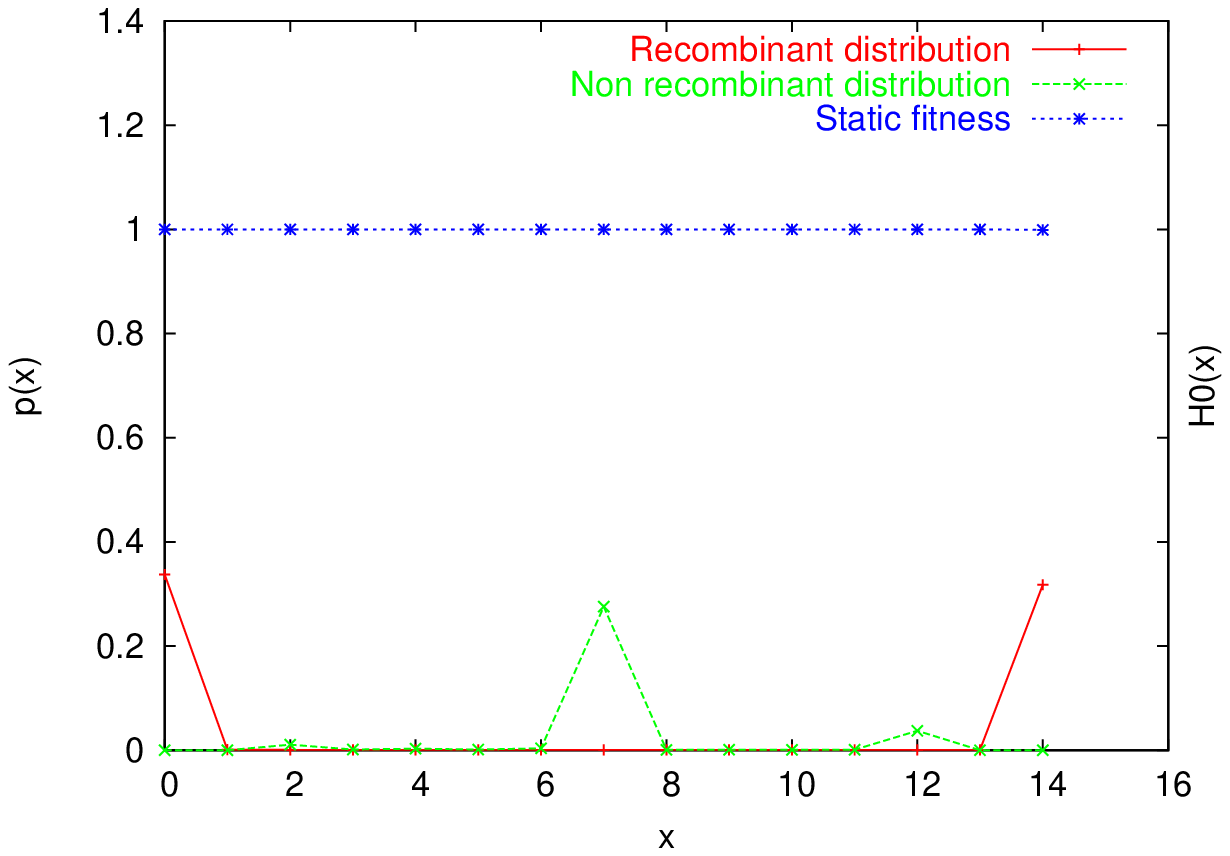} &
\includegraphics[scale=0.65]{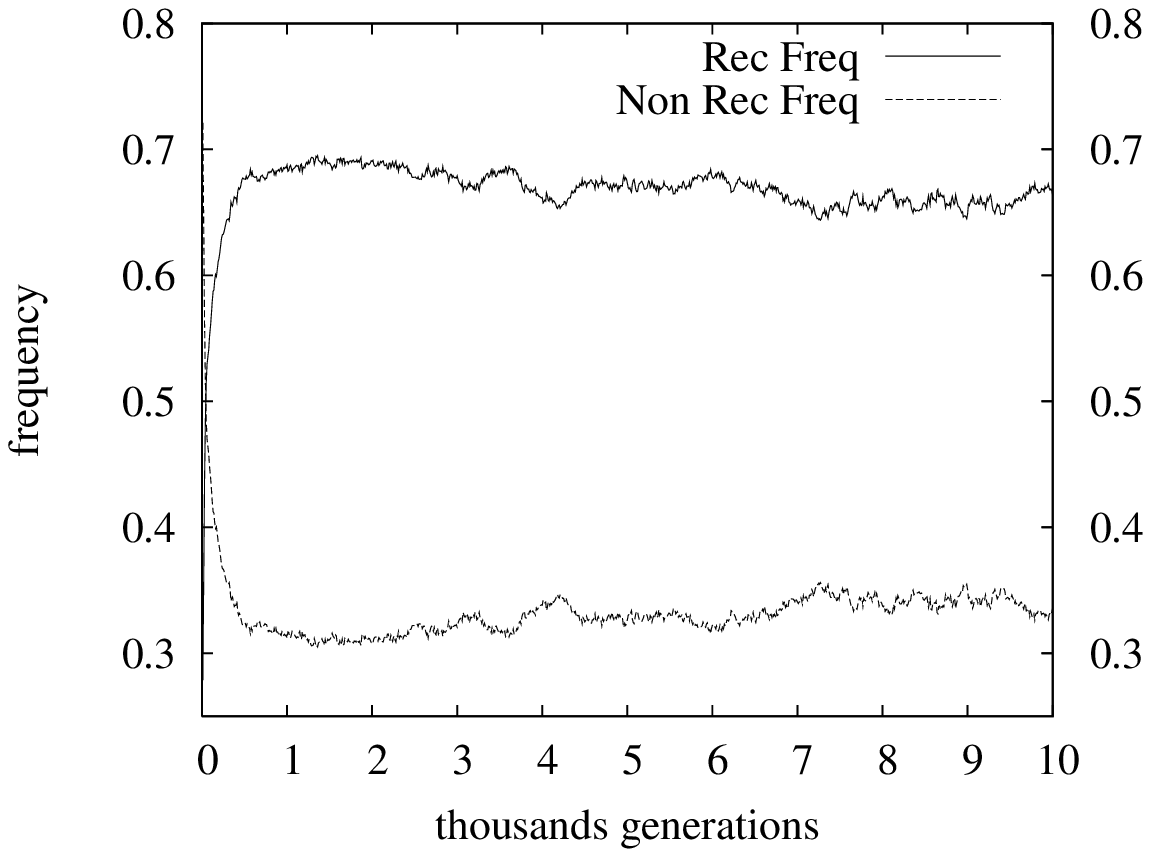}
\end{tabular}

 \caption{Regime of low mutation and small initial frequency of recombinants: the recombinants dominate both ends of the phenotypic space. Weak  competition ($J=1$, $\alpha = 2$, $R = 4$) and  flat static fitness ($\beta=100$, $\Gamma=14$). Left panel: the final distribution; right panel: frequency plots of recombinants and non recombinants. Mating range: $\Delta = 0$. Initial frequency of recombinants: 0.5; initial distribution parameters: $p= q =0.5$, initial population size $N_0 = 1000$, carrying capacity $K = 10000$. Annealing parameters: $\mu_0 =10^{-5}$, $ \mu_{\infty} = 10^{-6}$, $\tau = 1000$, $\delta = 100$. Total evolution time: 10000 generations.}

\label{fig6:final_distr}

\end{figure}

The simulation in a regime of low mutation was repeated for a larger initial frequency of recombinants $M = 0.5$. Competition induces both distributions to become wider, but recombination induces the recombinants to quickly reach the $x=0$ and $x=14$ positions whereas non recombinants cannot go beyond $x=1$ and $x=13$. The extinction of the latter two peaks relieves competition at the center of the phenotypic space thus allowing the appearance of a peak of non recombinants in $x=7$. The final distribution and the frequency plots of recombinants and non recombinants are shown in Figure~\ref{fig6:final_distr}.

\begin{figure}[ht!]
\begin{center}
\begin{tabular}{c}
\includegraphics[scale=0.7]{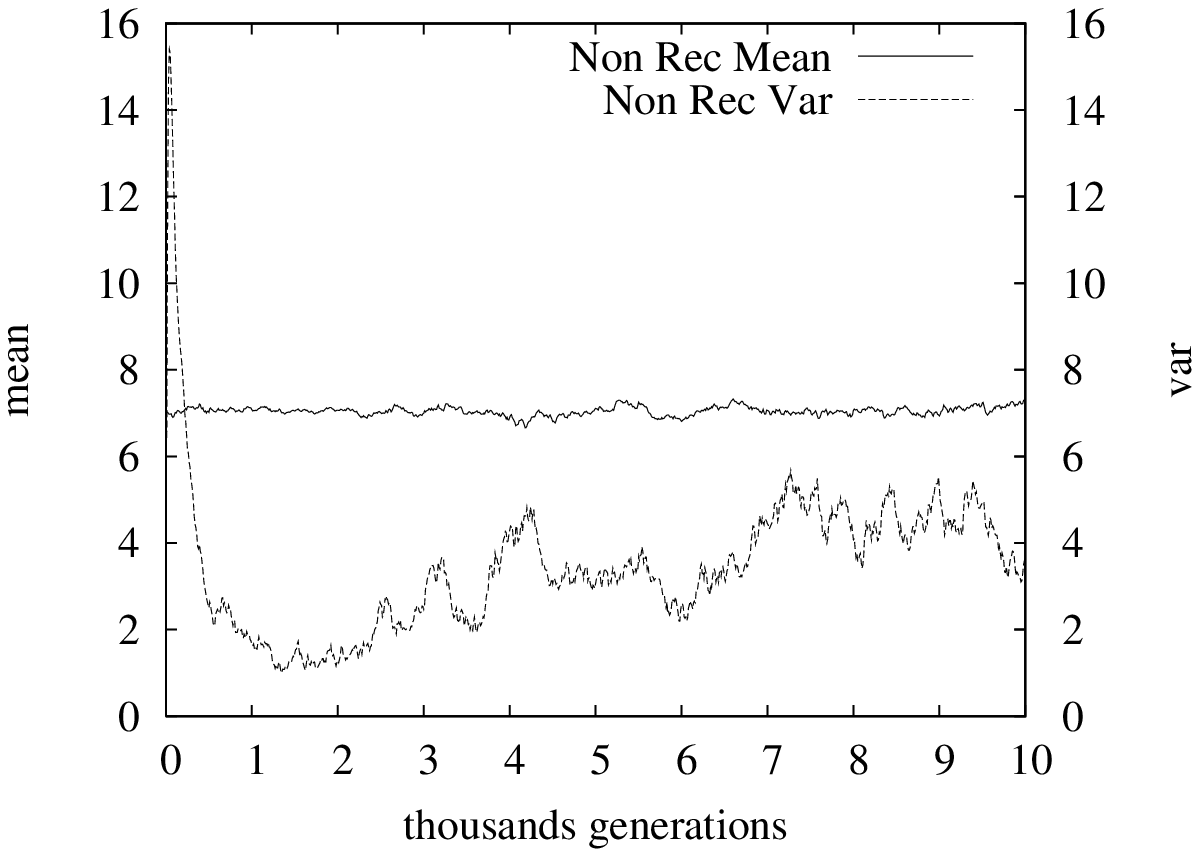} \\
\includegraphics[scale=0.7]{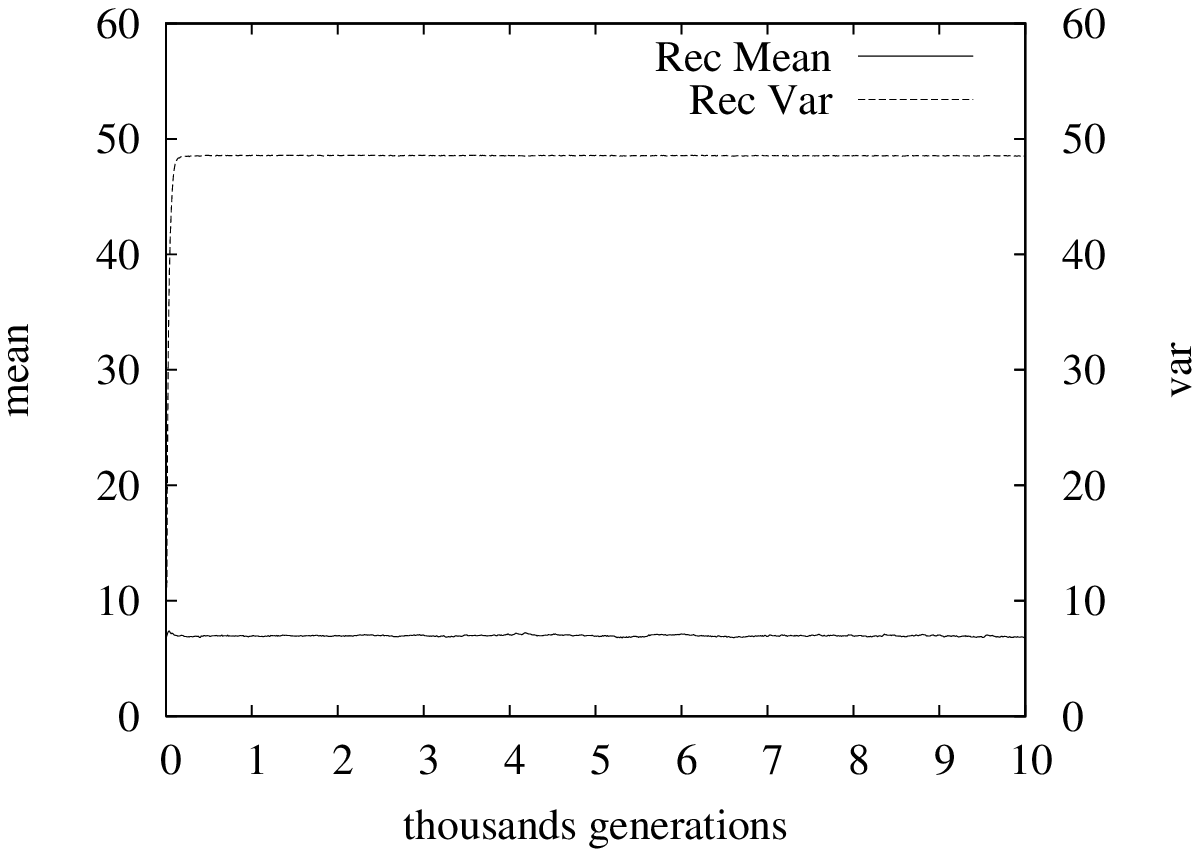} \\
\end{tabular}
\end{center}
  \caption{Regime of low mutation and large initial frequency of recombinants: the recombinants dominate both ends of the phenotypic space. Plots of variance and mean of recombinant and non recombinant distributions. Weak  competition ($J=1$, $\alpha = 2$, $R = 4$) and  flat static fitness ($\beta=100$, $\Gamma=14$). Mating range: $\Delta = 0$. Initial frequency of recombinants: 0.5; initial distribution parameters: $p= q =0.5$, initial population size $N_0 = 1000$, carrying capacity $K = 10000$. Annealing parameters: $\mu_0 =10^{-5}$, $ \mu_{\infty} = 10^{-6}$, $\tau = 1000$, $\delta = 100$. Total evolution time: 10000 generations.}

\label{fig6:var-mean}

\end{figure}

The description of the evolutionary dynamics is completed by the plots of mean and variance of recombinants and non recombinants shown in Figure~\ref{fig6:var-mean}. The variance of non recombinants raises rapidly in the first few generations and then it drops abruptly after the extinction of the peaks in $x=1$ and $x=13$. The wide oscillations in the variance are due to the fact that, apart for the large peak of non recombinants in $x=7$ there are other two eccentrical small peaks whose frequencies widely oscillate because of random sampling effects. The mean on the other hand is almost constant at $x = 7$ because the deformations of the distribution are symmetrical. The variance of recombinants conversely increases monotonically as the distribution first widens and then splits in two peaks in $x=0$ and $x=14$.

\subsubsection{The role of assortativity}

\begin{figure}[ht!]

\begin{tabular}{ccc}
\hspace{-2 cm} & \includegraphics[scale=0.65]{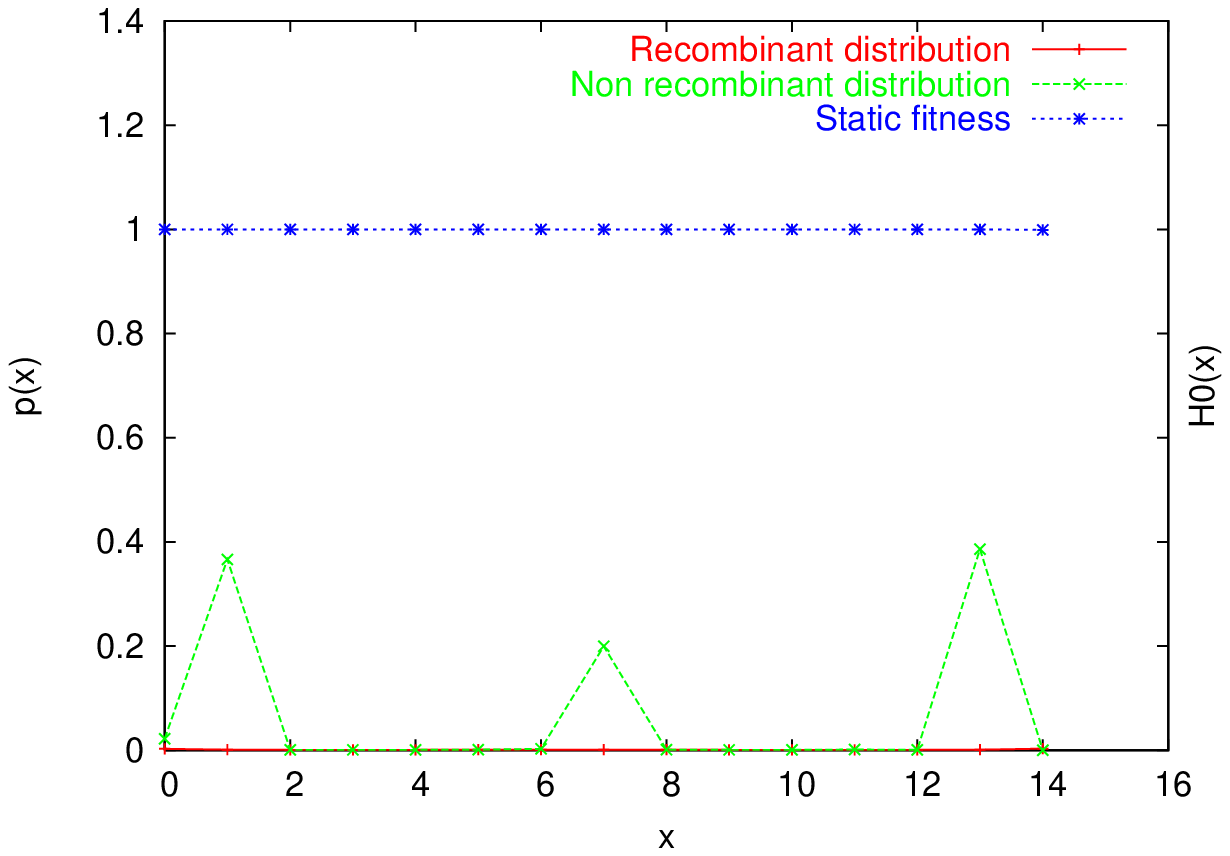} &
\includegraphics[scale=0.65]{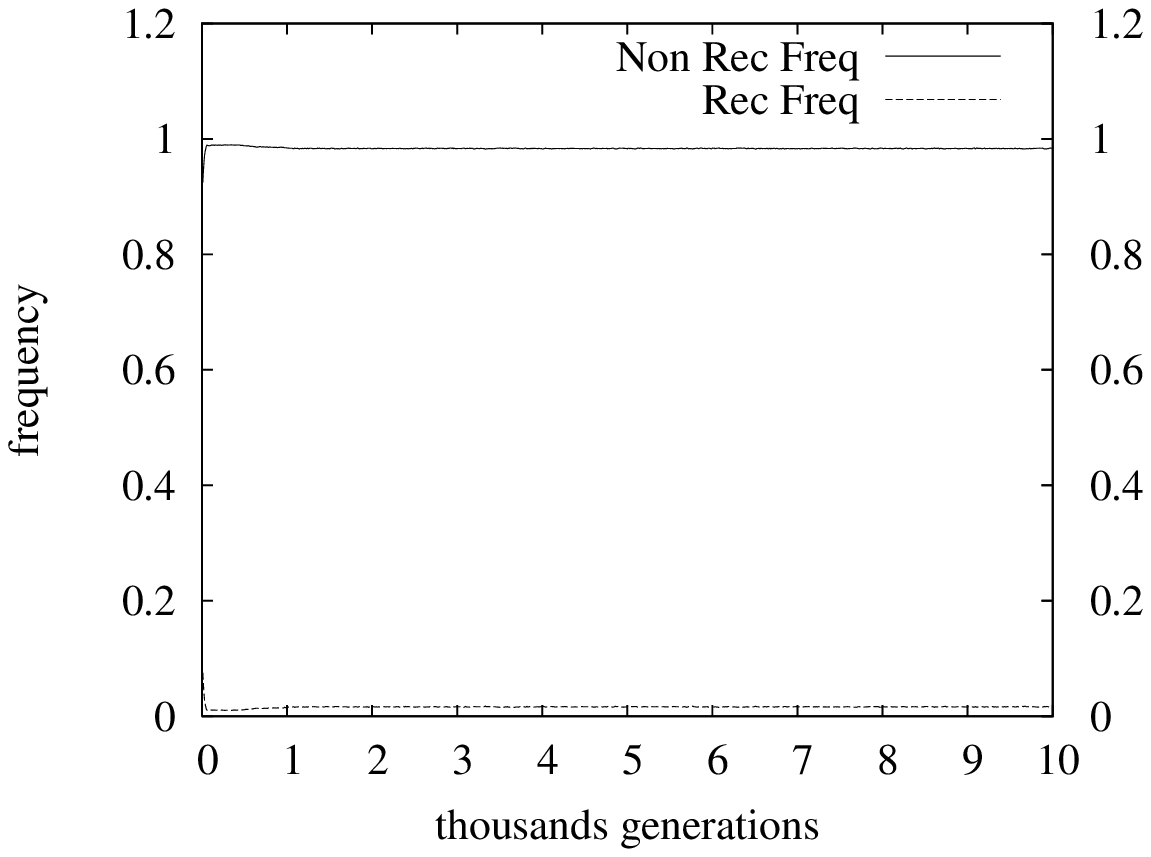} \\
\hspace{-2 cm} & \includegraphics[scale=0.65]{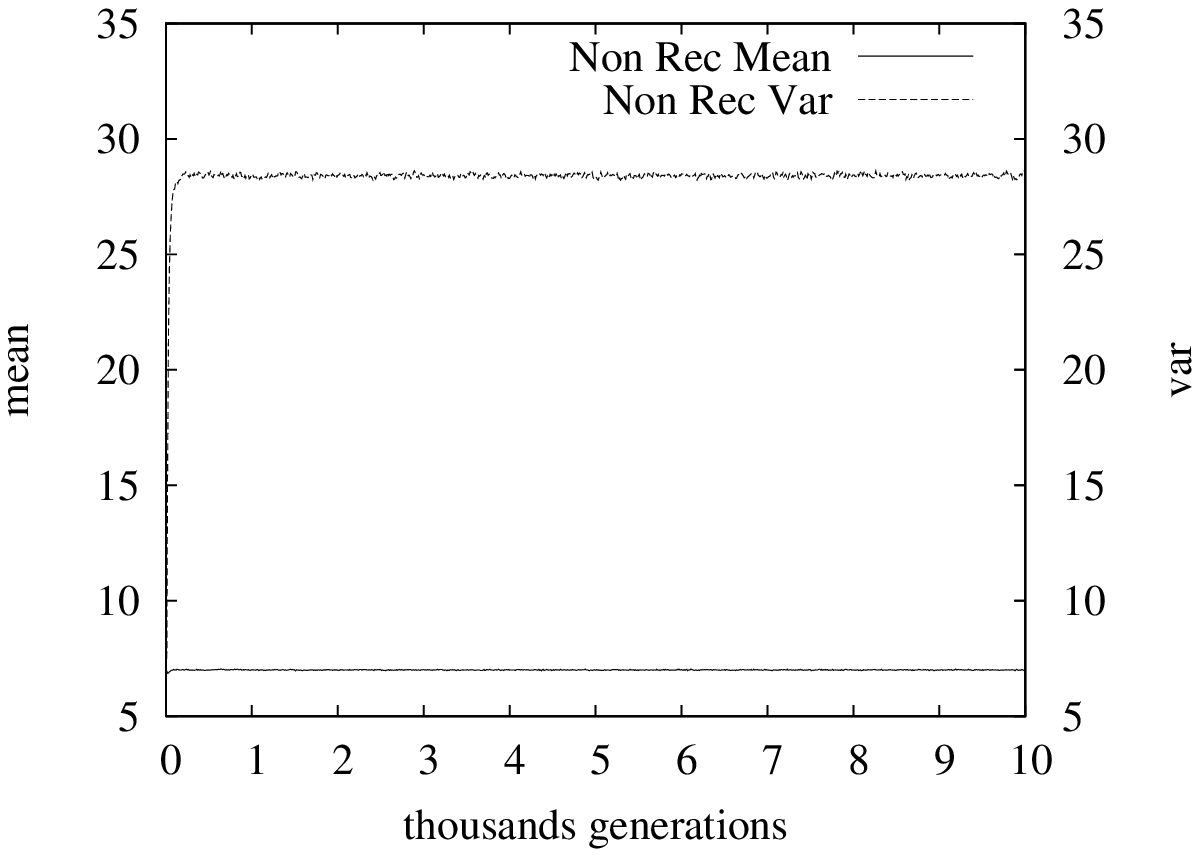} &
\includegraphics[scale=0.65]{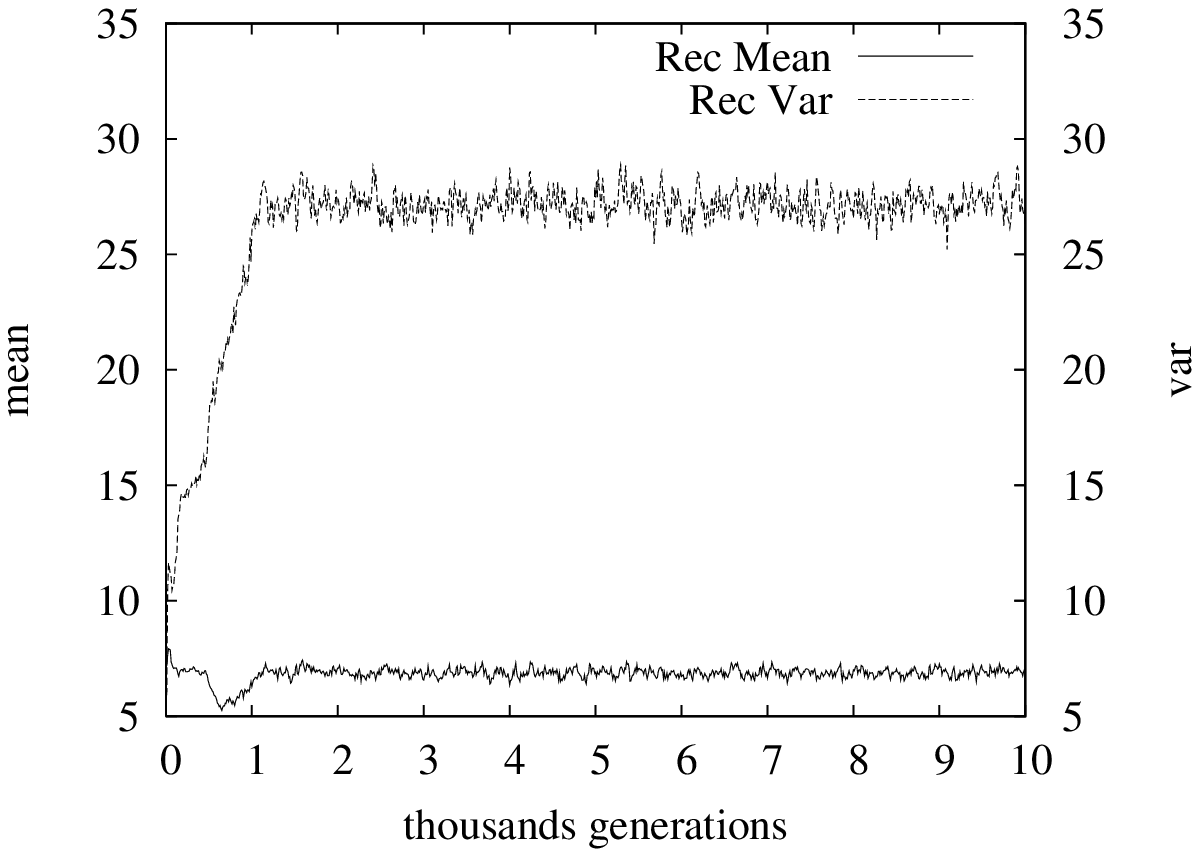} \\

\end{tabular}

\caption{Regime of random mating ($\Delta = 14$): top left: final distribution; top right: frequency plots; bottom left: mean and variance of non recombinant distribution; bottom right: mean and variance of recombinant distribution. Weak  competition ($J=1$, $\alpha = 2$, $R = 4$) and  flat static fitness ($\beta=100$, $\Gamma=14$). Initial frequency of recombinants: 0.2; initial distribution parameters: $p = q= 0.5$, initial population size $N_0 = 1000$, carrying capacity $K = 10000$. Annealing parameters: $\mu_0 =5\times 10^{-4}$, $ \mu_{\infty} = 10^{-6}$, $\tau = 1000$, $\delta = 100$. Total evolution time: 10000 generations.}

\label{fig7}

\end{figure}

In all the simulations discussed in Section~\ref{flat-fit} in the presence of competition, assortativity was maximal as $\Delta$ was set to zero which means that matings are allowed only among individuals with the same phenotype. The role of assortativity can be made more clear by performing simulations in a regime of random mating: $\Delta = 14$. The results of this simulation can be compared to those of the simulations performed to study the effect of low mutation (Section~\ref{flat-fit-mut}) where $\Delta = 0$. In that case, choosing an initial distribution characterized by $p = q = 0.5$, $M = 0.2$, the recombinants never became extinct and they typically coexisted with non recombinants at both ends of the phenotypic space. This could happen because $\Delta = 0$ prevents the scattering of the offsprings among several phenotypes thus allowing the survival of small colonies of recombinants until they reach favourable locations in the phenotypic space. Conversely, if we set $\Delta = 14$ the offsprings are scattered over several phenotypes and many tiny subpopulations ensue that are easily eradicated by random sampling effects. This explains why in these simulations the recombinants usually become extinct and the typical asymptotic distribution features three peaks of non recombinants in $x=1$, $x=7$, $x=13$ so as to minimize competition. All the transformations of the distributions are symmetrical so that the mean of both recombinants and non recombinants is always very close to 7; the variance  conversely increases monotonically as a result of competition that induces the widening of the distributions; it can be noticed that the variance of recombinants shows wider oscillations as compared to that of non recombinants as a consequence of random sampling effects in a very small population. In Figure~\ref{fig7} we show the plots of mean and variance, the final distribution and the frequency plots.

As a conclusion, the simulations  show that both competition and assortativity favour the evolutionary success of
recombinants. In particular, high competition enhances their velocity of movement in the phenotipic space,
while assortativity allows survival of recombinants even when their initial frequency is very small. The simulations also show that the evolutionary success of recombinants depends on the naturally low mutation rates that limit the velocity of movement of non recombinants in the phenotypic space. When mutations affect qualitative traits in fact, they are usually lethal and the DNA repair mechanisms evolved so as to keep the rate as small as possible.

\subsection{Steep static fitness landscape }

\subsubsection{Absence of competition }

If competition is absent and the static fitness landscape is steep, the distribution of both recombinants and
non recombinants (initially centered at $x=7$) move towards the fitness maximum at $x=0$. The high mutation rate ($\mu_0 = 10^{-4}$, $\mu_{\infty} = 10^{-6}$) provides non recombinants with a velocity of movement in the phenotypic space comparable to that of recombinants. As a consequence the non recombinants usually reach the $x=0$ position earlier than recombinants that are thus led to extinction. In some runs however, it may happen that recombinants and non recombinants reach the $x=0$ position in the same time or that recombinants arrive a little earlier than non recombinants. In any case, the recombinants do not have enough time to establish a large colony and when the non recombinants reach $x=0$, thanks to their higher fertility, they soon overwhelm recombinants that tend to disappear. 

\begin{figure}[ht!]

\begin{tabular}{ccc}
\hspace{-2 cm} & \includegraphics[scale=0.65]{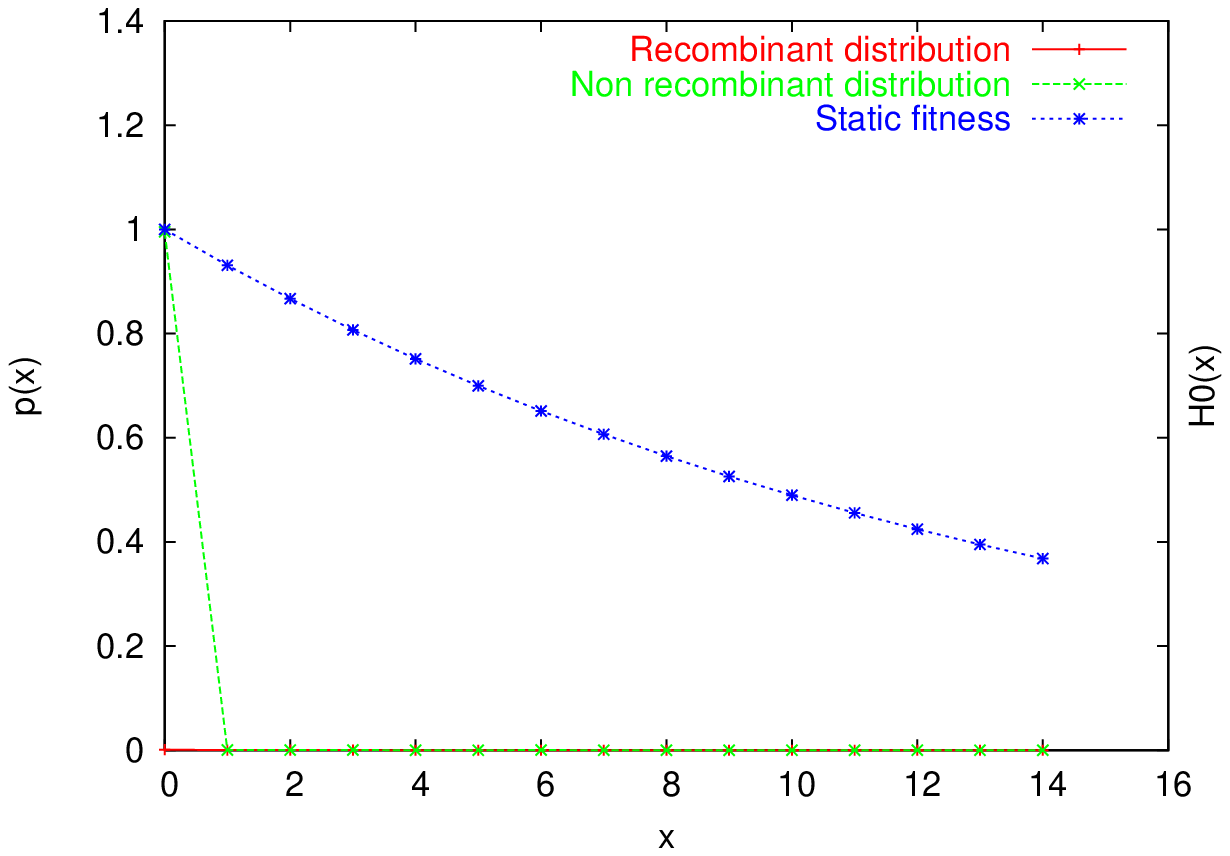} &
\includegraphics[scale=0.65]{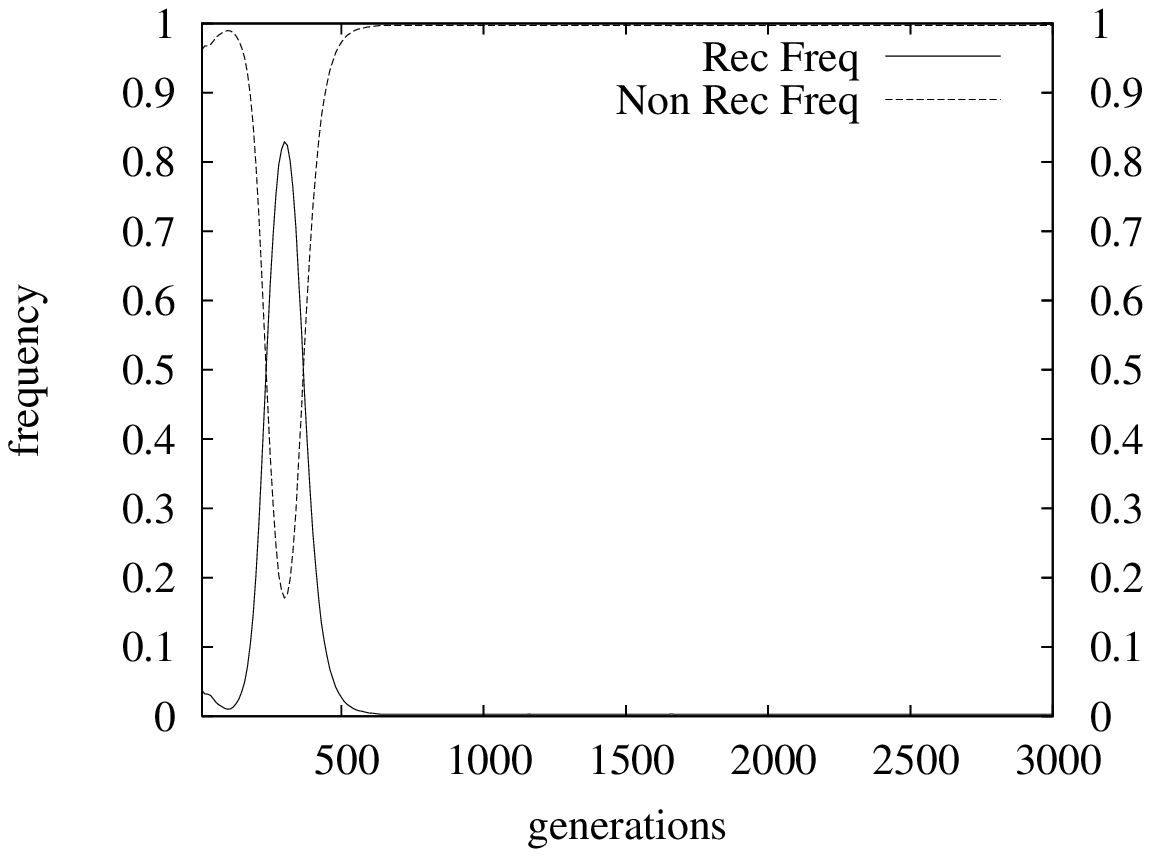} \\
\hspace{-2 cm} & \includegraphics[scale=0.65]{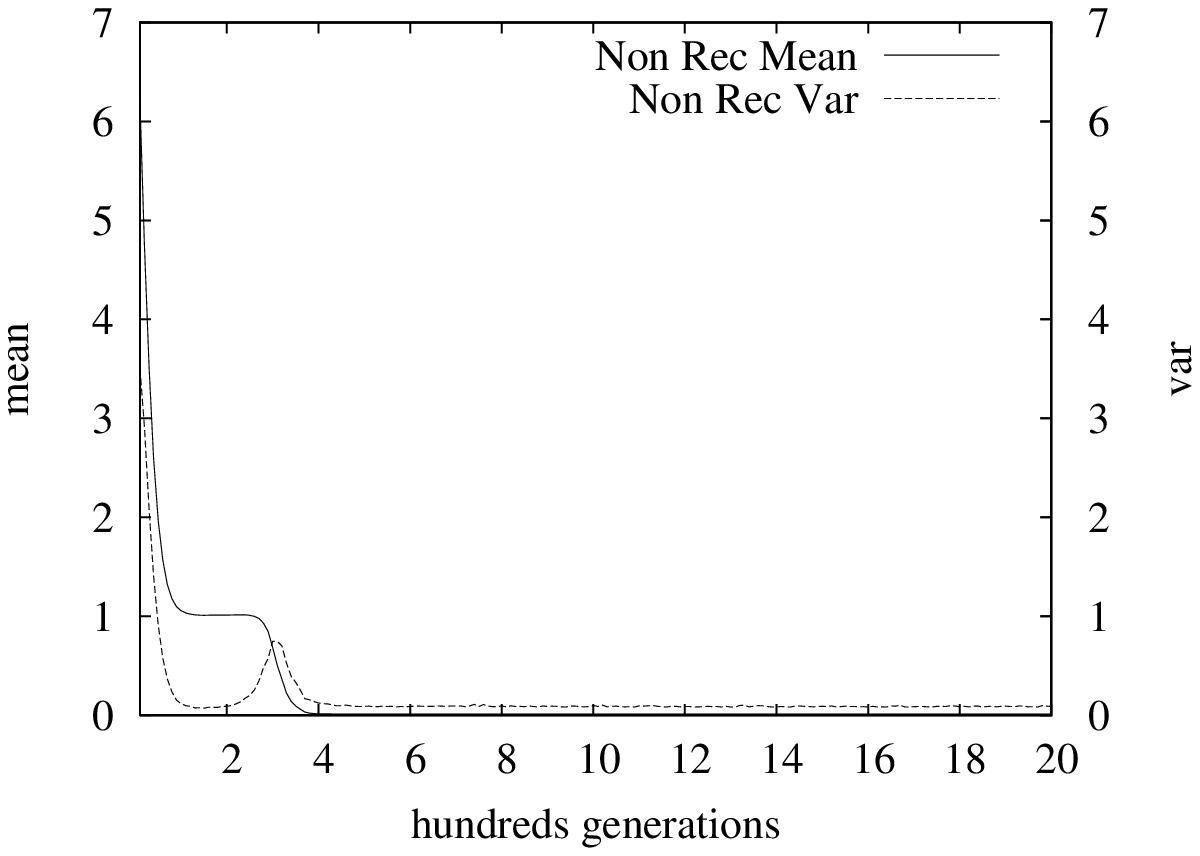} &
\includegraphics[scale=0.65]{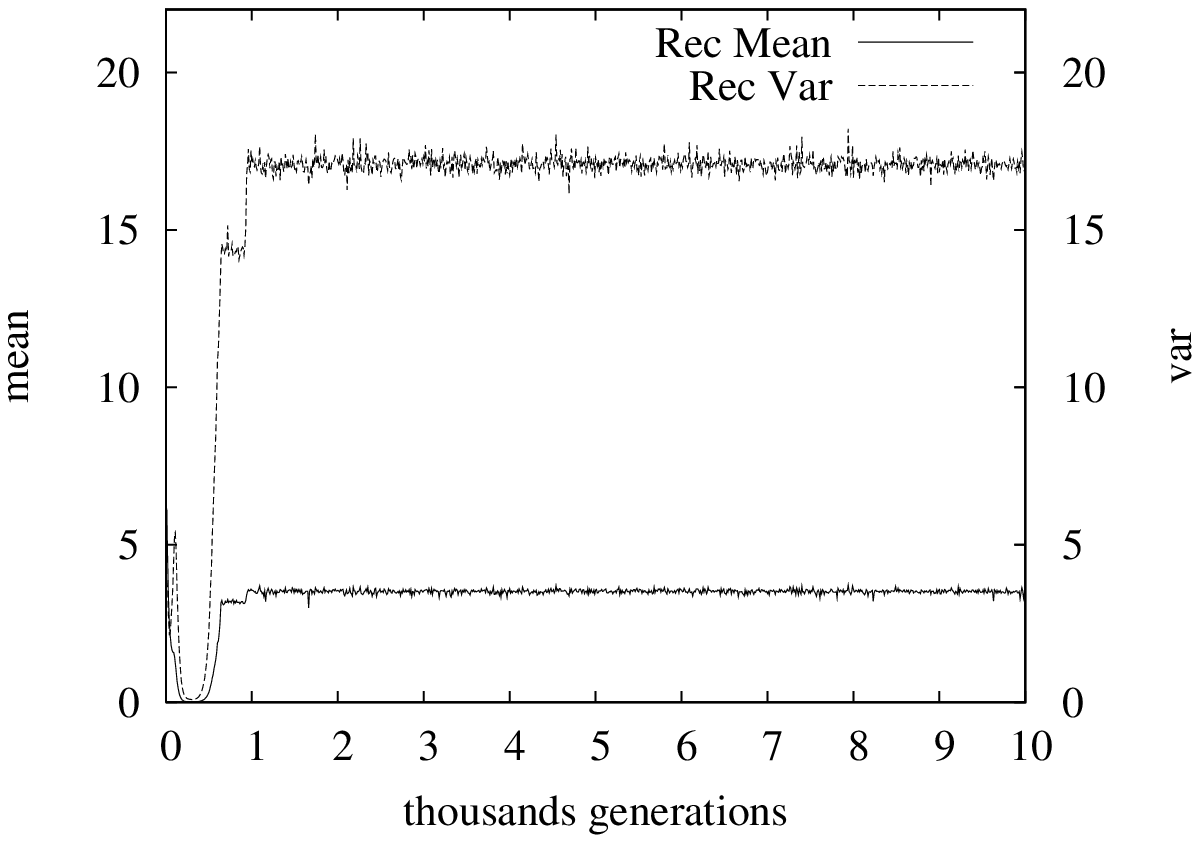} \\

\end{tabular}

\caption{Steep static fitness ($\beta=1$, $\Gamma=14$) and absence of competition ($J=0$): top left: final distribution; top right: frequency plots; bottom left: mean and variance of non recombinant distribution; bottom right: mean and variance of recombinant distribution. Mating range: $\Delta = 14$. Initial frequency of recombinants: 0.1; initial distribution parameters: $p = q= 0.5$, initial population size $N_0 = 1000$, carrying capacity $K = 10000$. Annealing parameters: $\mu_0 = 10^{-4}$, $ \mu_{\infty} = 10^{-6}$, $\tau = 1000$, $\delta = 100$. Total evolution time: 10000 generations; for the sake of clarity only 2000 and 3000 generations were displayed in the plots of mean and variance of non recombinants and in the frequency plots respectively.  }
\label{fig:8}

\end{figure}

In Figure~\ref{fig:8} we report  a simulation in which recombinants reach $x=0$ first. The mean of non recombinants decreases in a step-like way as they establish a delta-peak in $x=1$ that later moves to $x=0$; their variance conversely decreases abruptly from a very high value related to the wide initial distribution to almost zero forming a little hump in correspondence to the shift from the peak in $x=1$ to that in $x=0$. The variance and mean of recombinants conversely, first decrease when they establish a peak in $x=0$ and then increase again when the peak in $x=0$ disappears and the few survivors become distributed according to a wide and flat distribution. Finally, the frequency of recombinants increases significantly when they occupy the $x=0$ and then decreases when they are expelled from this position.     

\paragraph{The role of mutation}

The simulation illustrated in Figure~\ref{fig:8} shows that when the initial mutation rate is as high as $10^{-4}$ the non recombinants are characterized by a high velocity of movement in the phenotypic space so that they are able to reach the $x=0$ position before or in the same time as the recombinants. On the other hand, if we set $\mu_0 = 10^{-5}$ the mobility of non recombinants is significantly reduced as compared to that of recombinants that can rely on the recombination mechanism. As a consequence, the recombinants can reach the $x=0$ position much earlier than non recombinants and are therefore able to establish there large colonies that exert a competition pressure strong enough to cause the extinction of the peaks of non recombinants in $x=1$ or $x = 2$. 

\begin{figure}[ht!]

\begin{tabular}{ccc}
\hspace{-2 cm} & \includegraphics[scale=0.65]{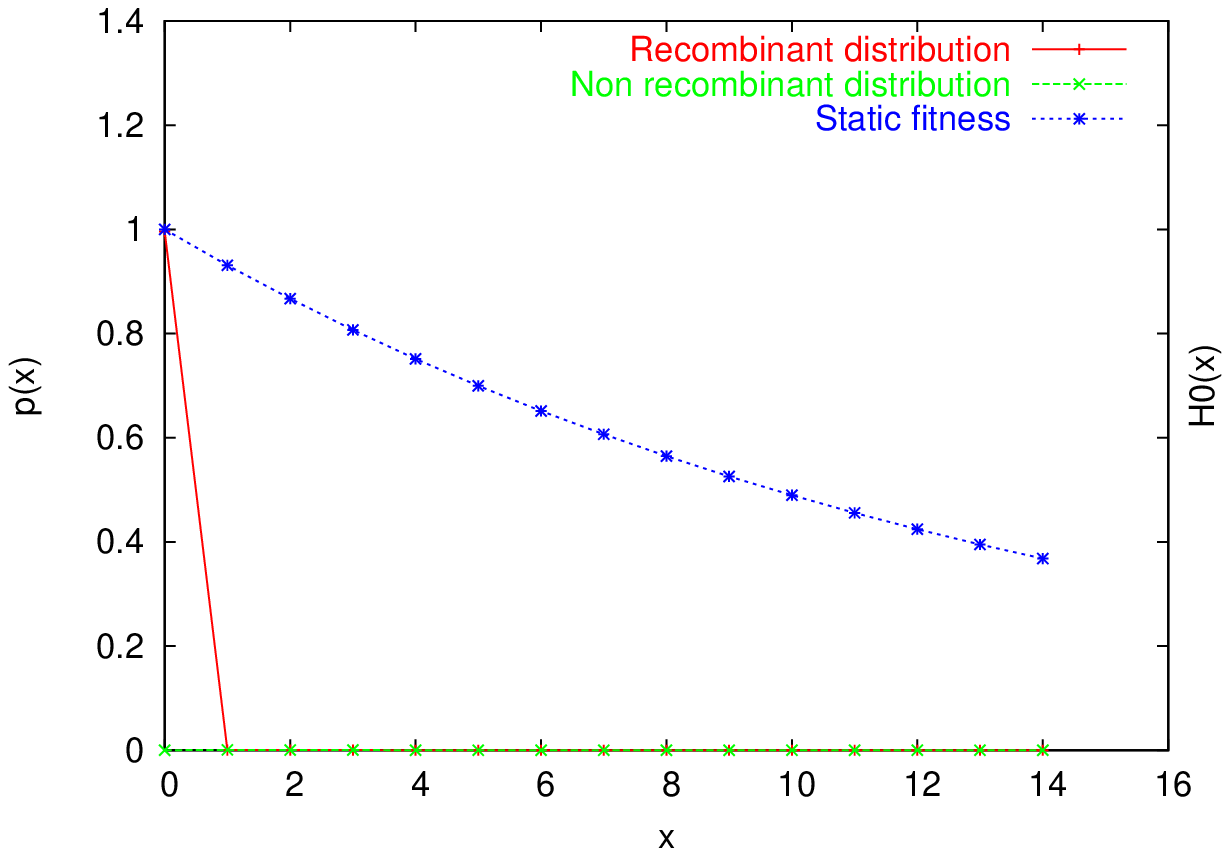} &
\includegraphics[scale=0.65]{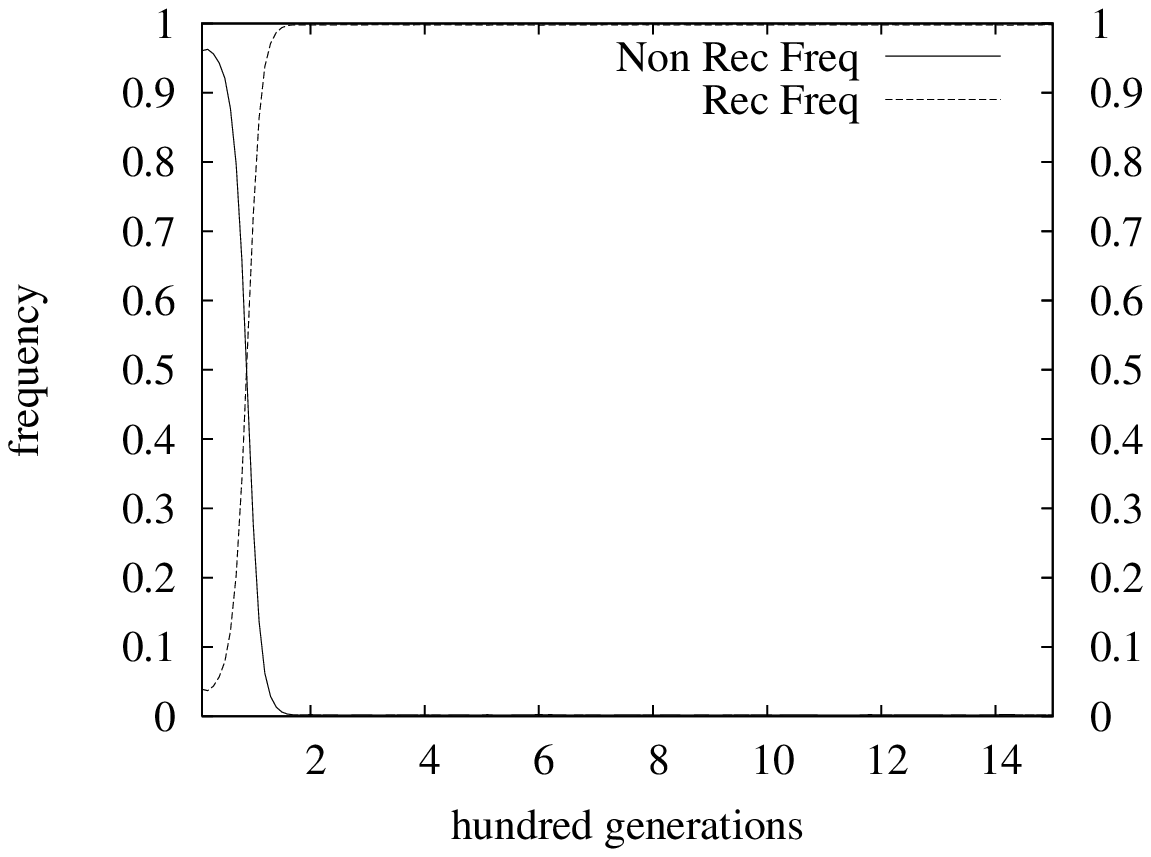} \\
\hspace{-2 cm} & \includegraphics[scale=0.65]{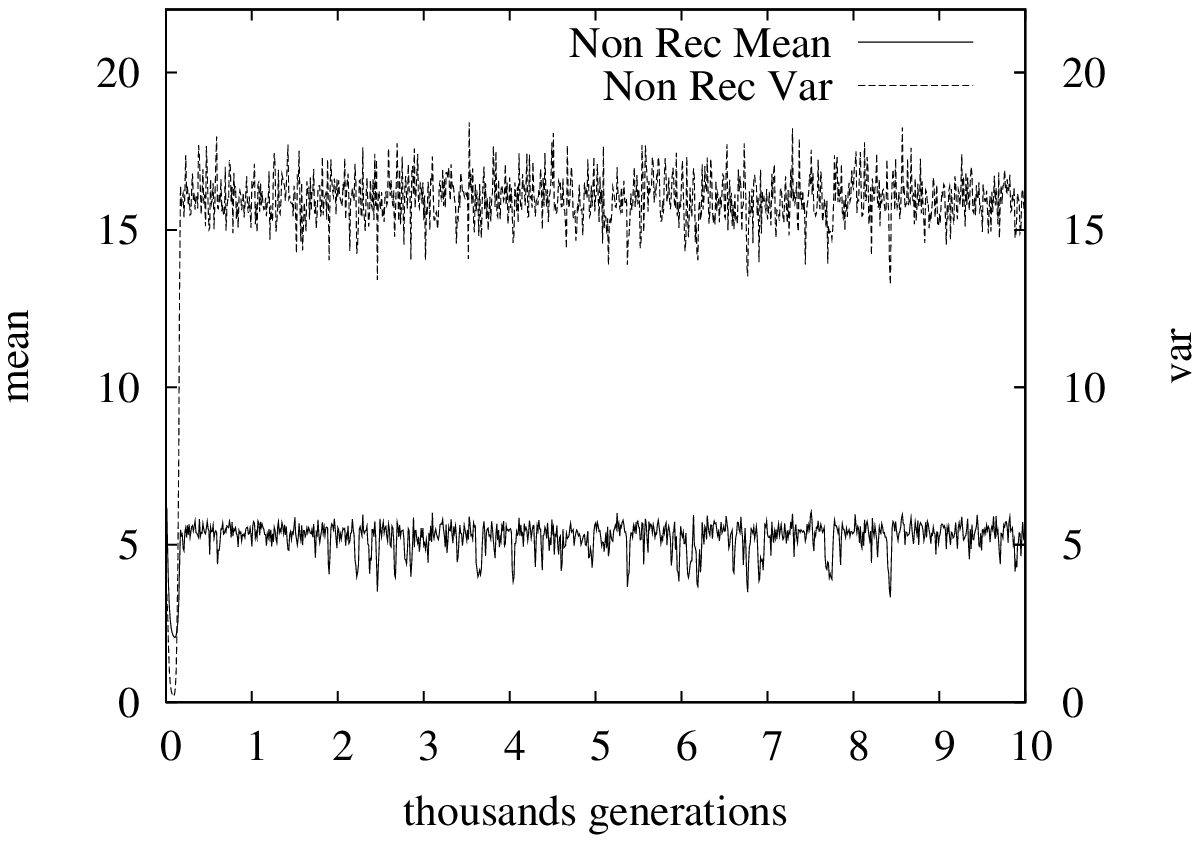} &
\includegraphics[scale=0.65]{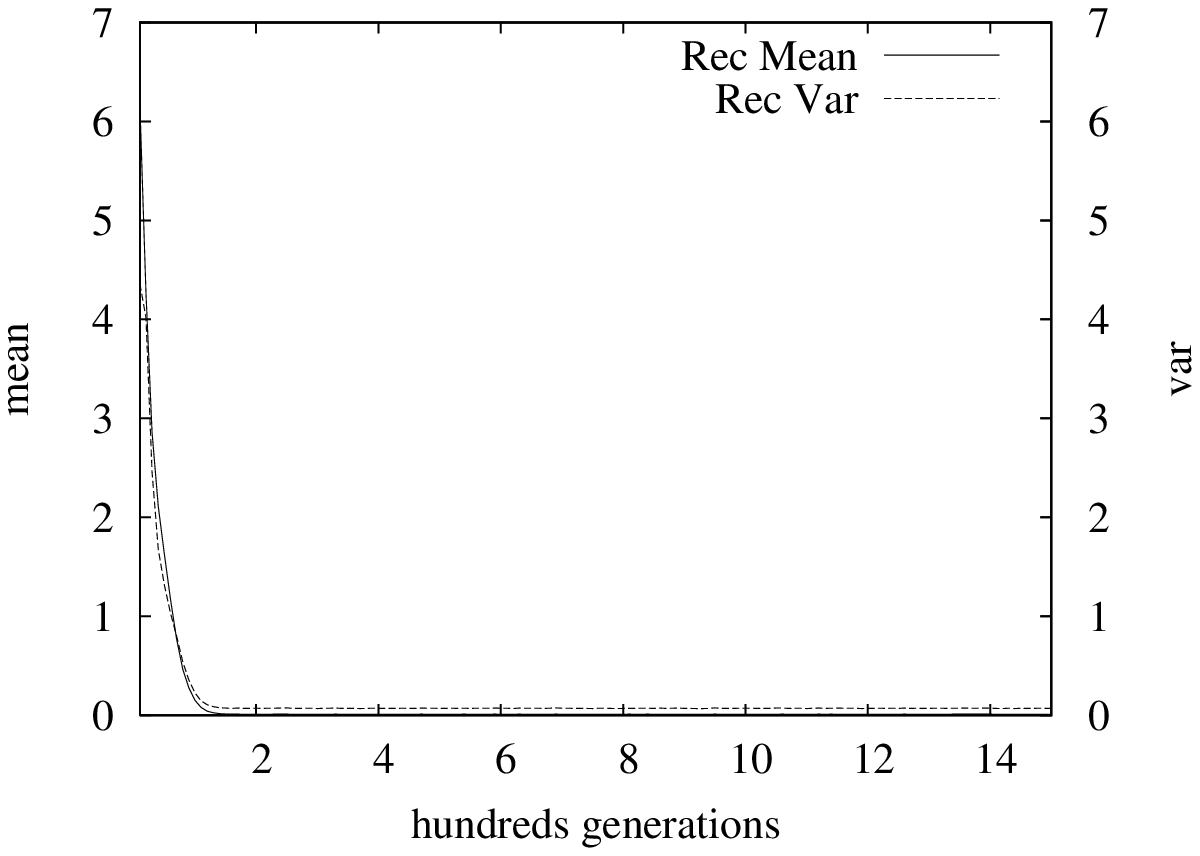} \\

\end{tabular}

\caption{Steep static fitness ($\beta=1$, $\Gamma=14$) and absence of competition ($J=0$) in a low mutation regime: top left: final distribution; top right: frequency plots; bottom left: mean and variance of non recombinant distribution; bottom right: mean and variance of recombinant distribution. Mating range: $\Delta = 14$. Initial frequency of recombinants: 0.1; initial distribution parameters: $p = q= 0.5$, initial population size $N_0 = 1000$, carrying capacity $K = 10000$. Annealing parameters: $\mu_0 = 10^{-5}$, $ \mu_{\infty} = 10^{-6}$, $\tau = 1000$, $\delta = 100$. Total evolution time: 10000 generations; for the sake of clarity only 1500 generations were displayed in the plots of mean and variance of  recombinants and in the frequency plots.  }
\label{fig:9}

\end{figure}

In Figure~\ref{fig:9} we display the results of a typical run. The mean and variance of non recombinants decrease abruptly when they form a peak in $x=2$. When this peak becomes extinct however, the non recombinant distribution becomes extremely wide and flat, with  mean and variance showing wide fluctuations because of random sampling effects in a small subpopulation. The variance and mean of recombinants on the other hand decrease monotonically as the distribution moves towards $x=0$ and in the same time narrows down until it becomes a delta-peak.

\paragraph{The role of assortativity}

\label{steep-no_comp-assort}

The role of assortativity is the same as observed in the case of a flat static fitness landscape. In Figure~\ref{fig:8} the random mating regime causes dispersion so that even if a little group of recombinants reach $x=0$ before the non recombinants, it is soon eradicated by random fluctuations. This explains why in that set of simulations the recombinants became the majority of the population only in 3 runs out of 10. If the simulation is repeated by setting $\Delta = 0$, there will be no scattering of offsprings so that the recombinants outperform the non recombinants in 6 runs out of 10. It can be therefore concluded that a high assortativity plays a role similar to that of a low initial mutation rate.  

The plots of frequency, mean and variance are extremely similar to those displayed in Figure~\ref{fig:9} and therefore they are not shown. In particular, also in this case it is possible to notice wide oscillations of mean and variance of recombinants after the extinction of the peak in $x=1$ due to random sampling effects in a tiny subpopulation.

\paragraph{The  role of initial frequency}

In Section~\ref{steep-no_comp-assort} we discussed the effect of assortativity and we concluded that high values such as $\Delta = 0$ prevent dispersion of offsprings enabling small groups of recombinants to survive until they reach the most favourable regions of the phenotypic space. On the other hand, if the initial frequency of recombinants is high, they can retain a size sufficiently large to resist eradication by random fluctuations even if a long mating range induces offspring dispersion. This situation can be observed if we repeat the simulation shown in Figure~\ref{fig:8} by changing the initial frequency of recombinants from $M=0.1$ to $M=0.5$. While with $M=0.1$ the recombinants become the dominant group only in 3 runs out of 10, with $M=0.5$ they outperform the non recombinants in 9 runs out of 10. 

The plots of frequency, mean and variance once again are very similar to those obtained in the cases of high assortativity ($\Delta = 0$) and low initial mutation rate ($\mu_0 = 10^{-5}$, Figure~\ref{fig:9}) so that they are not shown. The mean and variance of non recombinants decrease when they form a sharp peak in $x=2$ and when this peak disappears mean and variance increase again and show wide oscillations due to random sampling effects in a small population. On the contrary, mean and variance of recombinants decrease monotonically until they vanish as a result of the creation of the peak in $x=0$.

\subsubsection{Presence of competition }

In this Section we study the effect of competition in a steep static fitness landscape, and in particular we will distinguish the case of weak, intermediate and strong competition. The last part of the Section will be devoted to the study of the interplay between competition and assortativity.

\paragraph{Weak competition}

Let us consider an example of  the case $M = 0.5$, $p = q = 0.5$, \emph{i.e.} recombinants and non recombinants have the same initial frequency and the initial distributions are both centered in the middle of the phenotypic space. In a regime of weak competition such as $J = 0.8$ both distributions move towards the maximum of static fitness but the $x=0$ position is first reached by the recombinants. As the peak of recombinants in $x=0$ becomes more and more populated it exerts a stronger and stronger competition on the peak of non recombinants in $x=2$ (or $x=1$) which is therefore led to extinction. The disappearance of the non recombinant peak in $x=2$ relieves competition in the middle of the phenotypic space therefore allowing the appearance of a new peak of non recombinants in $x=7$ or $x=8$. The central peak of non recombinants is stabilized by competition: the central region of the phenotypic space in fact becomes a favourable position because the low value of the static fitness is compensated by the low level of competition experienced in that location.

\begin{figure}[ht!]

\begin{tabular}{ccc}
\hspace{-2 cm} & \includegraphics[scale=0.65]{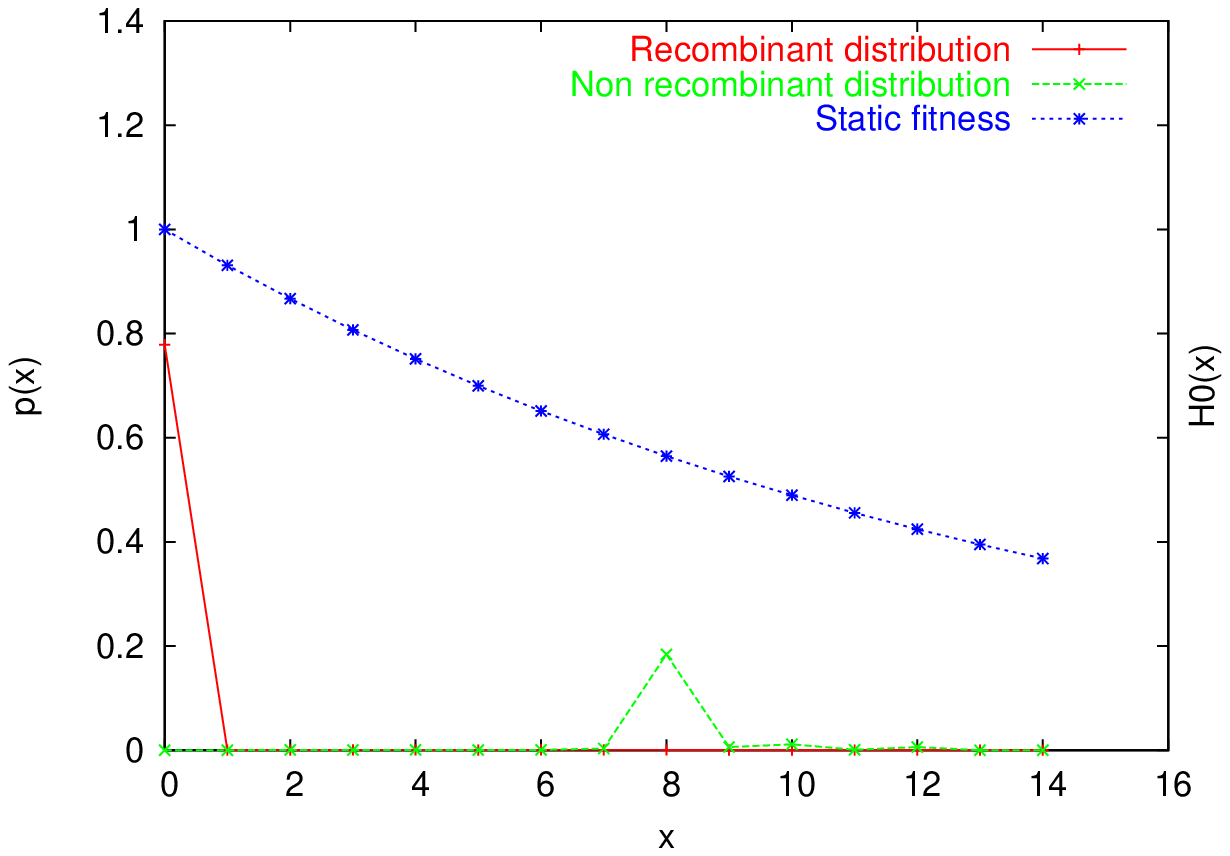} &
\includegraphics[scale=0.65]{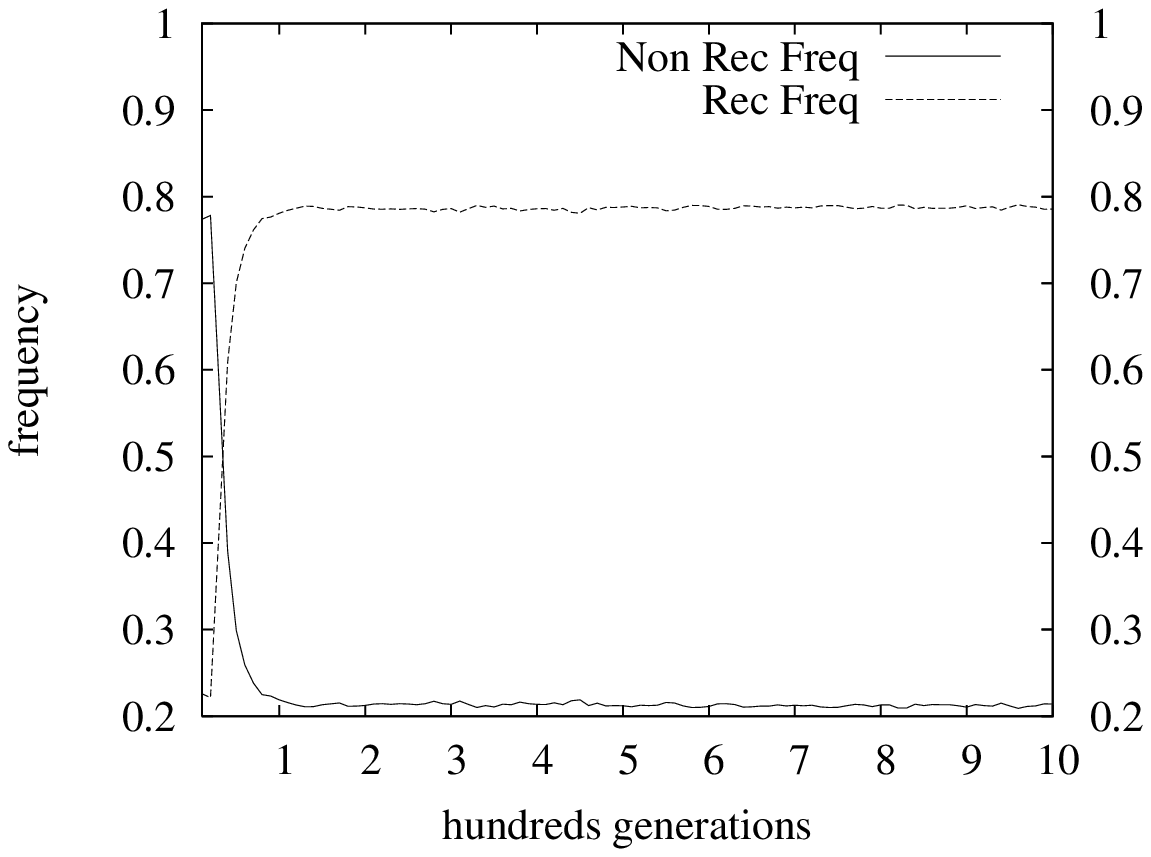} \\
\hspace{-2 cm} & \includegraphics[scale=0.65]{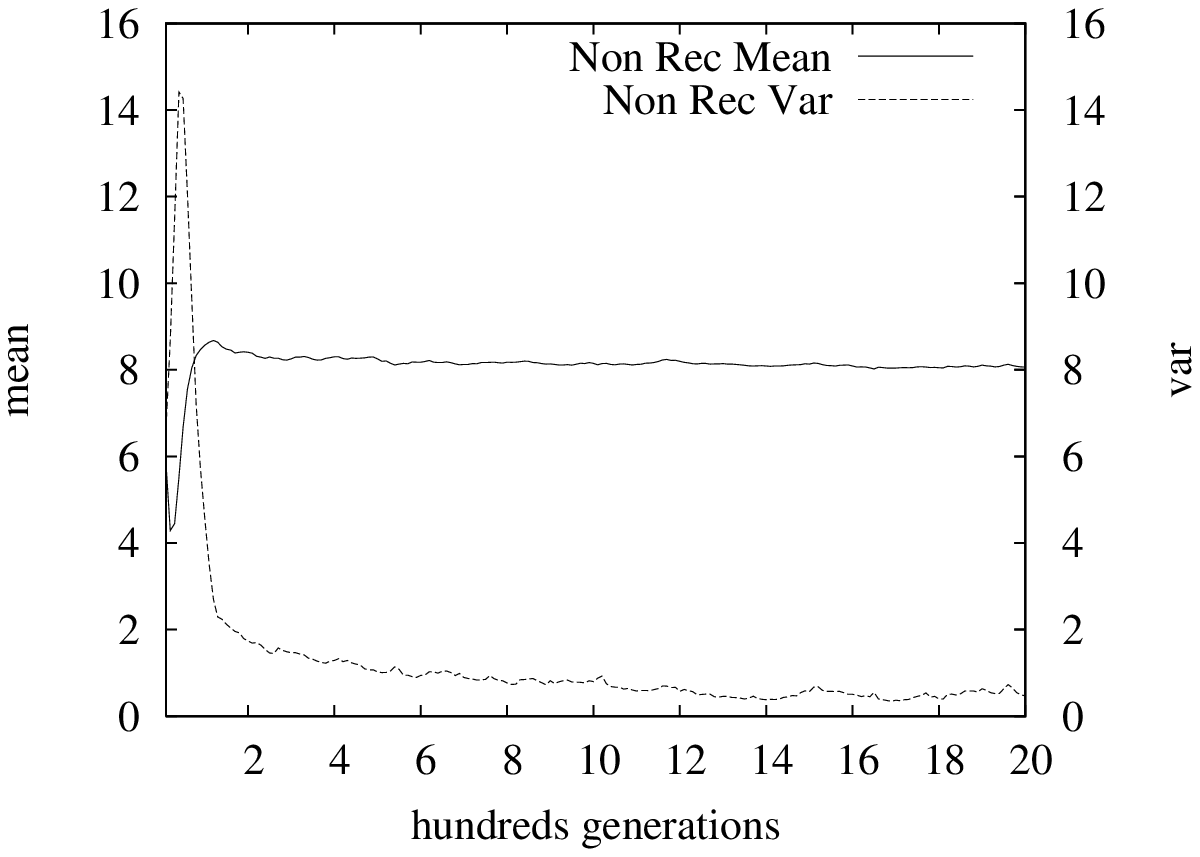} &
\includegraphics[scale=0.65]{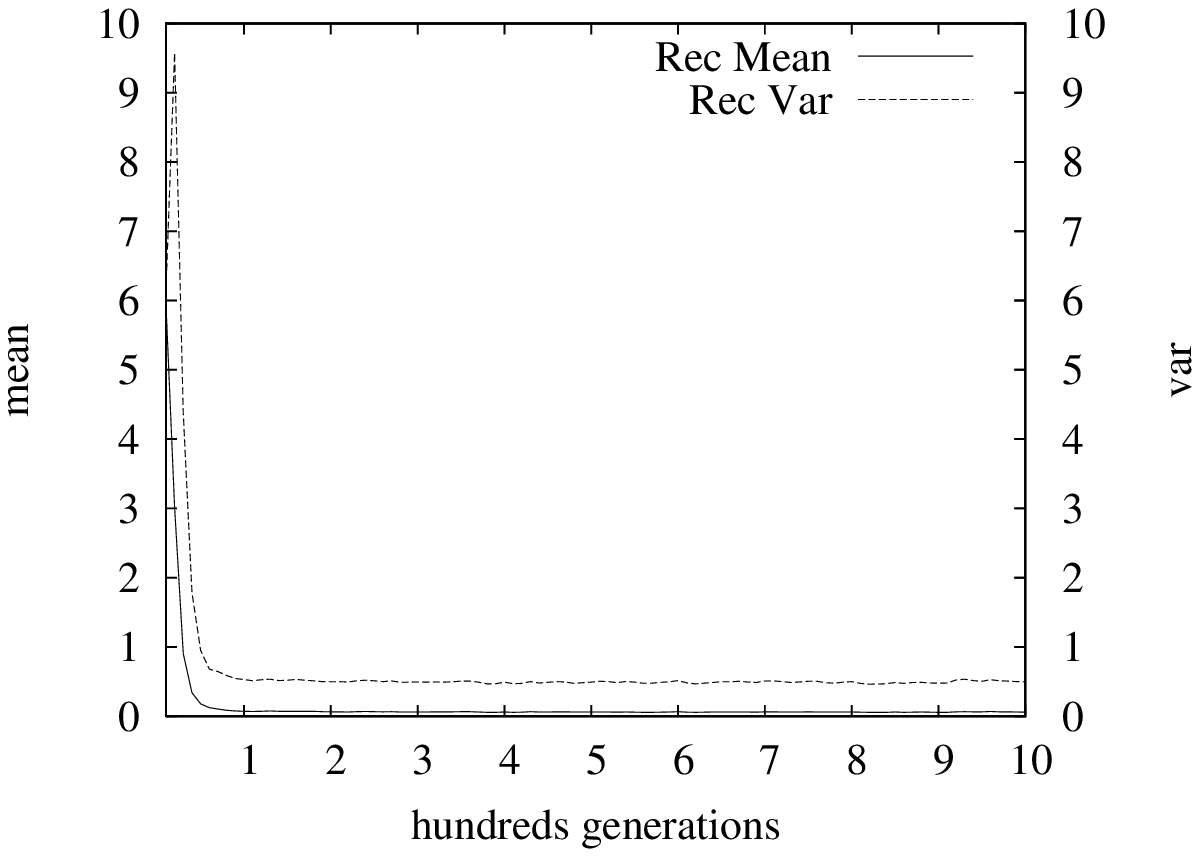} \\

\end{tabular}

\caption{Steep static fitness ($\beta=1$, $\Gamma=14$) and weak competition ($J=0.8$) in a random mating regime ($\Delta = 14$): top left: final distribution; top right: frequency plots; bottom left: mean and variance of non recombinant distribution; bottom right: mean and variance of recombinant distribution. Initial frequency of recombinants: 0.5; initial distribution parameters: $p = q= 0.5$, initial population size $N_0 = 1000$, carrying capacity $K = 10000$. Annealing parameters: $\mu_0 = 10^{-5}$, $ \mu_{\infty} = 10^{-6}$, $\tau = 1000$, $\delta = 100$. Total evolution time: 10000 generations; for the sake of clarity only 1000 generations were displayed in the plots of mean and variance of  recombinants and in the frequency plots, while 2000 generations were considered in the plots of mean and variance of non recombinants.  }
\label{fig:12}

\end{figure}

In Figure~\ref{fig:12} we show the plots of frequency, mean and variance as well as the final distribution in a typical case. The mean of non recombinants tends to increase when the peak  in $x=2$ becomes extinct and the subpopulation becomes concentrated in a peak in $x=8$; the variance accordingly, after an initial increase due to the widening of the distribution, drops abruptly as the peak in $x=8$ is very sharp. Conversely, mean and variance of non recombinants decrease monotonically as the distribution moves towards $x=0$ and turns into a delta-peak.

\paragraph{Intermediate competition}

\begin{figure}[ht!]

\begin{tabular}{ccc}
\hspace{-2 cm} & \includegraphics[scale=0.65]{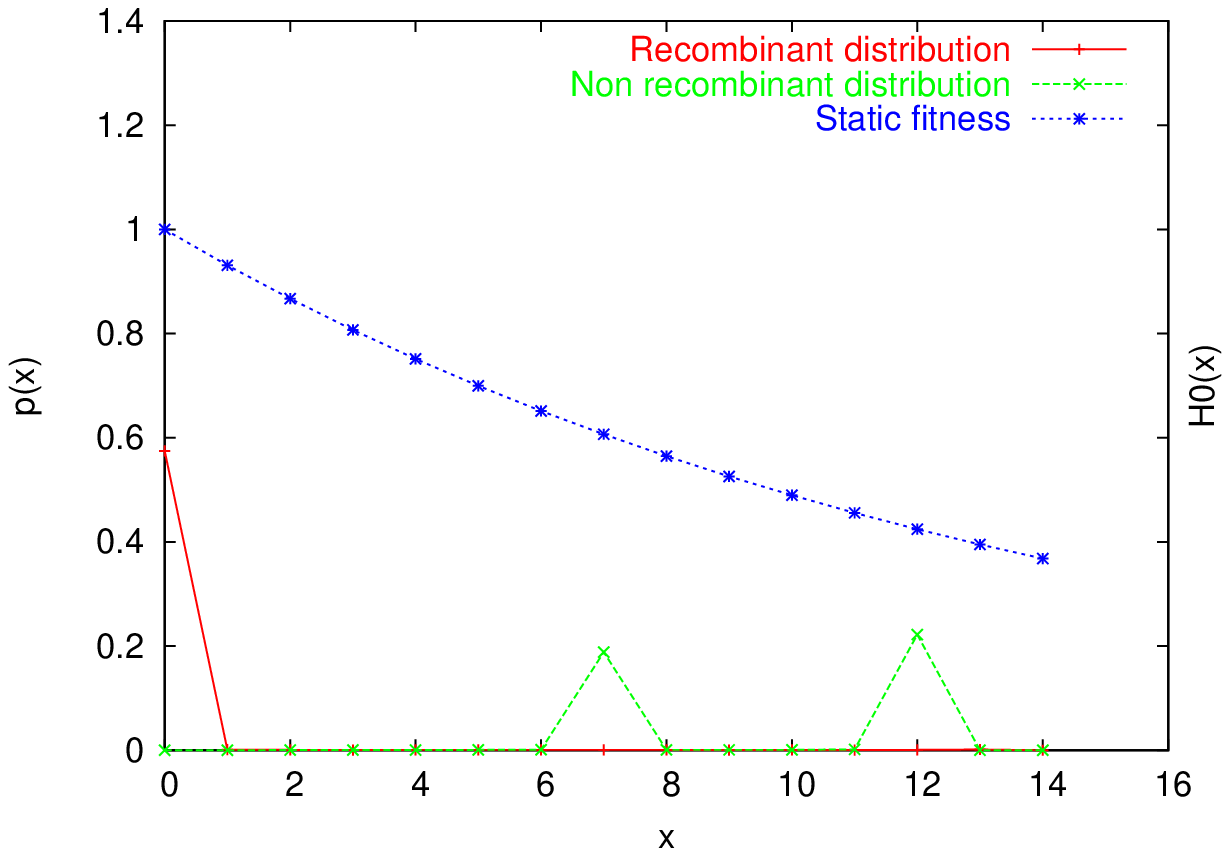} &
\includegraphics[scale=0.65]{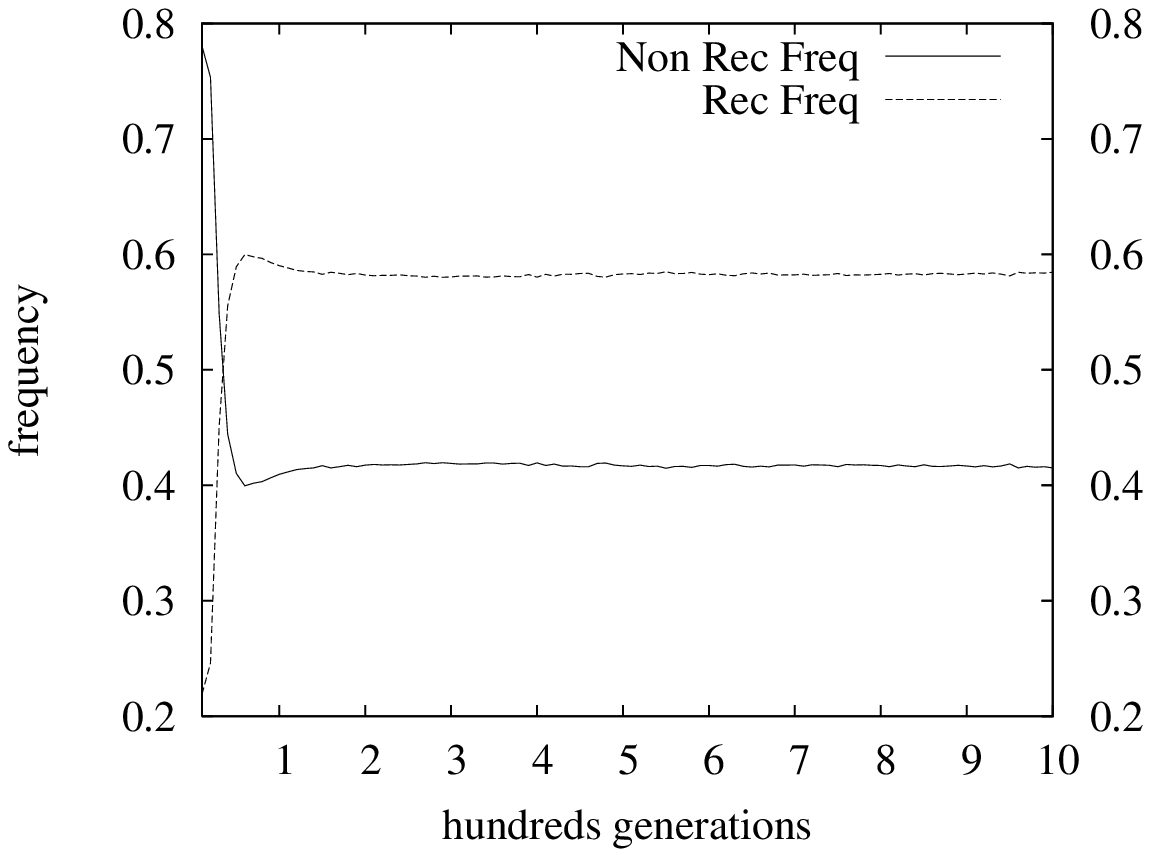} \\
\hspace{-2 cm} & \includegraphics[scale=0.65]{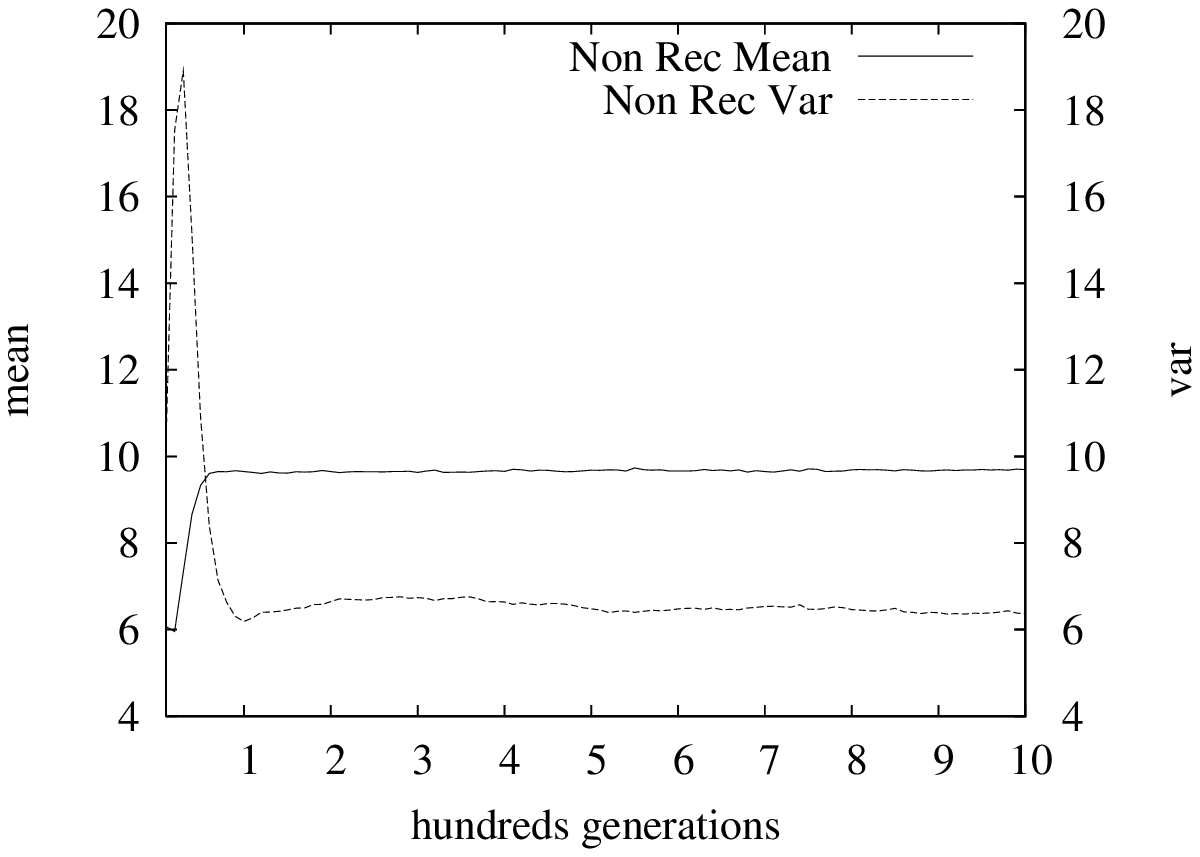} &
\includegraphics[scale=0.65]{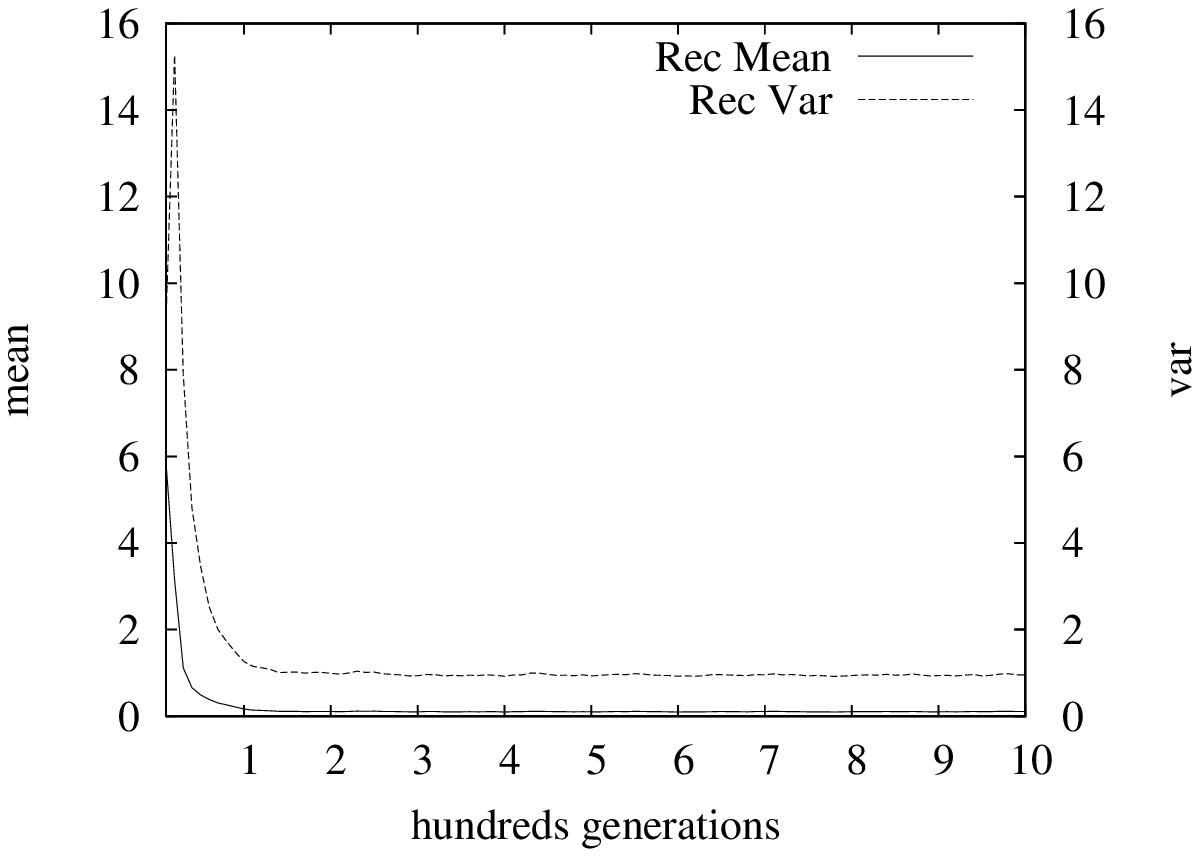} \\

\end{tabular}

\caption{Steep static fitness ($\beta=1$, $\Gamma=14$) and intermediate competition ($J=1.8$) in a random mating regime ($\Delta = 14$): top left: final distribution; top right: frequency plots; bottom left: mean and variance of non recombinant distribution; bottom right: mean and variance of recombinant distribution. Initial frequency of recombinants: 0.5; initial distribution parameters: $p = q= 0.5$, initial population size $N_0 = 1000$, carrying capacity $K = 10000$. Annealing parameters: $\mu_0 = 10^{-5}$, $ \mu_{\infty} = 10^{-6}$, $\tau = 1000$, $\delta = 100$. Total evolution time: 10000 generations; for the sake of clarity only 1000 generations were displayed in the plots of mean and variance of  recombinants and non recombinants and in the frequency plots.  }
\label{fig:13}

\end{figure}

As a consequence of the competition pressure and the steepness of the static fitness, the non recombinant distribution splits in two peaks near the opposite ends of the phenotypic space, while the recombinant distribution moves towards $x=0$ (strictly speaking there is a peak of recombinants also near $x=14$ but this is negligible as compared to the peak in $x=0$). As the recombinants form a delta peak in $x=0$, the non recombinant peak near this end of the phenotypic space is driven to extinction. The higher competition pressure now induces the appearance of a second peak of non recombinants at the center of the phenotypic space. The plots of frequency, variance and mean are very similar to those found in the case $J = 0.8$ and they are shown in Figure~\ref{fig:13}. The mean of non recombinants increases after the disappearance of the peak near $x=0$ while the variance decreases because the two surviving peaks are not far from the mean. On the contrary, mean and variance of  recombinants decrease monotonically as already discussed in the case $J = 0.8$.

\paragraph{Strong competition}

\begin{figure}[ht!]

\begin{tabular}{ccc}
\hspace{-2 cm} & \includegraphics[scale=0.65]{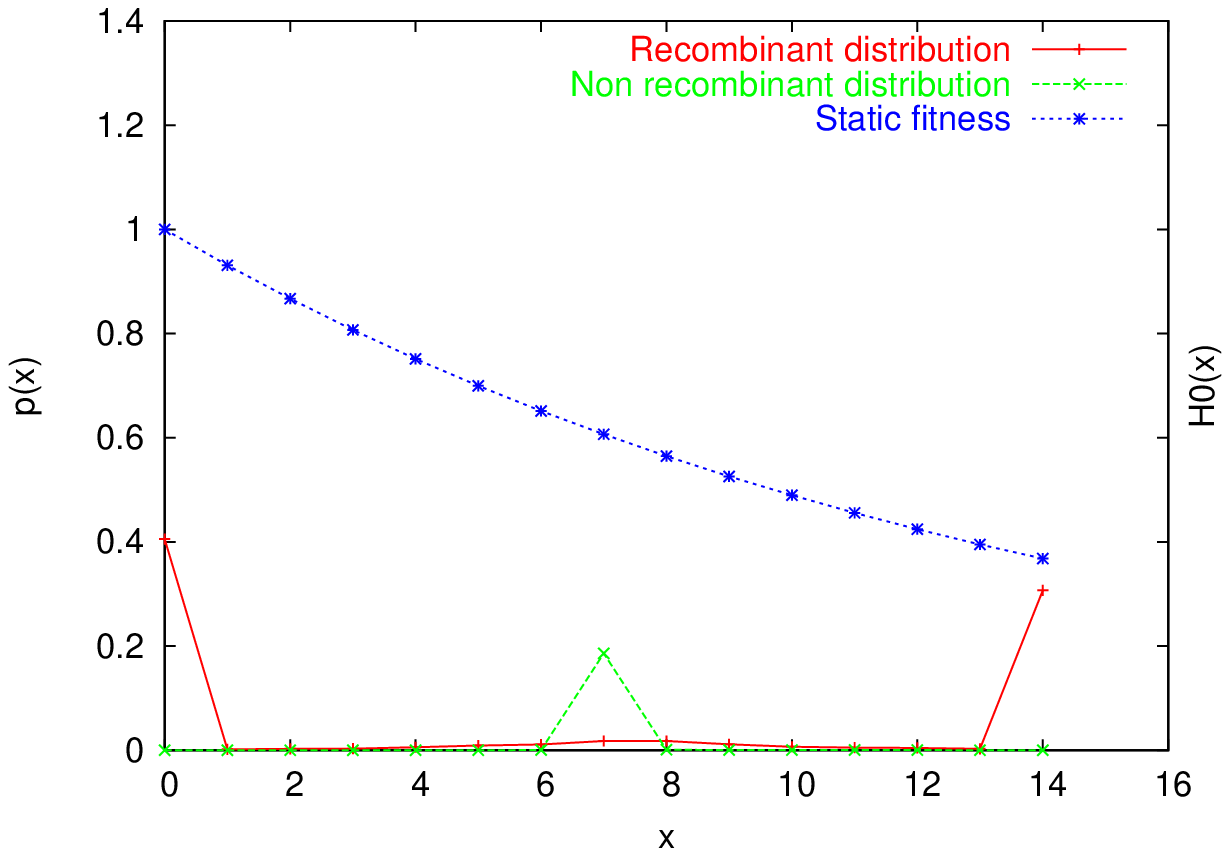} &
\includegraphics[scale=0.65]{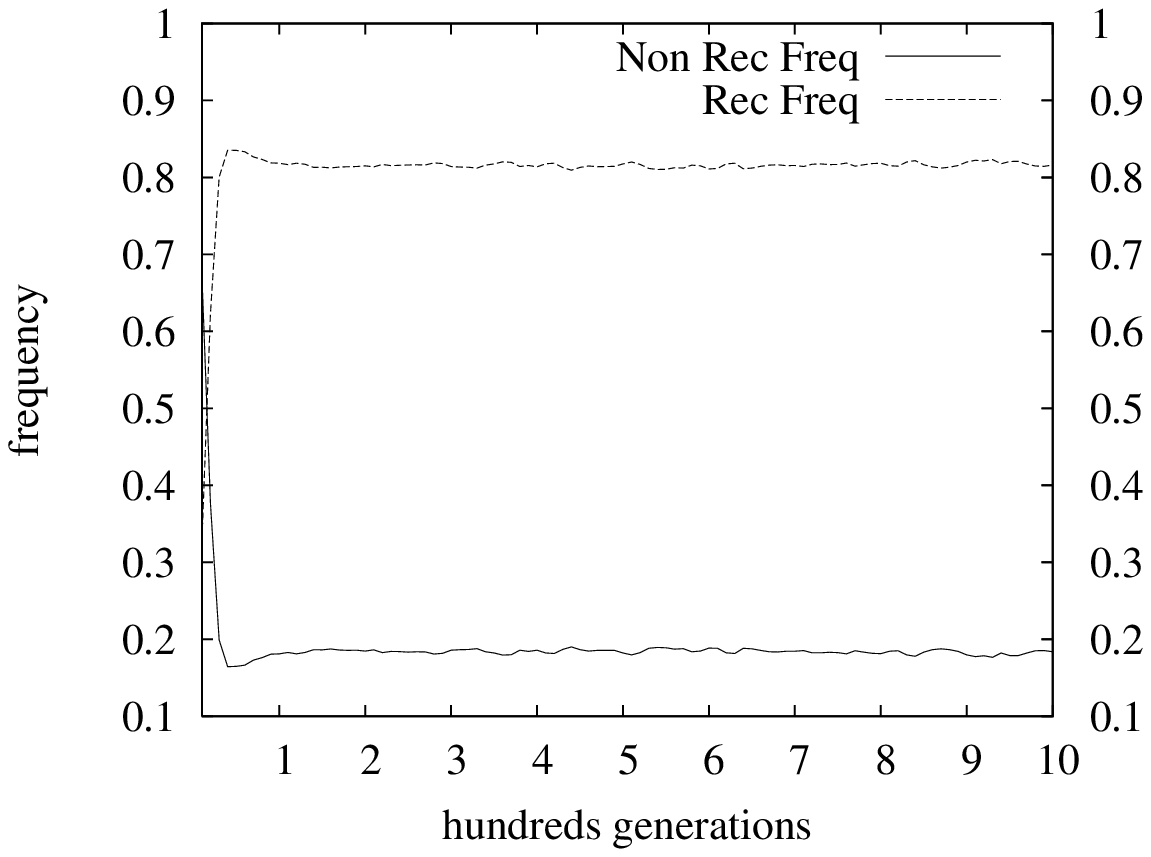} \\
\hspace{-2 cm} & \includegraphics[scale=0.65]{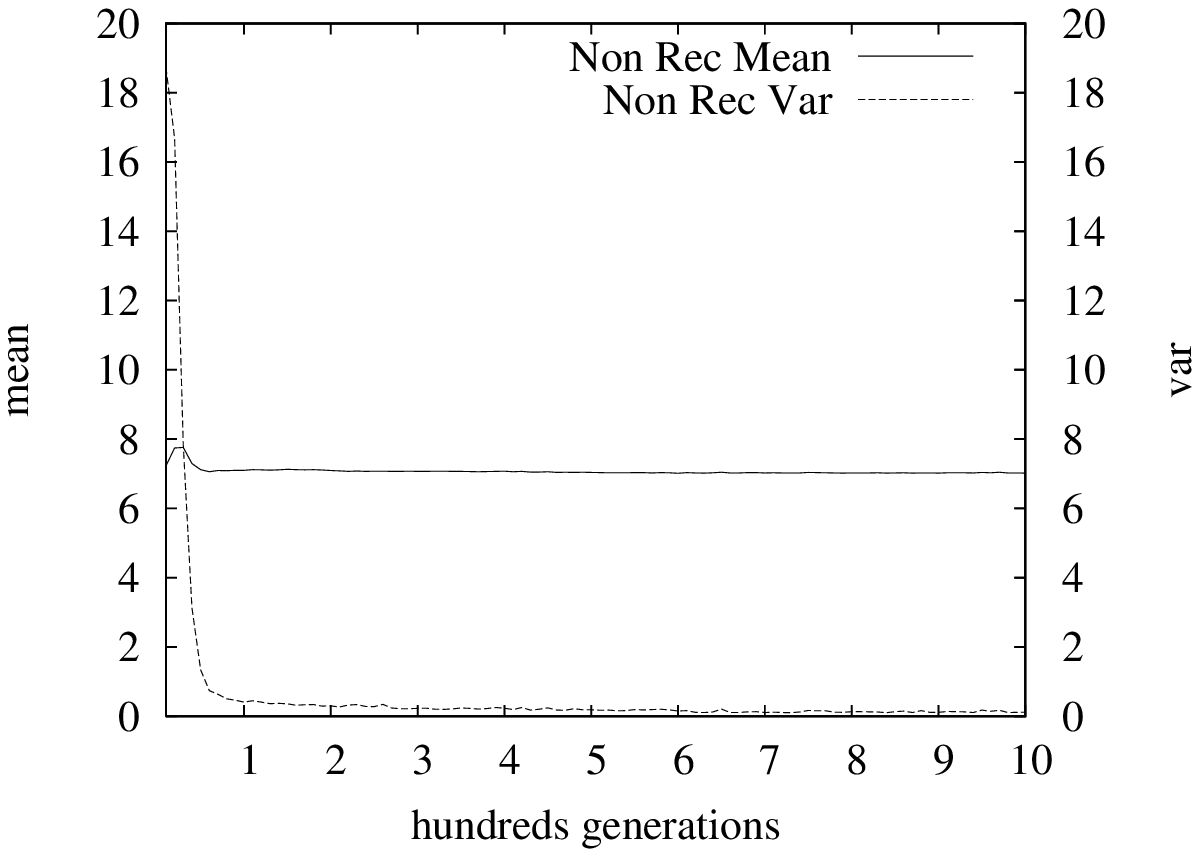} &
\includegraphics[scale=0.65]{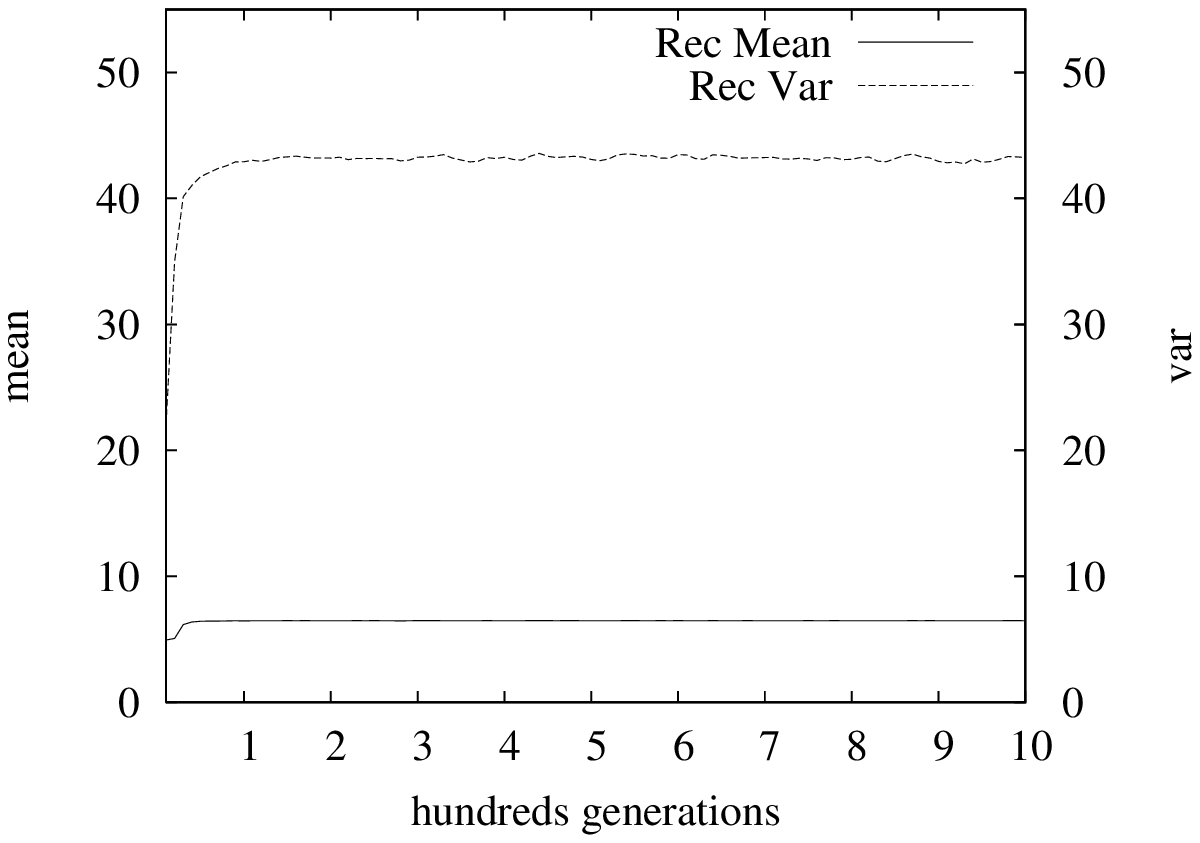} \\

\end{tabular}

\caption{Steep static fitness ($\beta=1$, $\Gamma=14$) and strong competition ($J=7$) in a random mating regime ($\Delta = 14$): top left: final distribution; top right: frequency plots; bottom left: mean and variance of non recombinant distribution; bottom right: mean and variance of recombinant distribution. Initial frequency of recombinants: 0.5; initial distribution parameters: $p = q= 0.5$, initial population size $N_0 = 1000$, carrying capacity $K = 10000$. Annealing parameters: $\mu_0 = 10^{-5}$, $ \mu_{\infty} = 10^{-6}$, $\tau = 1000$, $\delta = 100$. Total evolution time: 10000 generations; for the sake of clarity only 1000 generations were displayed in the plots of mean and variance of  recombinants and non recombinants and in the frequency
 plots.  }
\label{fig:14}

\end{figure}

In a regime of strong competition both distributions rapidly split in two peaks at the opposite ends of the phenotypic space. In particular, the recombinants quickly reach the $x=0$ and $x=14$ positions where competition is minimal and they force to extinction the peaks of non recombinants in $x=1-2$ and in $x=12-13$. The extinction of the latter two peaks relieves competition in the middle of the phenotype space where a new peak of non recombinants appears, usually in $x=7$. It can also be noticed that the peaks of recombinants in $x=0$ and $x=14$ are linked by a very flat and wide bell-shaped distribution fed by the offsprings yielded by the $0\times 14$ crossings that occur in a regime of random mating. The mean of non recombinants is constantly equal to 7  because all the deformations of the distribution are symmetrical; the variance conversely decreases almost to zero because in the end only a sharp peak of non recombinants survives in $x=7$. The variance of the recombinants on the other hand increases monotonically to a very large value ($var \cong 43$) related to the fact that the peaks in $x=0$ and $x=14$ in the final distribution are very far from the mean (that is constantly equal to 7). The plots of frequency, mean and variance are shown in Figure \ref{fig:14}.

\paragraph{The role of assortativity}

This simulation shows that even using an intermediate level of competition ($J = 1.8$), the introduction of a regime of maximal assortativity ($\Delta = 0$) yields a stationary distribution similar to that attained with very high levels of competition and random mating ($J=7$, $\Delta = 14$). The simulation with $J=1.8$, $\Delta = 0$ in fact, yields two peaks of recombinants in $x=0$ and $x=14$ and a peak of non recombinants spanning phenotypes $x=7$ and $x=8$. For the sake of comparison, recall that the simulation with $J = 1.8$, $\Delta = 14$ leads to a single peak of recombinants in $x=0$ and two peaks of non recombinants in $x=7$ and $x=12$. It can be therefore concluded that a high value of assortativity leads to an evolutionary pattern similar to the one determined by a high competition level.

\begin{figure}[ht!]

\begin{tabular}{ccc}
\hspace{-2 cm} & \includegraphics[scale=0.65]{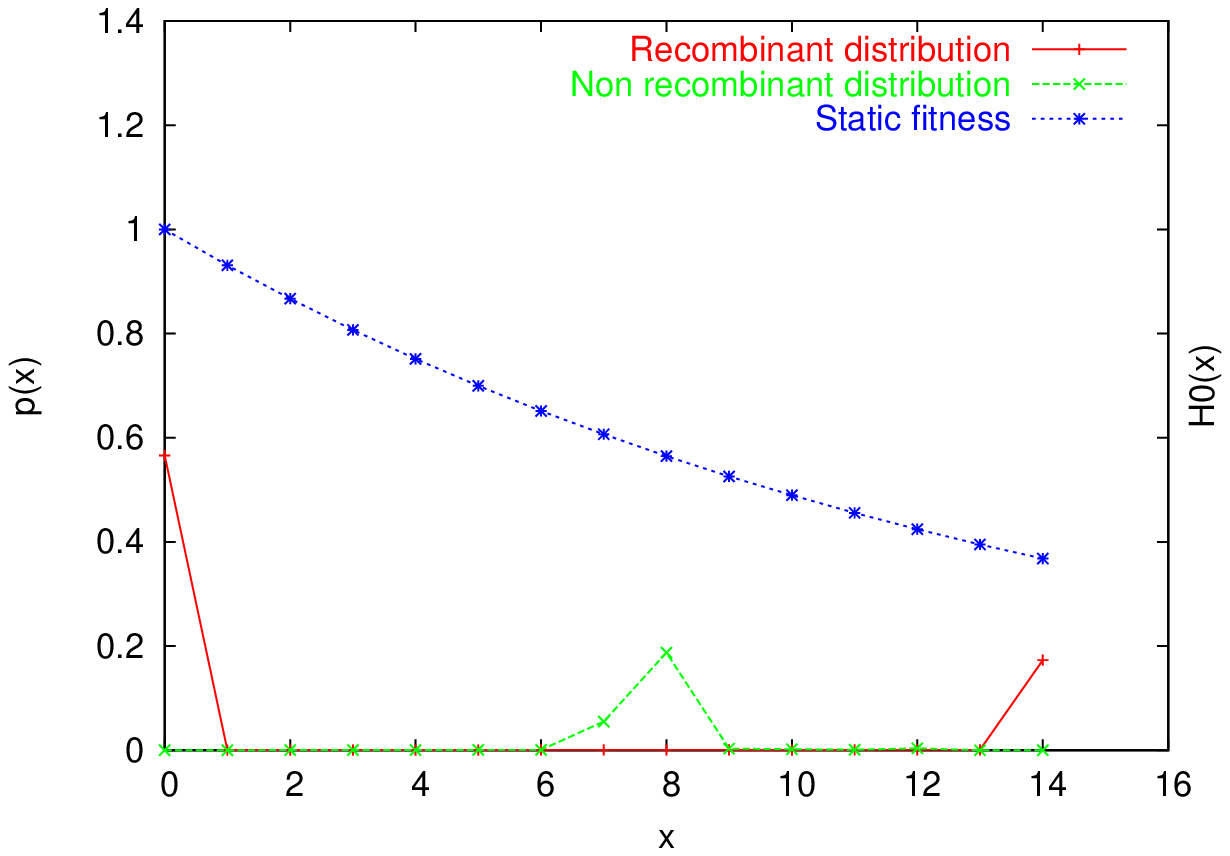} &
\includegraphics[scale=0.65]{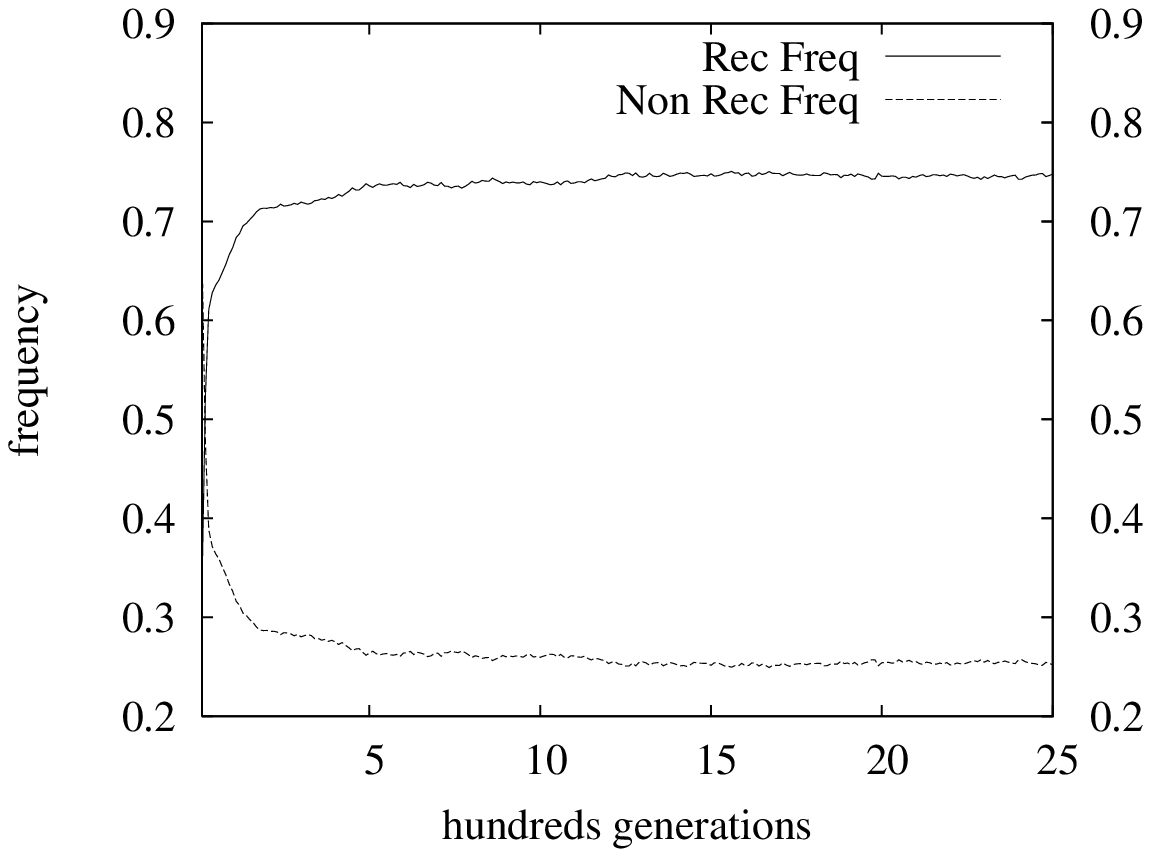} \\
\hspace{-2 cm} & \includegraphics[scale=0.65]{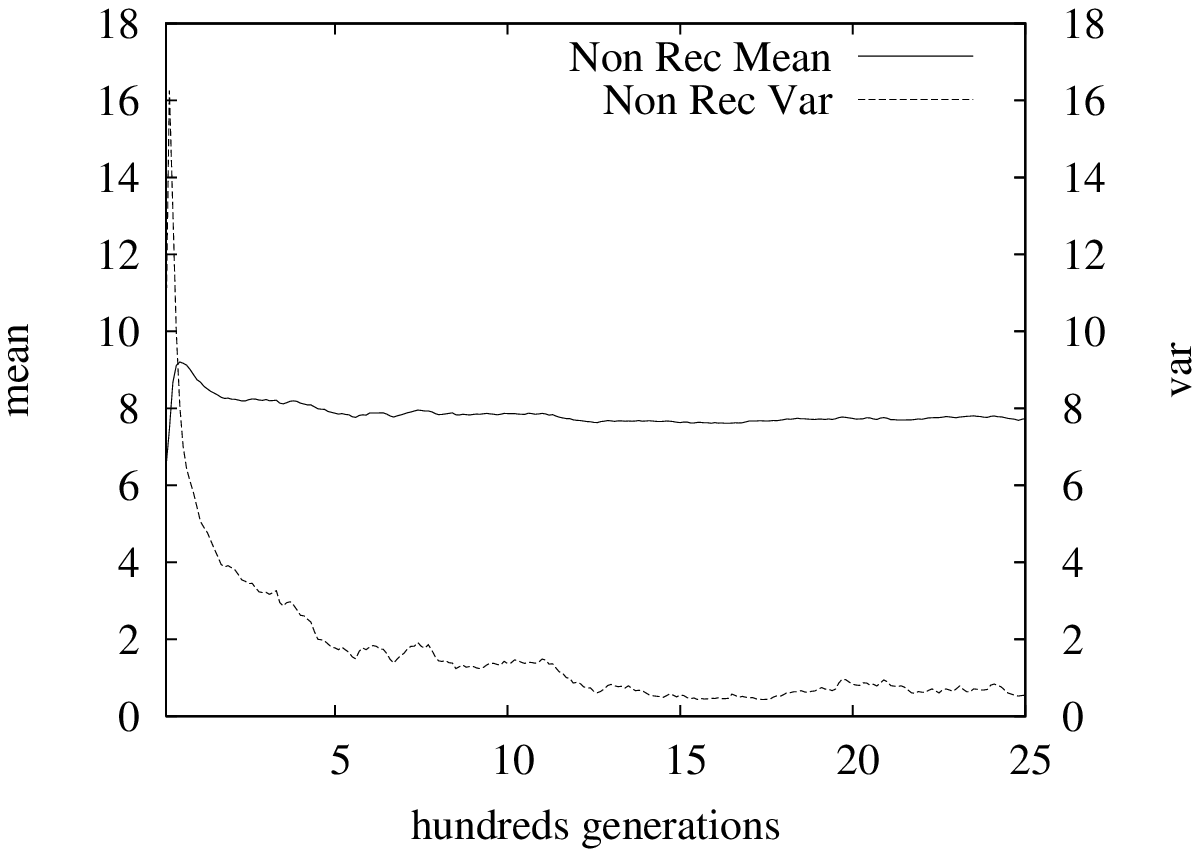} &
\includegraphics[scale=0.65]{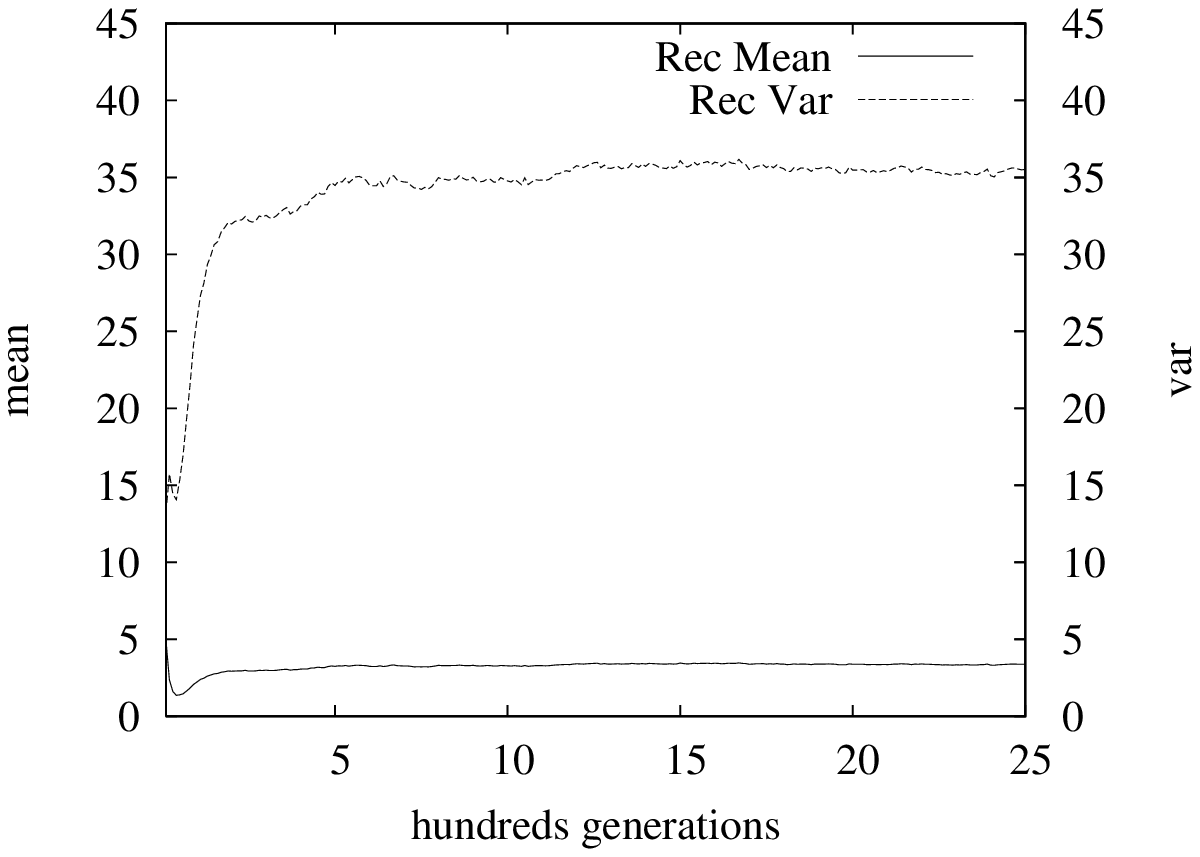} \\

\end{tabular}

\caption{Steep static fitness ($\beta=1$, $\Gamma=14$) and intermediate competition ($J=1.8$) in a maximal assortativity  regime ($\Delta = 0$): top left: final distribution; top right: frequency plots; bottom left: mean and variance of non recombinant distribution; bottom right: mean and variance of recombinant distribution. Initial frequency of recombinants: 0.5; initial distribution parameters: $p = q= 0.5$, initial population size $N_0 = 1000$, carrying capacity $K = 10000$. Annealing parameters: $\mu_0 = 10^{-5}$, $ \mu_{\infty} = 10^{-6}$, $\tau = 1000$, $\delta = 100$. Total evolution time: 10000 generations; for the sake of clarity only 2500 generations were displayed in the plots of mean and variance of  recombinants and non recombinants and in the frequency plots.  }
\label{fig:15}

\end{figure}

From a dynamical point of view, during the simulation both the distribution of recombinants and non recombinants tend to become bimodal, but the recombinants are the first to reach the $x=0$ and $x=14$ positions thus determining the extinction of the peaks of non recombinants in $x=1-2$ and $x=12-13$; after that a new peak of non recombinants appears in the central region of the phenotypic space. In the example that we are showing the mean of non recombinants tends to increase as the first peak that becomes extinct is the one nearer to $x=0$. The variance of non recombinants conversely decreases as their final distribution only features a very sharp peak at $x=7-8$. The mean of recombinants stabilizes on a constant value between 4 and 5 because the peak in $x=0$ is more populated than the one in $x=14$ and the mean is thus closer to the former than to the latter. The variance, finally reaches a very high value because the two peaks are very far from the mean. The plots relative to this simulation are shown in Figure \ref{fig:15}.

\subsection{Influence of genome length}

\subsubsection{Flat static fitness landscape}

In all the simulations performed so far we set a genome length $L=14$. We now investigate the evolutionary dynamics with $L=28$ to study the influence of genome length on the evolution of recombinants and non recombinants. As the genome length was doubled, the mating range $\Delta$, the competition range $R$ and the steepness parameter $\Gamma$ of the static fitness have also been doubled as compared to the simulations of the previous sections.

The first simulation we discuss must be compared to the one shown in Figures~\ref{fig2:fin-distr} and~\ref{fig2:var-mean}. The initial distribution is composed of a delta peak of recombinants in $x=0$ accounting for $90 \%$ of the population and a delta peak of non recombinants in $x=28$. If we set the same low competition level chosen in the simulation with $L=14$ ($\mu_0 = 5 \times 10^{-4}$, $\mu_{\infty} = 10^{-6}$, $\tau = 1000$, $\delta = 100$) two new peaks of recombinants and non recombinants depart from the old ones in $x=0$ and $x=28$ respectively, and move towards the center of the phenotypic space where competition is lower. Due to the very low mutation level, however, the new peaks cannot go very far from the old ones and the final distribution will be characterized by two peaks of recombinants near $x=0$ and two peaks of non recombinants near $x=28$. In agreement with this pattern, the values of  mean and variance of non recombinants are initially 28 and 0 respectively; after that the mean decreases and the variance increases. The mean and variance of non recombinants conversely both start from zero and then increase monotonically to a higher stable level. This simulation clearly shows that a longer genome length determines an enlargement of the phenotypic space so that higher mutation levels are required to provide both recombinants and non recombinants with a mobility sufficiently high to traverse wider phenotypic distances.

\begin{figure}[ht!]

\begin{tabular}{ccc}
\hspace{-2 cm} & \includegraphics[scale=0.65]{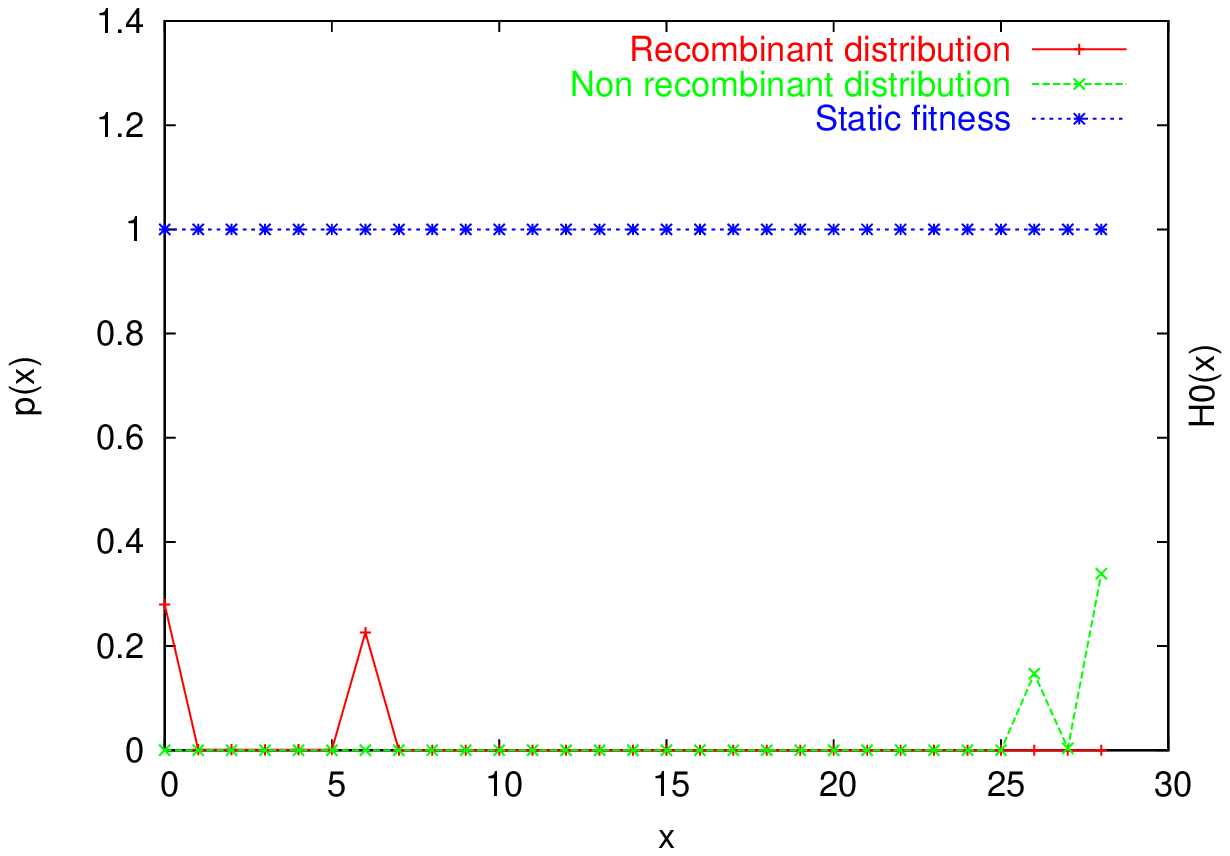} &
\includegraphics[scale=0.65]{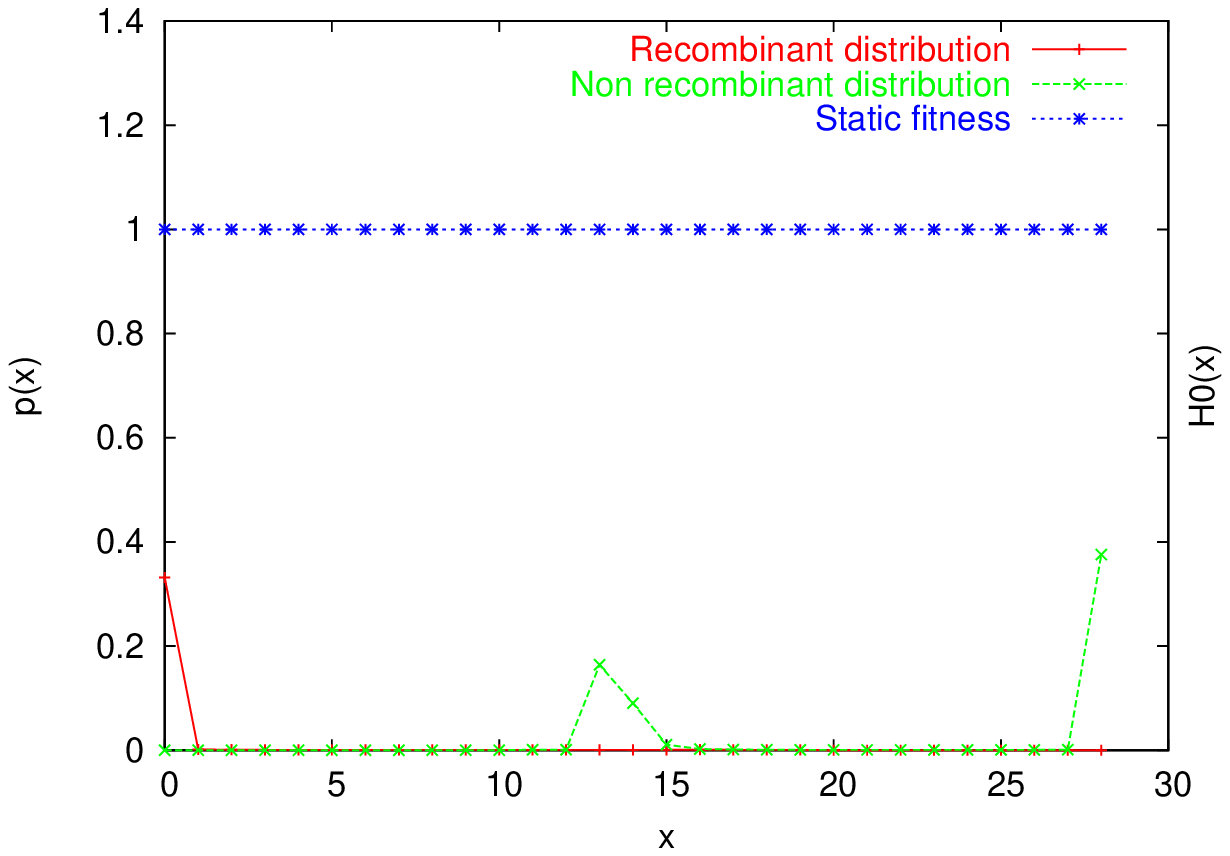} \\
\end{tabular}

\caption{Effects of genome length and mutation levels: final distributions . Left panel: low mutation levels
$\mu_0 = 5 \times 10^{-4}$, $\mu_{\infty} = 10^{-6}$, $\tau = 1000$ and $\delta = 100$; right panel: high
mutation level: $\mu_0 =10^{-2}$, $\mu_{\infty} = 10^{-6}$, $\tau = 2000$ and $\delta = 200$. Flat static
fitness $\beta = 100$, $\Gamma = 28$ and weak competition $J=1$, $\alpha = 2$, $R = 8$. Initial distribution: $p = 0$, $q = 1$, $M = 0.9$, $N_0 = 1000$, $K = 10000$. Mating range $\Delta = 0$, genome length $L = 28$.
Total evolution time: 10000 generations.      }

\label{fig3_L28_distr}

\end{figure}

\begin{figure}[ht!]

\begin{tabular}{ccc}
\hspace{-2 cm} & \includegraphics[scale=0.65]{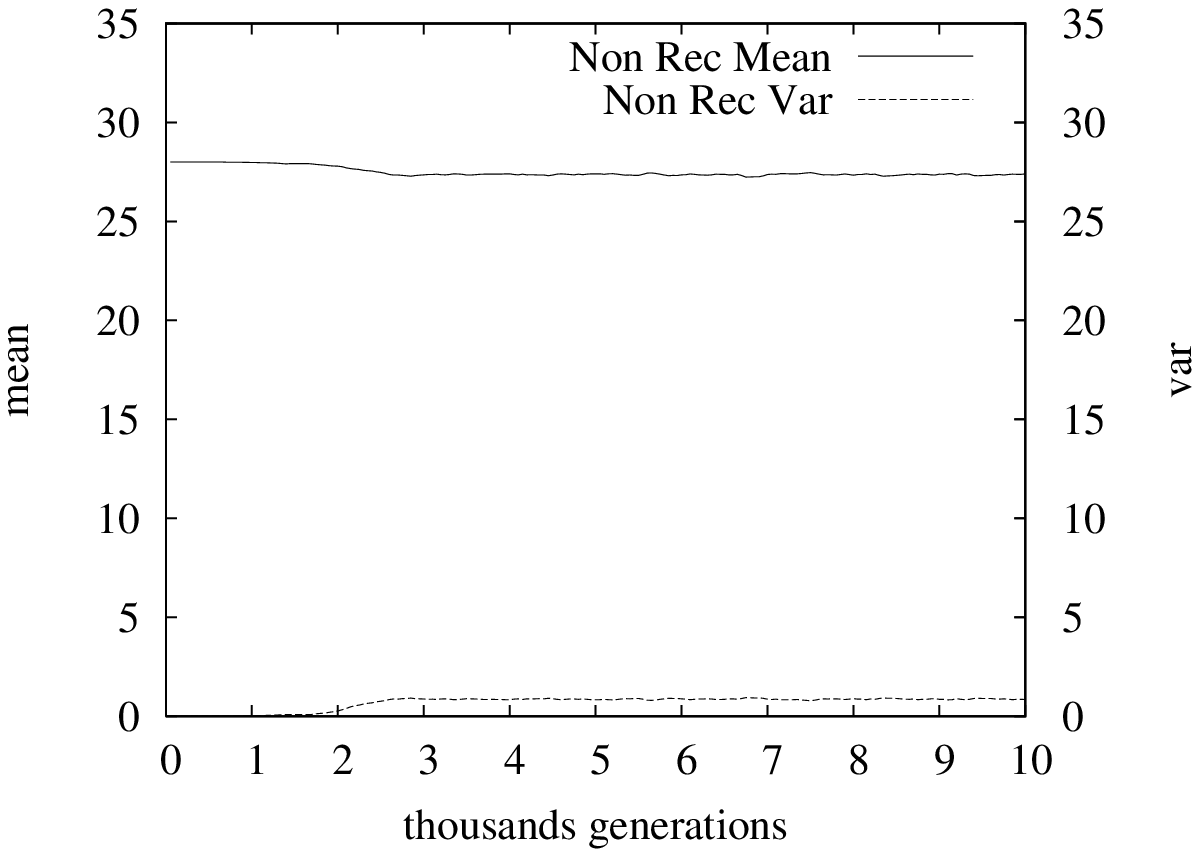} &
\includegraphics[scale=0.65]{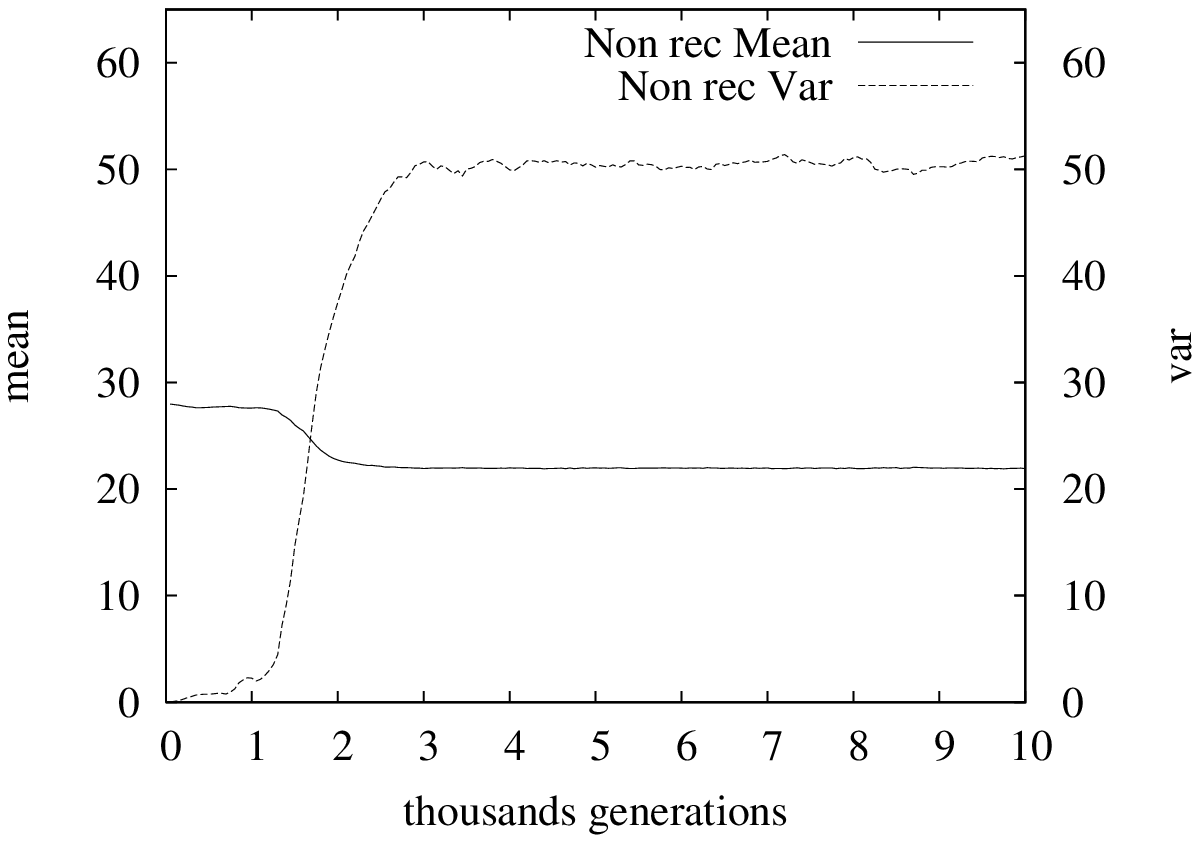} \\
\hspace{-2 cm} & \includegraphics[scale=0.65]{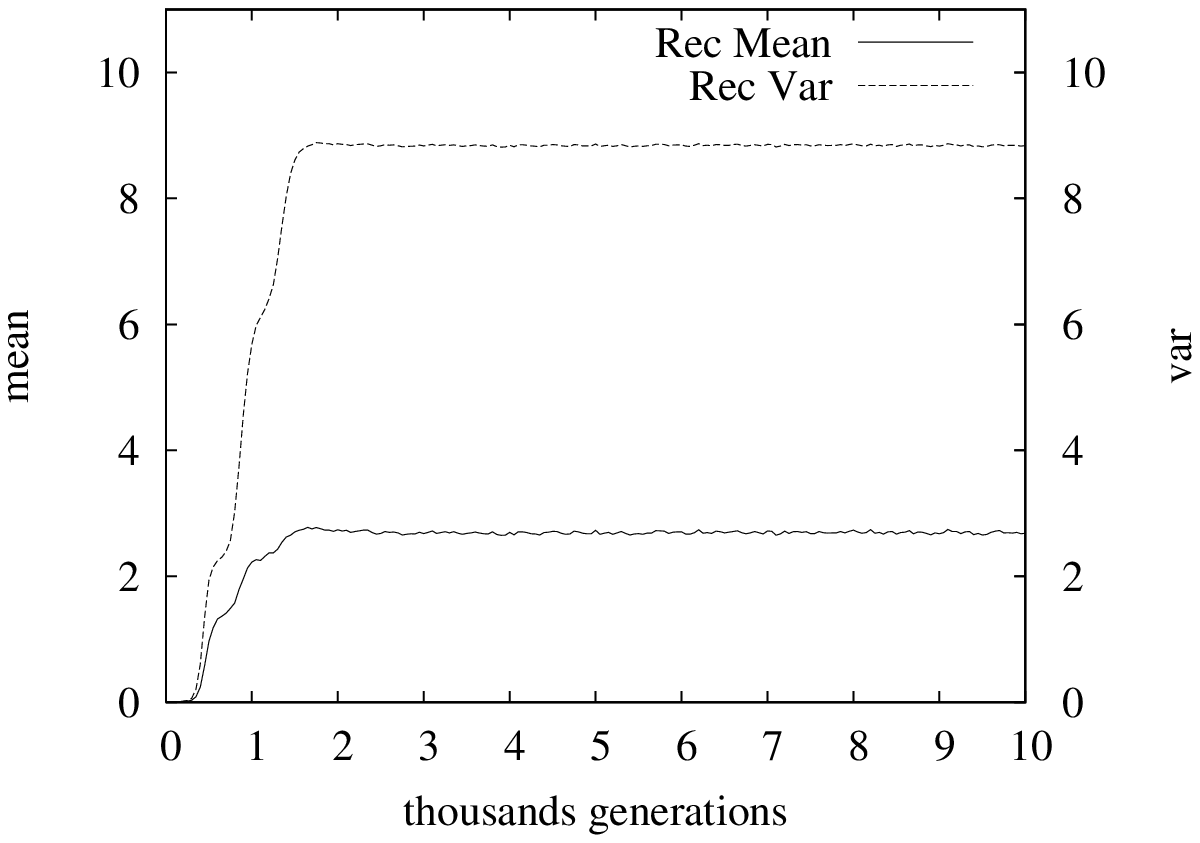} &
\includegraphics[scale=0.65]{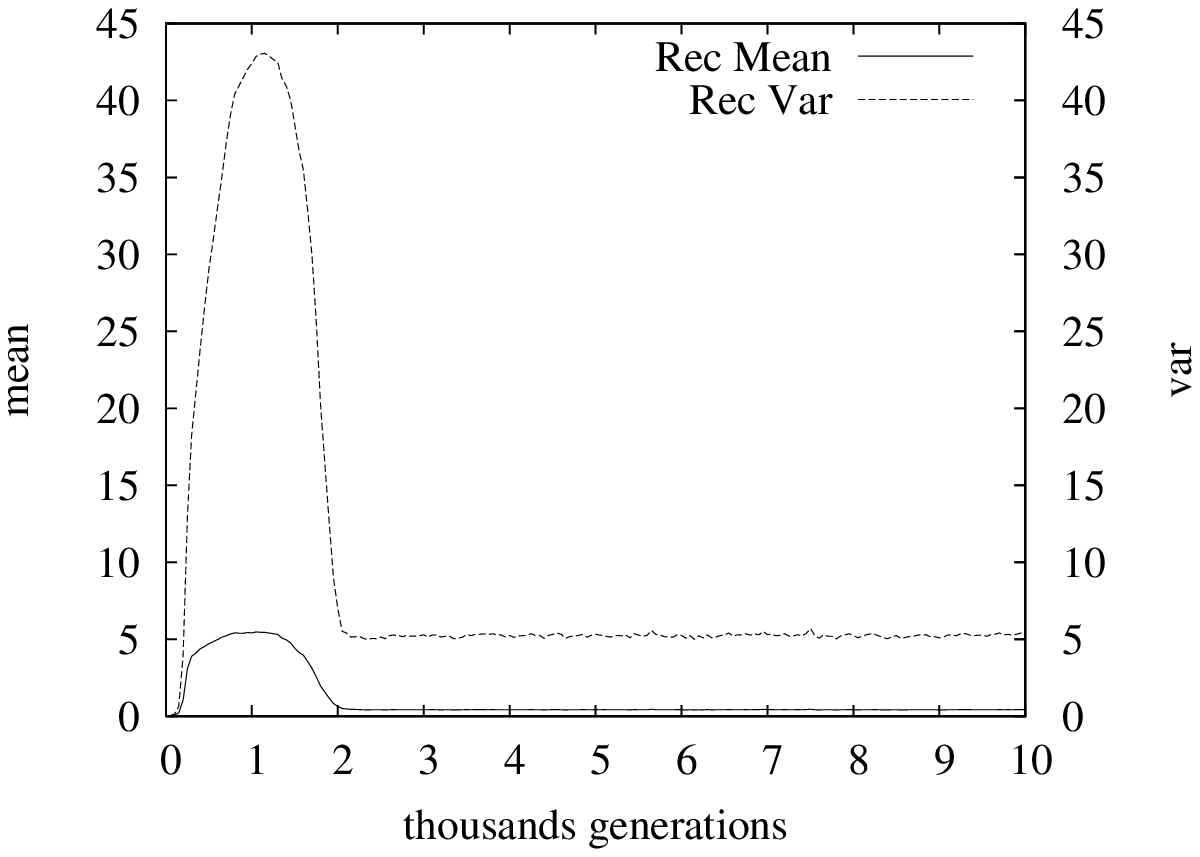} \\
\end{tabular}

\caption{Effects of genome length and mutation levels. Left column: low mutation levels $\mu_0 = 5 \times 10^{
-4}$, $\mu_{\infty} = 10^{-6}$, $\tau = 1000$ and $\delta = 100$; right column: high mutation level: $\mu_0 =
10^{-2}$, $\mu_{\infty} = 10^{-6}$, $\tau = 2000$ and $\delta = 200$. From top to bottom, first row: plots of mean and variance of non recombinants; second row: plots of mean and variance of recombinants. Flat static
fitness $\beta = 100$, $\Gamma = 28$ and weak competition $J=1$, $\alpha = 2$, $R = 8$. Initial distribution: $p = 0$, $q = 1$, $M = 0.9$, $N_0 = 1000$, $K = 10000$. Mating range $\Delta = 0$, genome length $L = 28$.
Total evolution time: 10000 generations.      }

\label{fig3_L28_var_mean}

\end{figure}

In fact, if we repeat the simulation with $\mu_0 = 10^{-2}$, $\mu_{\infty} = 10^{-6}$, $\tau = 2000$ and $\delta = 200$, the two new peaks of recombinants and non recombinants move towards the center of the phenotypic space and when they are close enough to each other the recombinant peak disappears as the lower fertility makes it  unable to resist the competition exerted by the non recombinant peak. The final distribution will be therefore characterized by a peak of recombinants in $x=0$ and two peaks of non recombinants in $x=28$ and in the center of the phenotypic space. This pattern is exactly what was found in the simulation with $L=14$ shown in Figures~\ref{fig2:fin-distr} and~\ref{fig2:var-mean}. As a consequence, the mean of non recombinants decreases from 28 to a lower stable level while the variance tends to increase. Mean and variance of recombinants, conversely, first increase up to a maximum as the newborn peaks moves towards the center of the phenotypic space, and then they drop abruptly when this peak becomes extinct. The final distributions and the plots of mean and variance of recombinants and non recombinants in the case of low and high mutation levels are shown in Figures~\ref{fig3_L28_distr} and~\ref{fig3_L28_var_mean}.

We now repeat the above simulation by changing the initial distribution such that $p = q = 0.5$. This experiment can be therefore compared with the simulation illustrated in Figures~\ref{fig3:examples} and~\ref{fig3:var-mean}. In this simulation the recombinant distribution splits in two bell-shaped curves moving towards the opposite ends of the phenotypic space until two delta peaks are established in $x=0$ and $x=28$; the non recombinants are confined to one or a few peaks in the center of the phenotypic space. If the mutation level is very low ($\mu_0 = 5 \times 10^{-4}$, $\mu_{\infty} = 10^{-6}$, $\tau = 1000$ and $\delta = 100$) this will be the final distribution. The variance of non recombinants decreases as their distribution becomes narrower and narrower but the mean remains roughly constant because the distribution changes in a symmetrical way. The variance of recombinants conversely rapidly reaches a very high level related to the creation of two delta peaks at the ends of the space.

\begin{figure}[ht!]
                                                                                                              \begin{tabular}{ccc}                                                                                          \hspace{-2 cm} & \includegraphics[scale=0.65]{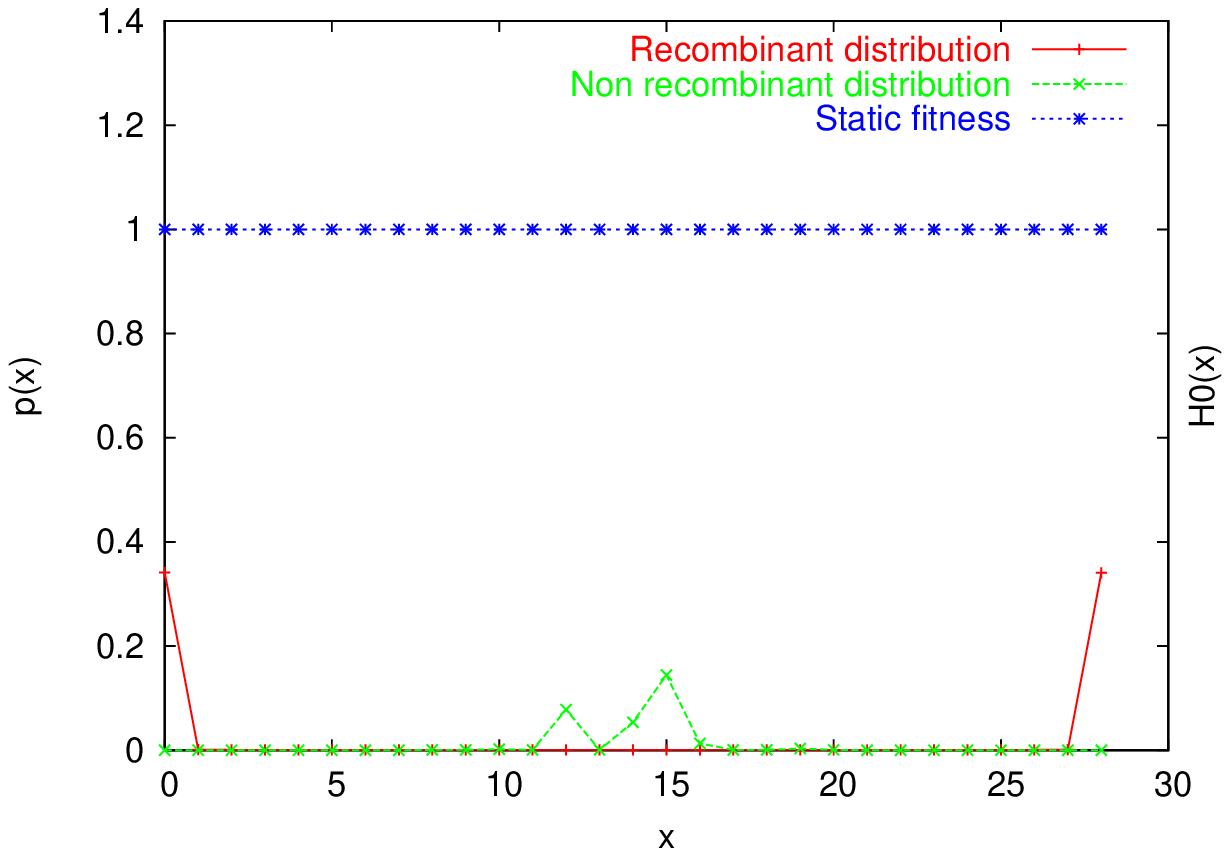} &
													      \includegraphics[scale=0.65]{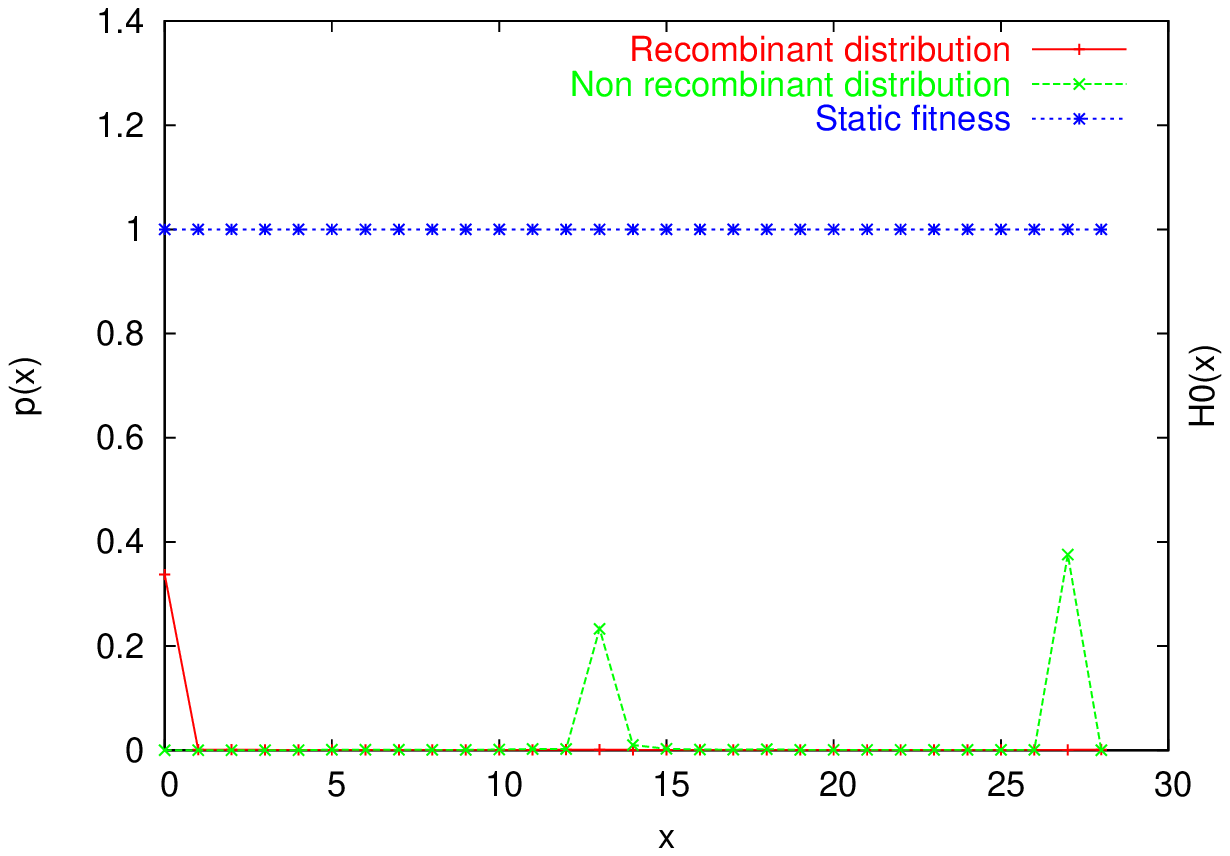} \\
													      \end{tabular}

													      \caption{Effects of genome length and mutation levels: final distributions. Left panel: low mutation levels $\mu_0 = 5 \times 10^{-4}$, $\mu_{\infty} = 10^{-6}$, $\tau = 1000$ and $\delta = 100$; right panel: high mutation level: $\mu_0 = 10^{-2}$, $\mu_{\infty} = 10^{-6}$, $\tau = 2000$ and $\delta = 200$. Flat static fitness $\beta = 100$, $\Gamma = 28$ and weak competition $J=1$, $\alpha = 2$, $R = 8$. Initial distribution: $p = 0.5$, $q = 0.5$, $M = 0.9$, $N_0 = 1000$, $K = 10000$. Mating range $\Delta = 0$, genome length $L = 28$. Total evolution time: 10000 generations.      }

													      \label{fig5_L28_distr}

													      \end{figure}

\begin{figure}[ht!]

\begin{tabular}{ccc}
\hspace{-2 cm} & \includegraphics[scale=0.65]{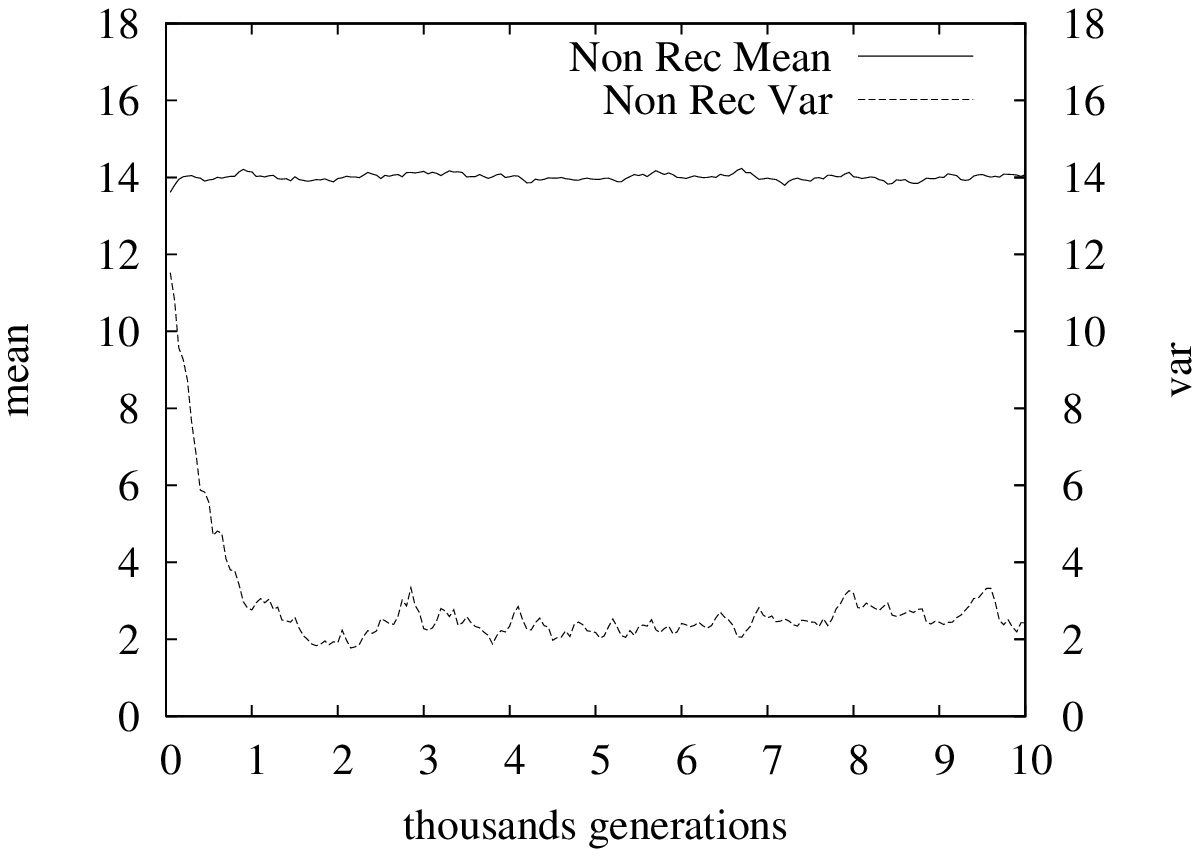} & \includegraphics[scale=0.65]{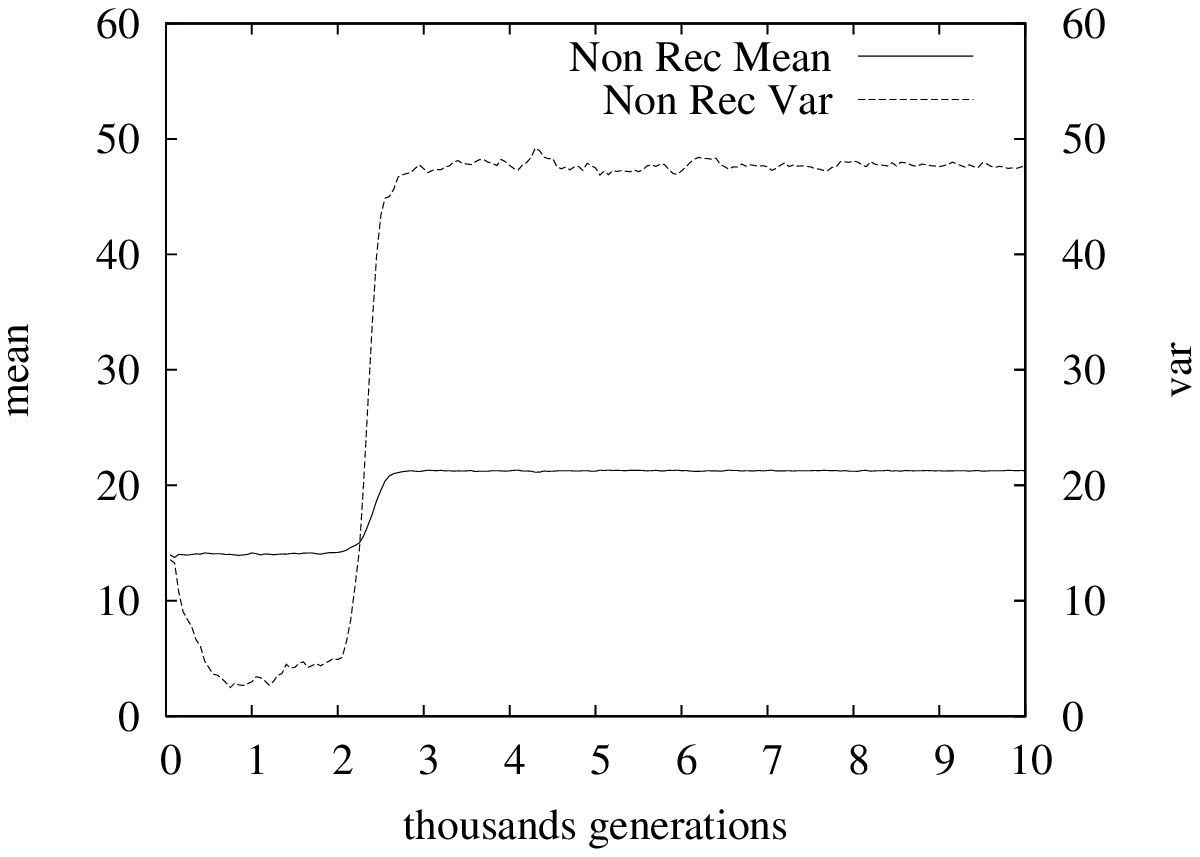} \\
\hspace{-2 cm} & \includegraphics[scale=0.65]{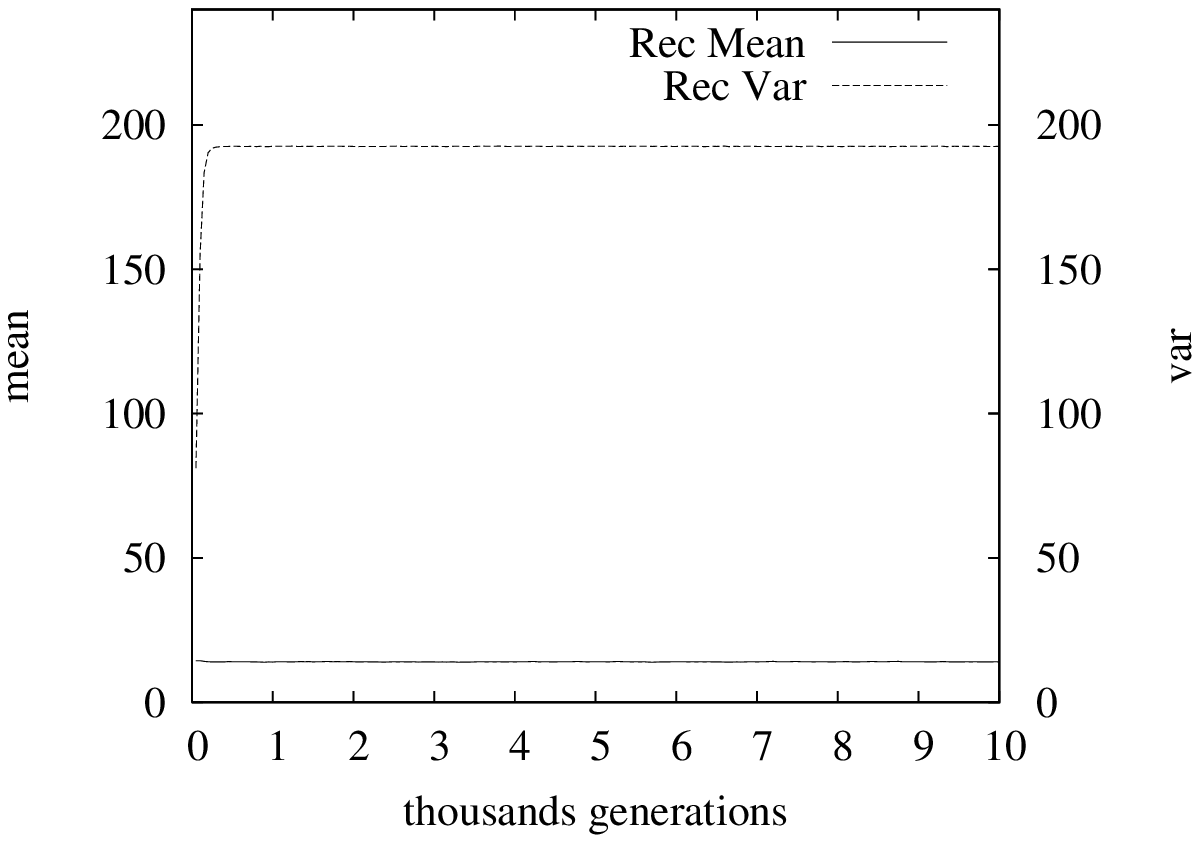} & \includegraphics[scale=0.65]{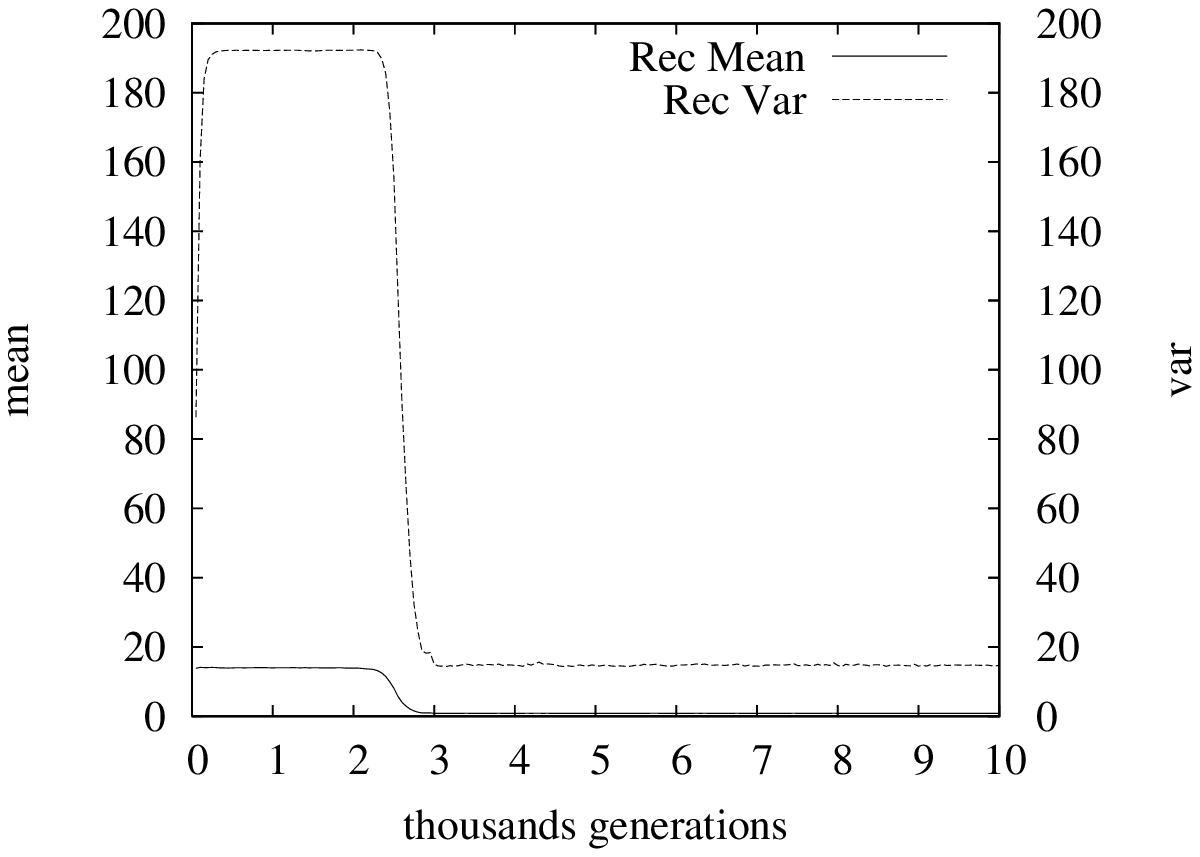} \\
\end{tabular}

\caption{Effects of genome length and mutation levels. Left column: low mutation levels $\mu_0 = 5 \times 10^{-4}$, $\mu_{\infty} = 10^{-6}$, $\tau = 1000$ and $\delta = 100$; right column: high mutation level: $\mu_0 = 10^{-2}$, $\mu_{\infty} = 10^{-6}$, $\tau = 2000$ and $\delta = 200$. From top to bottom, first row:  plots of mean and variance of non recombinants; second row: plots of mean and variance of recombinants. Flat static fitness $\beta = 100$, $\Gamma = 28$ and weak competition $J=1$, $\alpha = 2$, $R = 8$. Initial distribution: $p = 0.5$, $q = 0.5$, $M = 0.9$, $N_0 = 1000$, $K = 10000$. Mating range $\Delta = 0$, genome length $L = 28$. Total evolution time: 10000 generations.      }

\label{fig5_L28_var_mean}

\end{figure}

If the mutation level is sufficiently high, however, the non recombinants in the central peak produce mutants that move towards the ends of the phenotypic space and owing to their higher fertility can cause a significant decrease or even extinction of the local peak of recombinants. If the new peak of non recombinants moves towards $x=28$ for example, the variance of non recombinants first decreases (the distribution reduces to a sharp peak in the center of the space) and then increases again and so does the mean. The variance of recombinants conversely first reaches a very high level (two delta peaks at the ends of the space) and then drops abruptly when one of the peaks becomes extinct; the mean will also decrease accordingly. The final distributions and the plots of mean and variance of recombinants and non recombinants in the case of low and high mutation levels are shown in Figures~\ref{fig5_L28_distr} and~\ref{fig5_L28_var_mean}.

We now discuss a simulation in a regime of random mating that can be compared with the one shown in Figure~\ref{fig7}. The simulation shows that when $L=28$ the random mating regime ($\Delta = 28$) is not sufficient to cause the extinction of recombinants. In particular, if the mutation level is low ($\mu_0 = 10^{-5}$, $\mu_{\infty} = 10^{-6}$, $\tau = 1000$, $\delta = 100$) at the beginning of the simulation the non recombinant distribution splits in two peaks moving towards the opposite ends of the phenotypic space. The low mutation rate on the other hand, causes the non recombinant peaks to move very slowly so that one of them may be passed past by a small group of recombinants that reaches one end of the phenotypic space (for example the $x=0$ end) and drives the closest peak of non recombinants to extinction. This extinction event relieves the competition pressure in the middle of the phenotypic space where a new peak of non recombinants arises. The final distribution therefore shows a peak of recombinants near or at one end and two peaks of non recombinants in the middle and at the other end. In the case where the peak of recombinants is at $x=0$ the variance of non recombinants first increases (formation of two peaks moving towards the ends of the space) and then decreases abruptly when the peak near $x=0$ is led to extinction, which also causes an increase in the mean. The variance and mean of recombinants on the other hand, first reach a very high level related to a very flat and wide distribution and then decrease monotonically after the appearance of the peak near $x=0$.

\begin{figure}[ht!]

\begin{tabular}{ccc}
\hspace{-2 cm} & \includegraphics[scale=0.65]{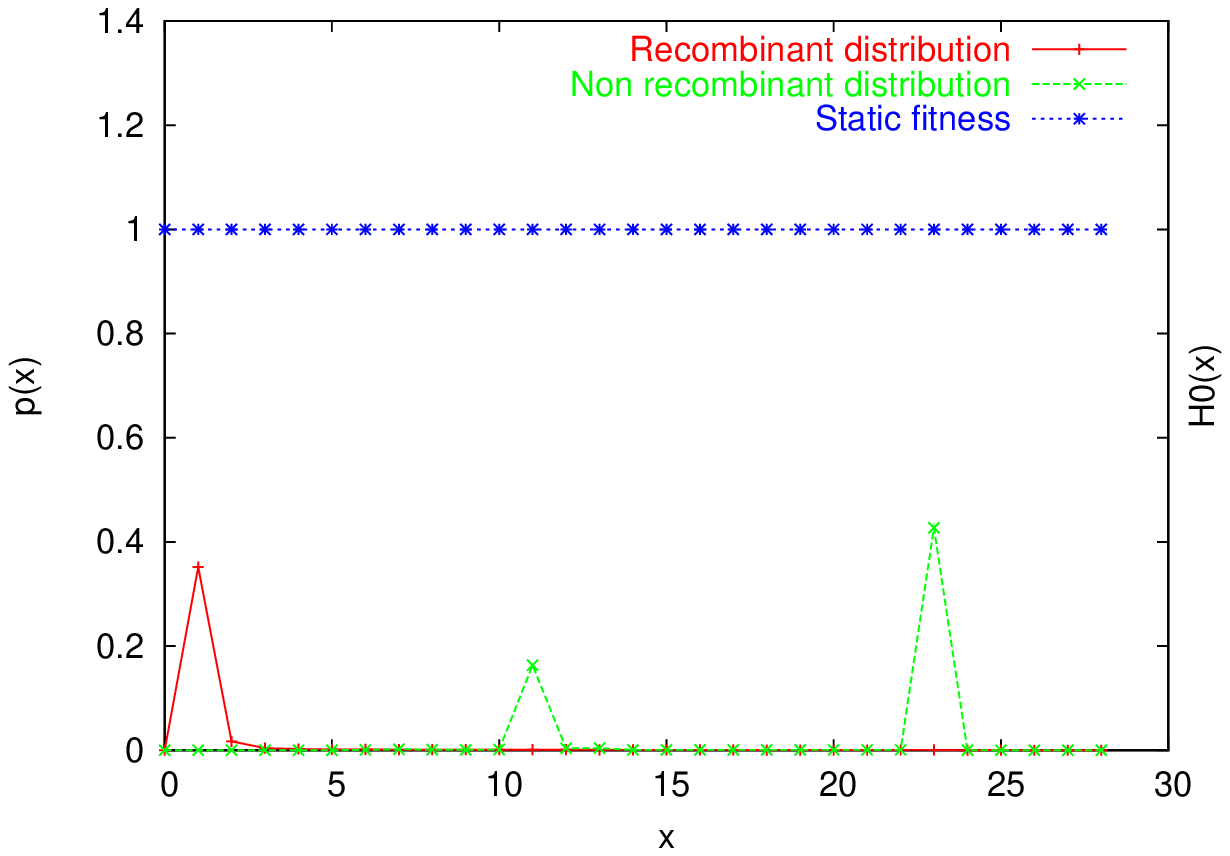} &
\includegraphics[scale=0.65]{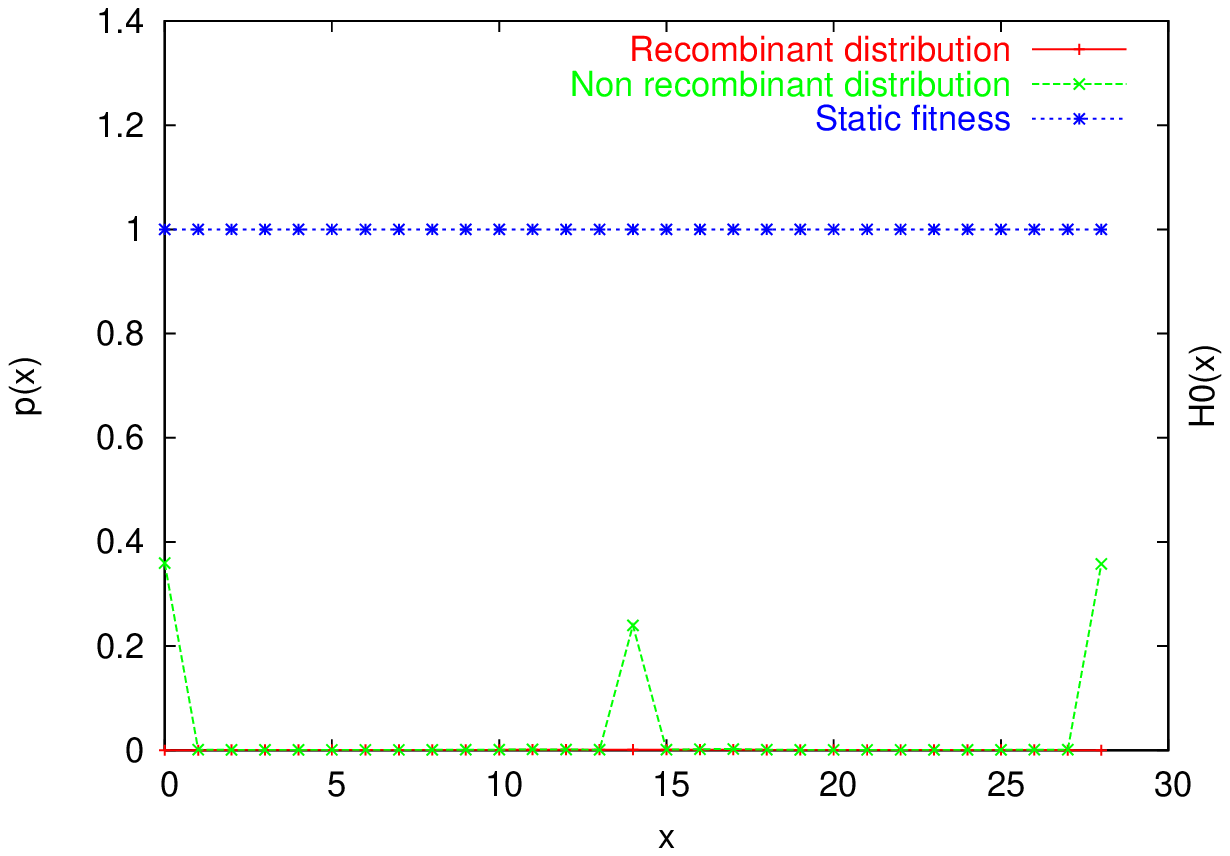} \\
\end{tabular}

\caption{Effects of genome length and mutation levels: final distributions  . Left panel: low mutation levels $\mu_0 = 10^{-5}$, $\mu_{\infty} = 10^{-6}$, $\tau = 1000$ and $\delta = 100$; right panel: high mutation level: $\mu_0 = 5 \times 10^{-3}$, $\mu_{\infty} = 10^{-6}$, $\tau = 1000$ and $\delta = 100$. Flat static fitness $\beta = 100$, $\Gamma = 28$ and weak competition $J=1$, $\alpha = 2$, $R = 8$. Initial distribution: $p = 0.5$, $q = 0.5$, $M = 0.2$, $N_0 = 1000$, $K = 10000$. Mating range $\Delta = 28$, genome length $L = 28$. Total evolution time: 10000 generations.      }

\label{fig13_L28_distr}

\end{figure}

\begin{figure}[ht!]

\begin{tabular}{ccc}
\hspace{-2 cm} & \includegraphics[scale=0.65]{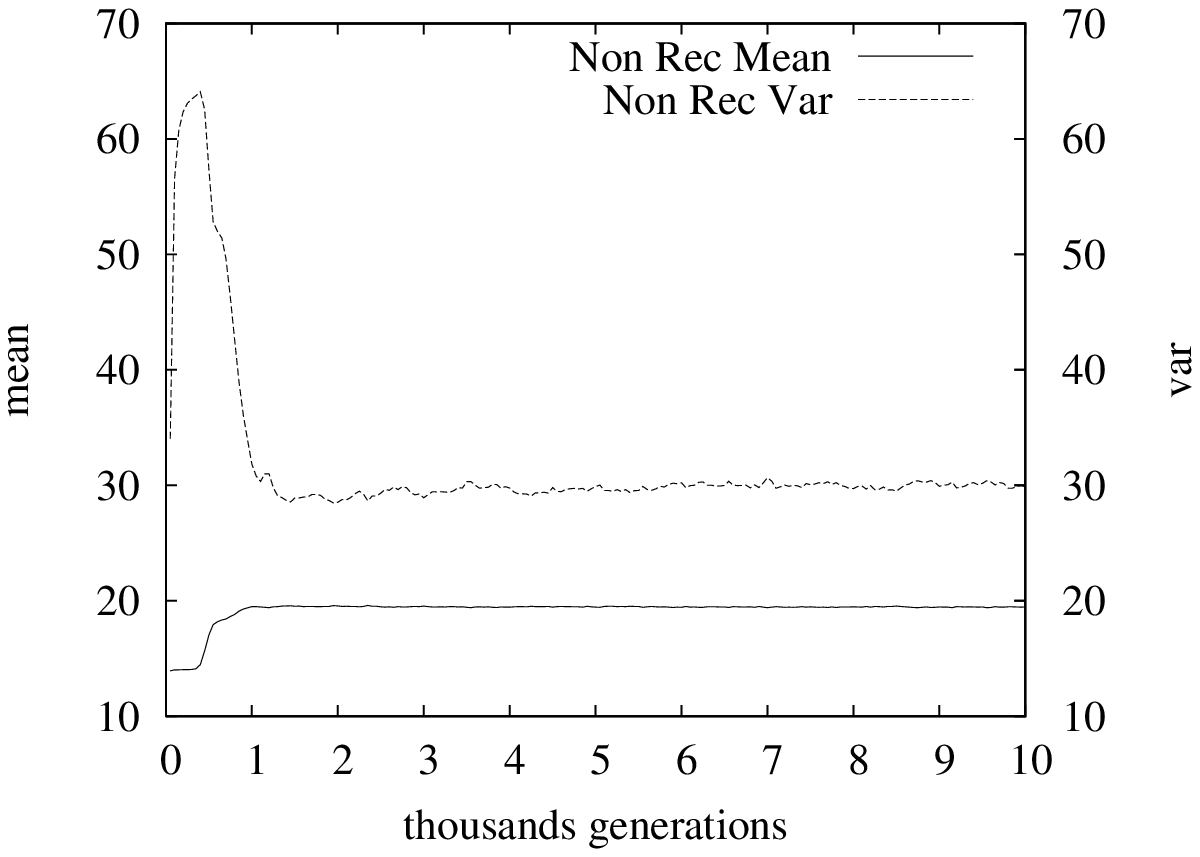} & \includegraphics[scale=0.65]{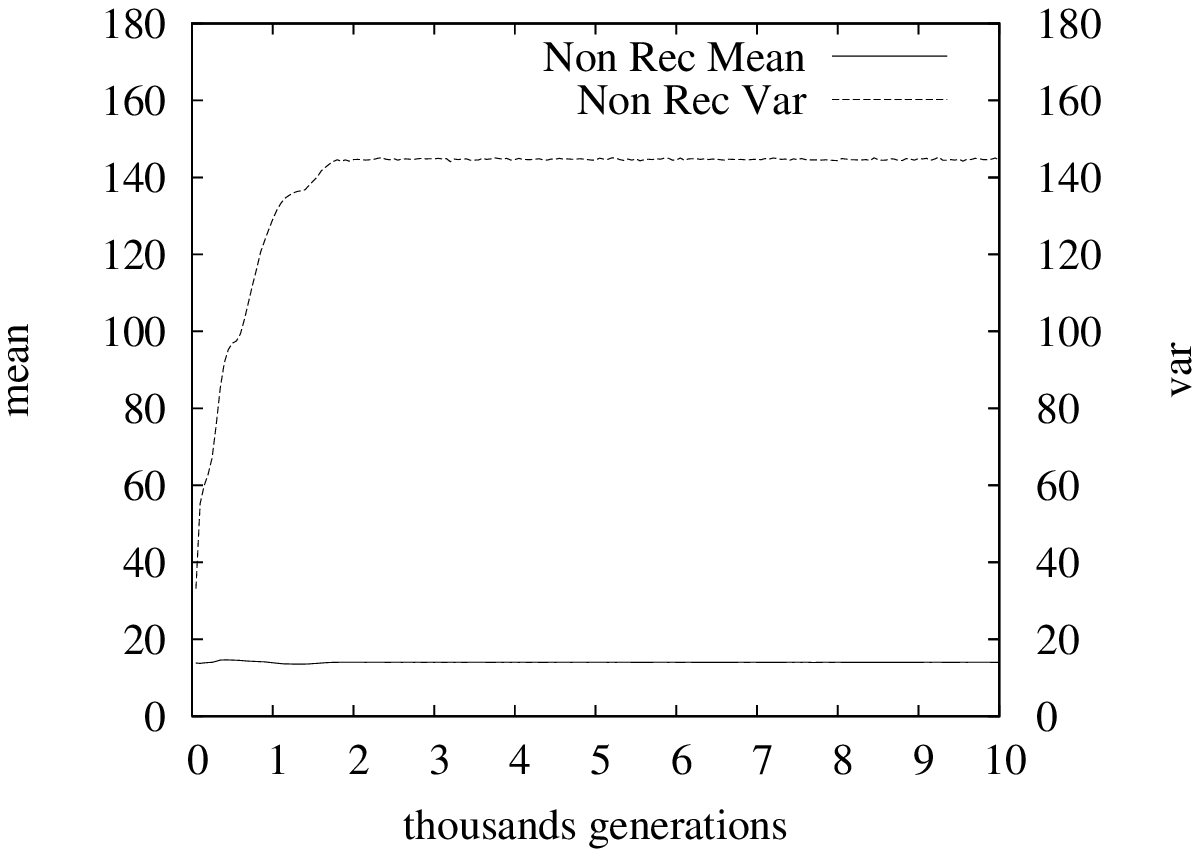} \\
\hspace{-2 cm} & \includegraphics[scale=0.65]{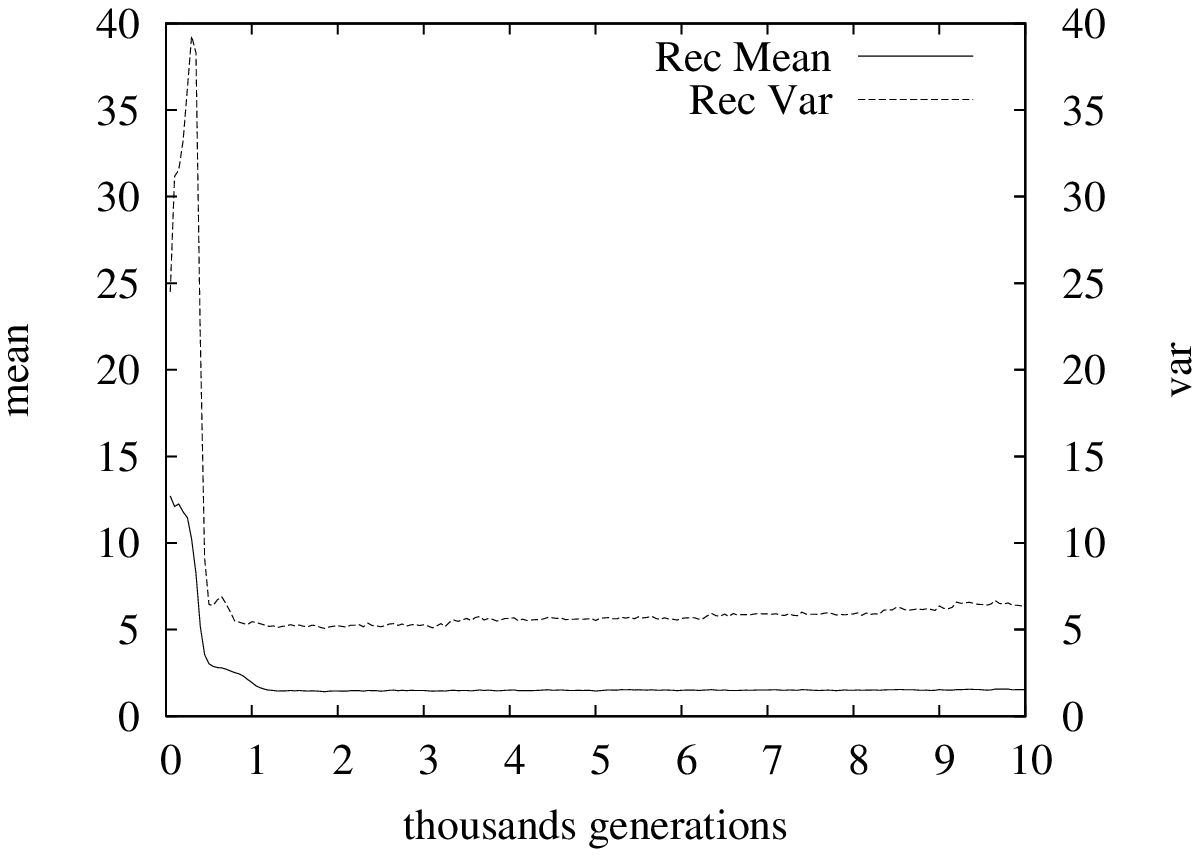} & \includegraphics[scale=0.65]{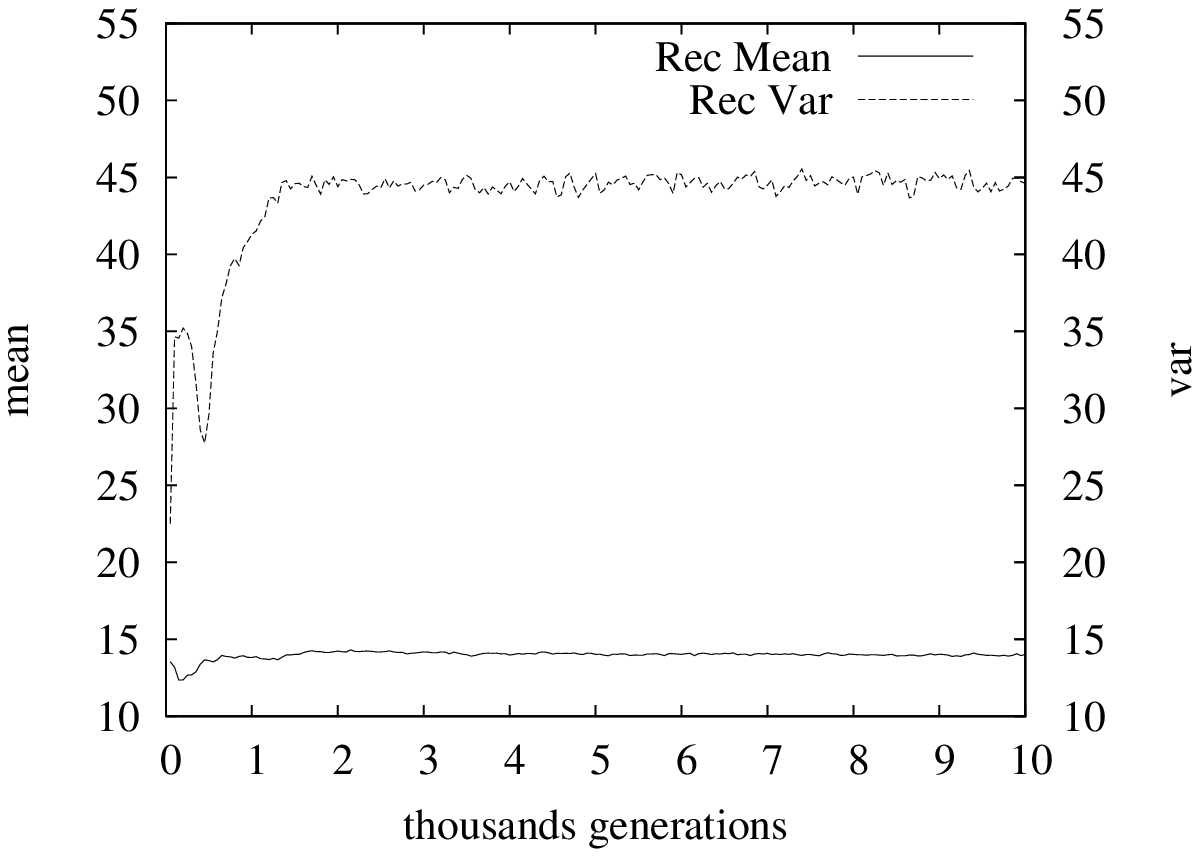} \\
\end{tabular}

\caption{Effects of genome length and mutation levels. Left column: low mutation levels $\mu_0 = 10^{-5}$, $\mu_{\infty} = 10^{-6}$, $\tau = 1000$ and $\delta = 100$; right column: high mutation level: $\mu_0 = 5 \times 10^{-3}$, $\mu_{\infty} = 10^{-6}$, $\tau = 1000$ and $\delta = 100$. From top to bottom, first row:  plots of mean and variance of non recombinants; second row: plots of mean and variance of recombinants. Flat static fitness $\beta = 100$, $\Gamma = 28$ and weak competition $J=1$, $\alpha = 2$, $R = 8$. Initial distribution: $p = 0.5$, $q = 0.5$, $M = 0.2$, $N_0 = 1000$, $K = 10000$. Mating range $\Delta = 28$, genome length $L = 28$. Total evolution time: 10000 generations.      }

\label{fig13_L28_var_mean}

\end{figure}

If we now repeat the simulation with a high mutation level ($\mu_0 =5 \times 10^{-3}$, $\mu_{\infty} = 10^{-6}$, $\tau = 1000$, $\delta = 100$) the pattern is different: the two peaks of non recombinants move quickly towards the ends of the space and when they are distant enough from each other, a third peak of non recombinants appears in the middle of the phenotypic space (as a consequence of the relief in competition pressure in that region). The variance of non recombinants therefore increases monotonically up to a high stable level while the mean remains constant because the distribution remains symmetrical throughout the whole process. The mean and variance of recombinants, on the other hand, tend to decrease at the beginning of the process when they attempt to create a small peak (immediately aborted) in the first half of the space; their distribution then becomes wide and flat thus leading to an increase in variance. The final distributions and the plots of mean and variance of recombinants and non recombinants in the case of low and high mutation levels are shown in Figures~\ref{fig13_L28_distr} and~\ref{fig13_L28_var_mean}.

\subsubsection{Steep static fitness landscape}
 
Let us consider a simulation in the case of absence of competition that can be compared to that illustrated in Figure~\ref{fig:8}. The simulation shows that if the mutation level is too low ($\mu_0 = 10^{-4}$, $\mu_{\infty} = 10^{-6}$, $\tau = 1000$ and $\delta = 100$) the non recombinants tend to become extinct. Both distributions in fact, move towards $x=0$ following the static fitness gradient, but when the non recombinants form a peak at $x=5-6$ they are passed past by the recombinants that establish a high peak in $x=0$ and force the non recombinants to extinction. The mean and variance of non recombinants reach a minimum when they form the peak at $x=5-6$, but when the peak disappears they rise again showing wide oscillations as the distribution becomes wide and flat and sensitive to random sampling effects due to the small size of the surviving group. The mean and variance of recombinants conversely decrease monotonically as the distribution becomes narrower and narrower while it moves towards $x=0$. 

\begin{figure}[ht!]

\begin{tabular}{ccc}
\hspace{-2 cm} & \includegraphics[scale=0.65]{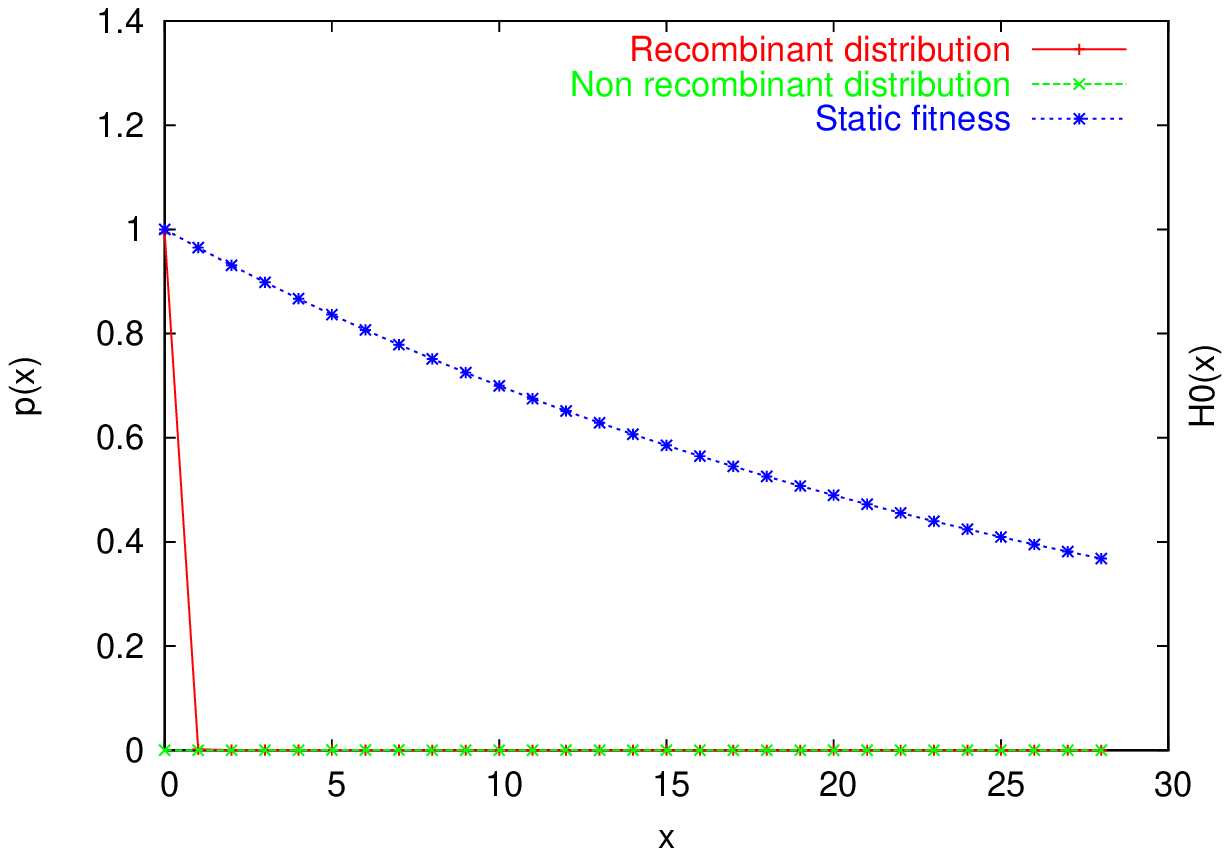} &
\includegraphics[scale=0.65]{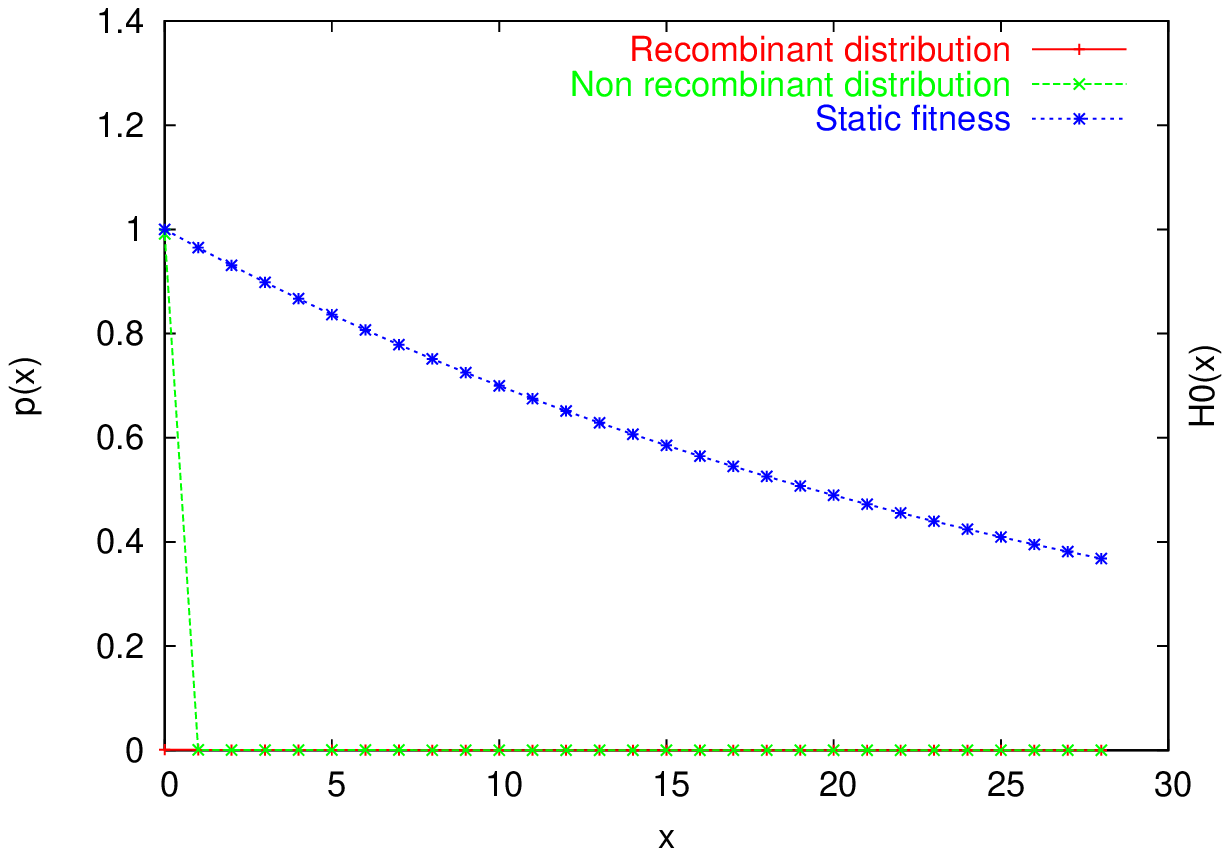} \\
\end{tabular}

\caption{Effects of genome length and mutation levels: final distributions. Left panel: low mutation levels
$\mu_0 = 10^{-4}$, $\mu_{\infty} = 10^{-6}$, $\tau = 1000$ and $\delta = 100$; right panel: high mutation
level: $\mu_0 = 5 \times 10^{-3}$, $\mu_{\infty} = 10^{-6}$, $\tau = 1000$ and $\delta = 100$. Steep static
fitness $\beta = 1$, $\Gamma = 28$ and absence of competition ($J=0$). Initial distribution: $p = 0.5$,
$q = 0.5$, $M = 0.1$, $N_0 = 1000$, $K = 10000$. Mating range $\Delta = 28$, genome length $L = 28$.
Total evolution time: 10000 generations.      }

\label{fig14_L28_distr}

\end{figure}

\begin{figure}[ht!]

\begin{tabular}{ccc}
\hspace{-2 cm} & \includegraphics[scale=0.65]{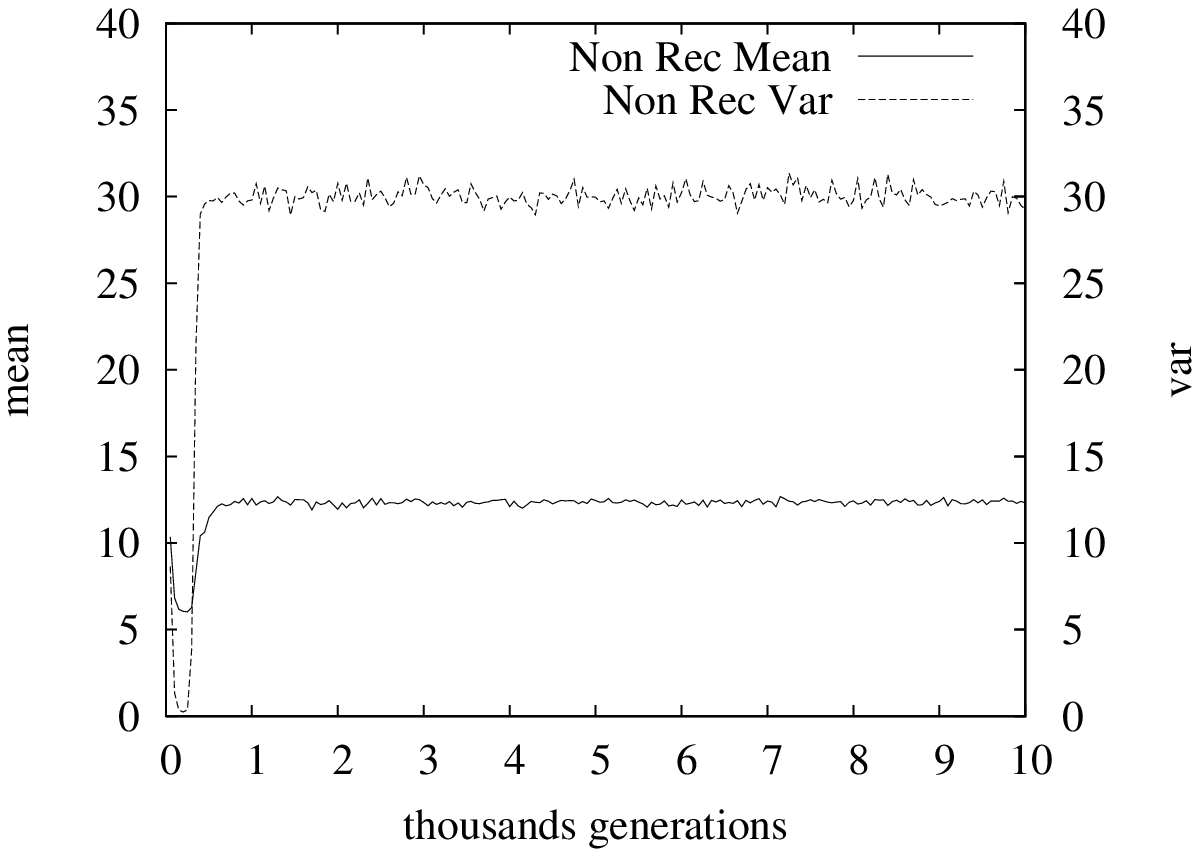} &
\includegraphics[scale=0.65]{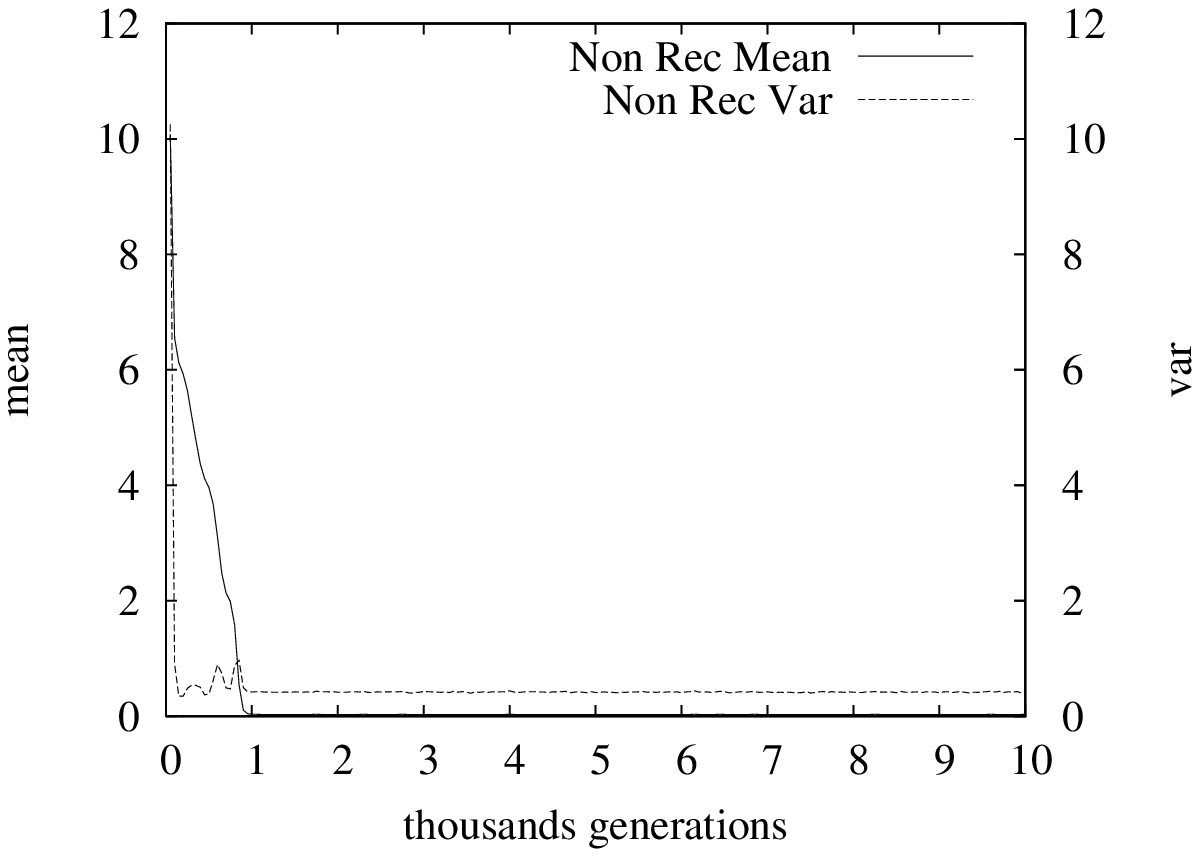} \\
\hspace{-2 cm} & \includegraphics[scale=0.65]{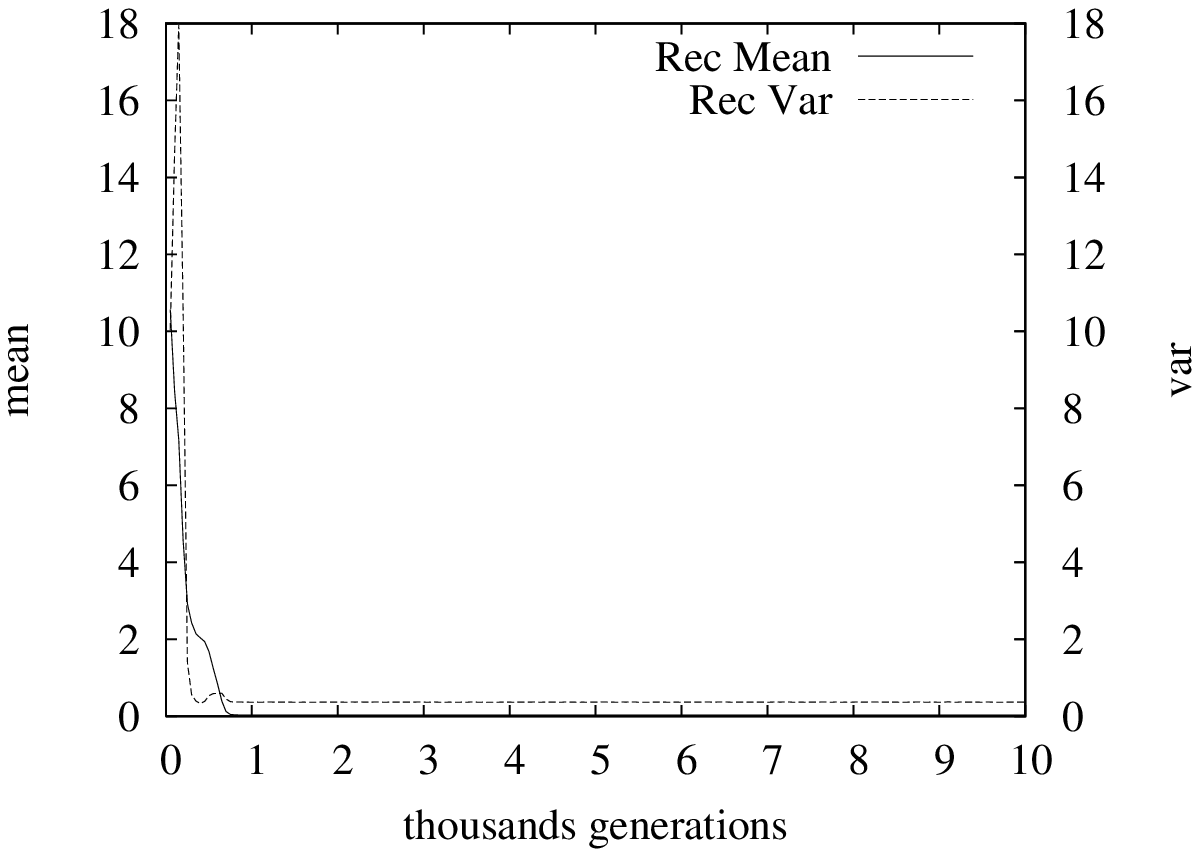} &
\includegraphics[scale=0.65]{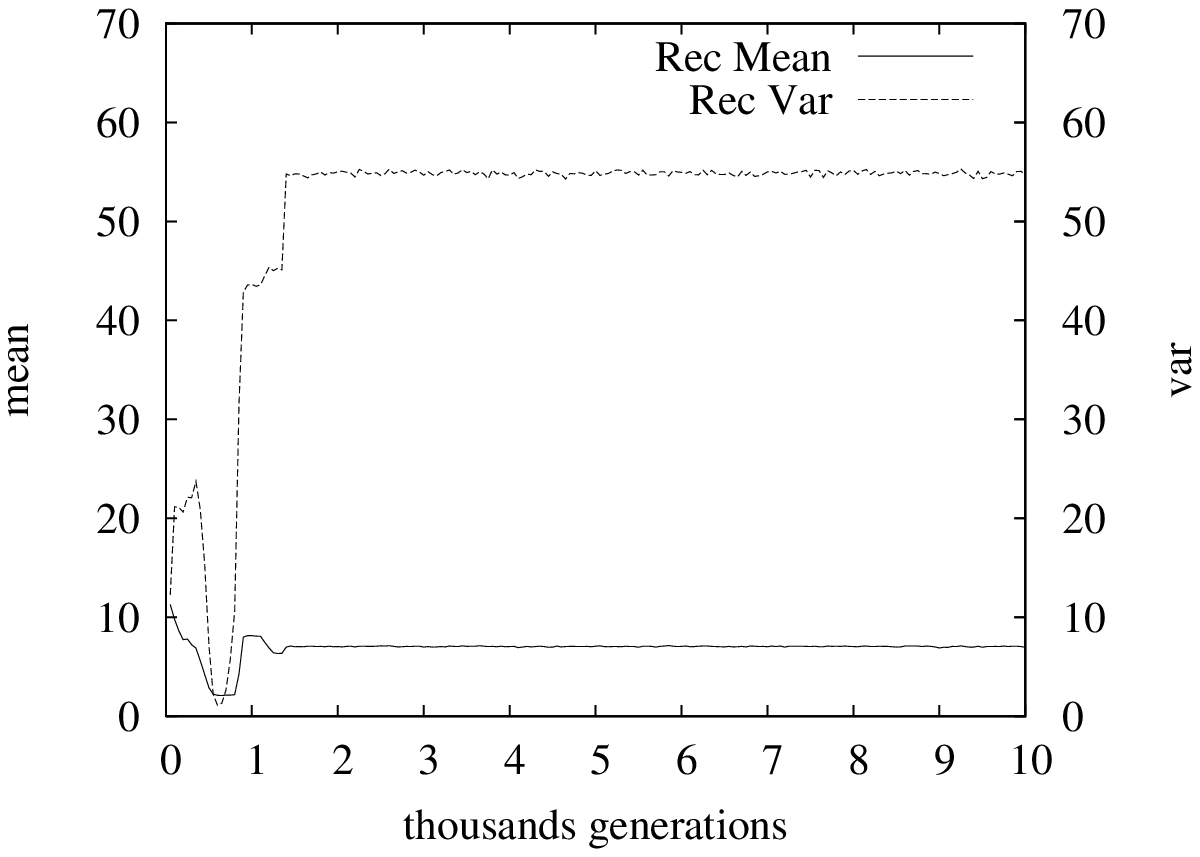} \\
\end{tabular}

\caption{Effects of genome length and mutation levels. Left column: low mutation levels $\mu_0 = 10^{-4}$,
$\mu_{\infty} = 10^{-6}$, $\tau = 1000$ and $\delta = 100$; right column: high mutation level:
$\mu_0 = 5 \times 10^{-3}$, $\mu_{\infty} = 10^{-6}$, $\tau = 1000$ and $\delta = 100$. From top to bottom, first row:  plots of mean and variance of non recombinants; second row: plots of mean and variance of recombinants. Steep static fitness $\beta = 1$, $\Gamma = 28$ and absence of competition ($J=0$). Initial distribution: $p = 0.5$, $q = 0.5$, $M = 0.1$, $N_0 = 1000$, $K = 10000$. Mating range $\Delta = 28$, genome length $L = 28$. Total evolution time: 10000 generations.      }

\label{fig14_L28_var_mean}

\end{figure}

If the mutation level is increased to $\mu_0 = 5 \times 10^{-3}$, $\mu_{\infty} = 10^{-6}$, $\tau = 1000$ and $\delta = 100$ the roles of recombinants and non recombinants are exchanged: the distributions once again shift towards $x=0$ but this time the non recombinants move faster and they reach $x=0$ before the recombinants that are thus led to extinction. Accordingly, the mean and variance of non recombinants decrease monotonically almost to zero; the mean decreases slower than the variance because a sharp peak first appears in the middle of the phenotypic space and then it moves gradually towards $x=0$. On the other hand, mean and variance  of recombinants first reach a minimum when a peak close to $x=0$ is formed, and then increase again when the peak disappears and the distribution becomes wide and flat. The final distributions and the plots of mean and variance of recombinants and non recombinants in the case of low and high mutation levels are shown in Figures~\ref{fig14_L28_distr} and~\ref{fig14_L28_var_mean}.

\begin{figure}[ht!]
\begin{center}
\begin{tabular}{c}                                                                                          
\includegraphics[scale=0.7]{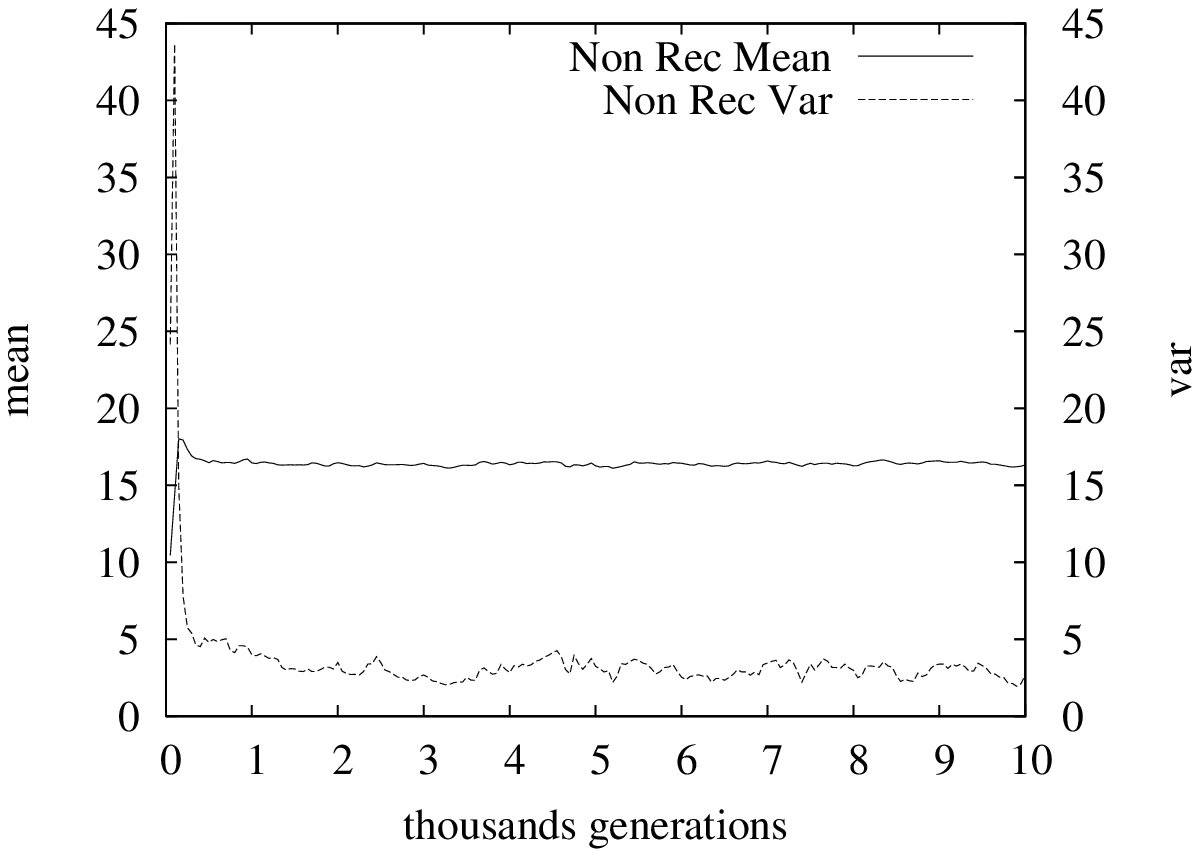} \\
\includegraphics[scale=0.7]{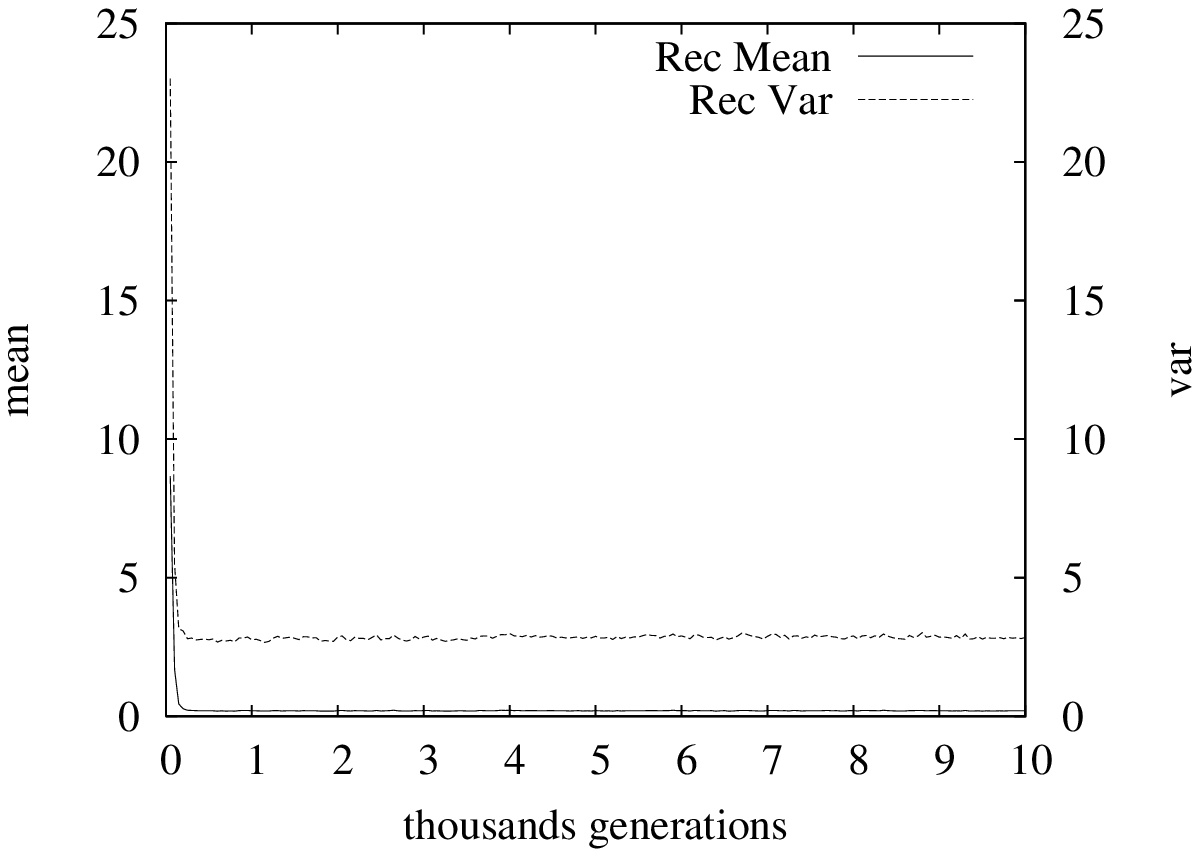} \\                 
\end{tabular} 
\end{center}
\caption{Plots of mean and variance of non recombinants (left) and recombinants (right) with high mutation levels ( $\mu_0 = 10^{-3}$, $\mu_{\infty} = 10^{-6}$, $\tau = 2000$ and $\delta = 200$) and weak competition ($J=0.8$, $\alpha = 2$, $R  = 8$). Steep static fitness $\beta = 1$, $\Gamma = 28$. Initial distribution: $p = 0.5$, $q = 0.5$, $M = 0.5$, $N_0 = 1000$, $K = 10000$. Mating range $\Delta = 28$, genome length $L = 28$. Total evolution time: 10000 generations.      }                                                                      \label{fig18_L28_var_mean}                                                                                                                                                                                                  \end{figure}

\begin{figure}[ht!]

\begin{center}
\includegraphics[scale=0.65]{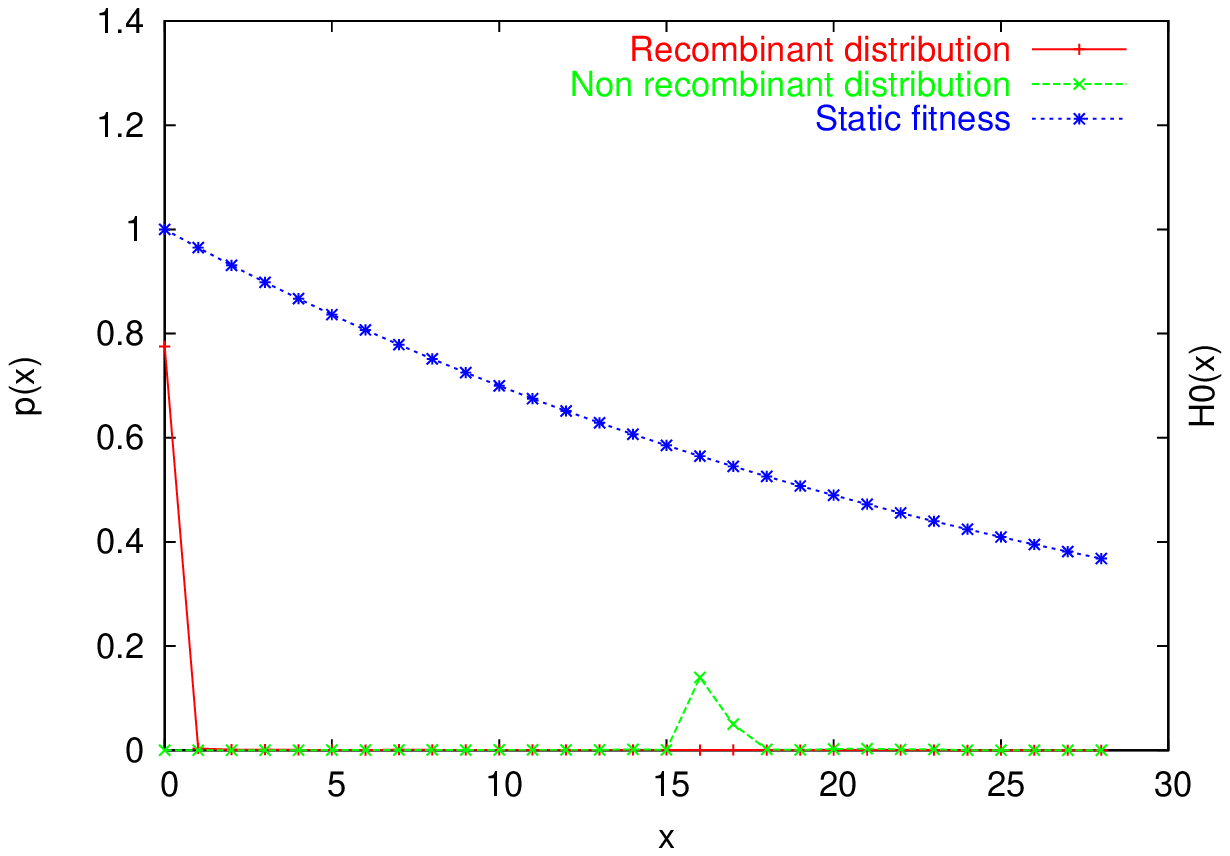} 
\end{center}

\caption{Final distribution with high mutation levels ( $\mu_0 = 10^{-3}$, $\mu_{\infty} = 10^{-6}$, $\tau = 2000$ and $\delta = 200$) and weak competition ($J=0.8$, $\alpha = 2$, $R  = 8$). Steep static fitness $\beta = 1$, $\Gamma = 28$. Initial distribution: $p = 0.5$, $q = 0.5$, $M = 0.5$, $N_0 = 1000$, $K = 10000$. Mating range $\Delta = 28$, genome length $L = 28$. Total evolution time: 10000 generations.}
 
\label{fig18_L28_distr}

\end{figure}

Let us now consider a simulation with weak competition comparable to the one illustrated in Figure~\ref{fig:12}. In this case the influence of competition is so strong that the evolutionary behavior is the same in a wide range of mutation levels. In this simulation the distributions of recombinants and non recombinants split in two curves moving towards the ends of the phenotypic space: the recombinants always reach the $x=0$ position first and they establish there a delta peak; the non recombinants on the other hand become dominant in the other direction and they form a peak in the middle of the phenotypic space at $x=15-16$. This particular pattern can be obtained both with the same low mutation level already employed with $L=14$ ($\mu_0 = 10^{-5}$, $\mu_{\infty} = 10^{-6}$, $\tau = 1000$ and $\delta = 100$) and with a high mutation level where, for consistency with the other examples discussed so far, we increased the initial mutation level by two orders of magnitude and we doubled the $\tau$ and $\delta$ parameters: $\mu_0 = 10^{-3}$, $\mu_{\infty} = 10^{-6}$, $\tau = 2000$ and $\delta = 200$. The final distribution and the plots of mean an variance relative to the latter mutation value are shown in Figures~\ref{fig18_L28_var_mean} and~\ref{fig18_L28_distr} and they are similar to those presented in Figure~\ref{fig:12}.

 \section{Phase diagrams}
 
 Up to now we have studied the evolution of recombination using single sets of parameters; we now perform systematic simulations to study the evolutionary dynamics in a wide range of parameters so as to bring to light a possible synergetic effect on the evolutionary success of recombinants.
 It is clear that this kind of research requires a huge number of
 simulations and one of the best ways to display the results is to plot the asymptotic frequency of
 recombinants as a function of both competition and assortativity. In particular, for each couple of values of competition and assortativity we perform 30 runs and we show the average  of the final frequency of recombinants.
 
 Before examining these diagrams, it is necessary to give a quantitative definition of assortativity.
 Assortativity is a measure of the bias of a phenotype towards mating with other similar phenotypes, and must
 be therefore inversely correlated with the mating range $\Delta$. This is why we have chosen to define
 assortativity as the difference between the phenotype length $L$ and the mating range $\Delta$:
 
 \[ \mathcal{A} = L - \Delta \]

 \subsection{Flat static fitness}

\begin{figure}[ht!]
\begin{center}
\includegraphics{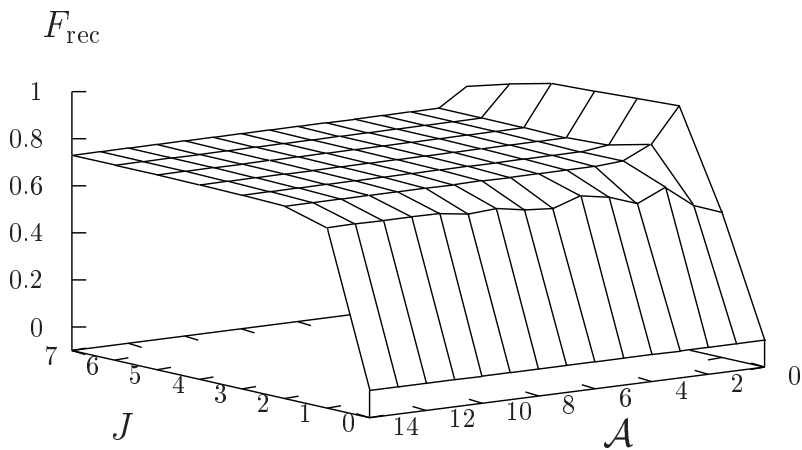} 
\end{center}

\caption{Tridimensional plot of the final frequency of recombinants as a function of competition and assortativity in a flat static fitness landscape ($\beta = 100$, $\Gamma = 14$). Initial frequency of recombinants: 0.5; initial distribution parameters: $p = q= 0.5$, initial population size $N_0 = 1000$, carrying capacity $K = 10000$. Annealing parameters: $\mu_0 = 10^{-5}$, $ \mu_{\infty} = 10^{-6}$, $\tau = 500$, $\delta = 50$. Total evolution time: 2500 generations. For each pair of parameters $(J,\mathcal{A})$ the final frequency of recombinants war averaged over 30 independent runs.  }

\label{phase_flat3D}

\end{figure}

If the static fitness landscape is flat, the surface of the recombinant frequency as a function of competition and assortativity, tends to rise in a smooth way as the competition is increased, until a plateau is reached for high competition values as shown in Figure~\ref{phase_flat3D}. In the presence of flat static fitness and absence of competition ($J=0$), all phenotypes have the same fitness and the evolutionary success depends only on fertility. This is why regardless of assortativity the recombinants pay the twofold price of recombination and they soon become extinct.

A very low level of competition such as $J=1$ determines a completely different scenario. The distribution of both recombinants and non recombinants splits in two peaks  and the final distribution depends on the ability of the two groups to reach first one or both ends of the phenotype space. In some instances the recombinants reach first only one end of the space where they establish a peak while two peaks of non recombinants will appear at the opposite end and in the middle of the space; the final frequency of recombinants will be about $35 \%$. A more interesting distribution arises when the recombinants reach both ends of the space earlier than non recombinants. In this situation recombinants and non recombinants coexist at both ends of the space with peaks of recombinants in $x=0$ and $x=14$ and peaks of non recombinants in $x=1$ and $x=13$. In a regime of random mating there will be also a flat bell-shaped distribution of recombinants produced by the $0 \times 14$ crossings and a small peak of non recombinants in $x=7$, so that the final frequency of recombinants will be about $65 \%$. The final frequency of recombinants slightly increases to $68 \%$ for $\mathcal{A} > 0$ because the colonies of recombinants in $x=0$ and $x=14$ are so big to force to extinction the non recombinant peaks in $x=1$ and $x=13$.

The range $J=2$ to $J=5$ corresponds to the plateau of the surface of recombinant frequency. The higher competition intensity enables the recombinants to reach first both ends of the phenotype space also in a regime or random mating. The final distribution therefore features two large peaks of recombinants in $x=0$ and $x=14$ and a smaller peak of non recombinants in $x=7$. If mating is random ($\mathcal{A} = 0$) there will also be a flat bell-shaped distribution of offsprings of the extreme phenotypes and the final frequency of recombinants will be about $92 \%$; conversely, if mating is assortative ($\mathcal{A} > 0$) the central bell-shaped distribution disappears and the final frequency of recombinants will be only about $ 75 \%$.

\subsection{Steep static fitness}

\begin{figure}[ht!]
\begin{center}
\includegraphics{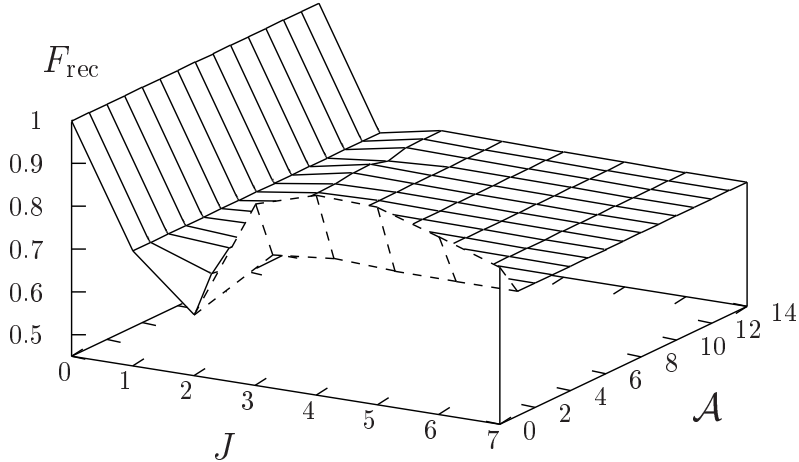}
\end{center}

\caption{Tridimensional plot of the final frequency of recombinants as a function of competition and assortativity in a steep static fitness landscape ($\beta = 1$, $\Gamma = 14$). Initial frequency of recombinants: 0.5; initial distribution parameters: $p = q= 0.5$, initial population size $N_0 = 1000$, carrying capacity $K = 10000$. Annealing parameters: $\mu_0 = 10^{-5}$, $ \mu_{\infty} = 10^{-6}$, $\tau = 500$, $\delta = 50$. Total evolution time: 2500 generations. For each pair of parameters $(J,\mathcal{A})$ the final frequency of recombinants war averaged over 30 independent runs.  }
\label{phase_steep3D}

\end{figure}

If the static fitness profile is steep, the  plot of the final frequency of recombinants as a function of
competition and assortativity shows high values when competition is weak and then it decreases to a roughly constant level as can be seen in Figure~\ref{phase_steep3D}. In absence of competition ($J=0$), in fact, both distributions move towards the maximum of static fitness in $x=0$, but this location is first attained by recombinants and the non recombinants become almost extinct; the final frequency of recombinants will be always higher than $99 \%$ regardless of assortativity. 

If a moderate competition such as $J=1$ is introduced, the distributions of recombinants and non recombinants split in two curves moving towards the ends of the phenotype space; the recombinants will thus form a peak in $x=0$ whereas the non recombinants form two peaks the first spanning $x=7$ and $x=8$ and the second at $x=12$. Competition therefore stabilizes the non recombinants peak: in fact the low level of competition experienced in the regions where the non recombinants establish their peaks counteracts the low static fitness.  Regardless of assortativity the final frequency of recombinants will be about $70 \%$.

The strong competition for $J \geq 3$ urges the recombinants to reach earlier than non recombinants both $x=0$ and $x=14$ where they establish their peaks, while the non recombinants will only form a small peak in the middle of the phenotype space. If mating is random ($\mathcal{A} = 0$) there will also be a flat bell-shaped distribution of recombinants produced by the $0 \times 14$ crossings, and as a result the final frequency of recombinants will be around $90-92 \%$. If mating is assortative, on the other hand, the central bell-shaped distribution of recombinants cannot be produced, therefore relieving competition in the middle of the phenotype space and allowing the peak of non recombinants to become more populated: the final frequency of recombinants will be about $75 \%$.   

Important information can be obtained by the study of sections of the  plot parallel to the
assortativity axes. Apart for some irregularities, these lines keep a constant level, showing that for any
given competition level, an increase in assortativity doesn't cause any significant increase in the final
level of recombinants so that in this case no synergistic effect can be detected. 

\vspace{0.5 cm}

\subsubsection{Assortativity and initial frequency of recombinants}

\begin{figure}[ht!]
\begin{center}
\includegraphics{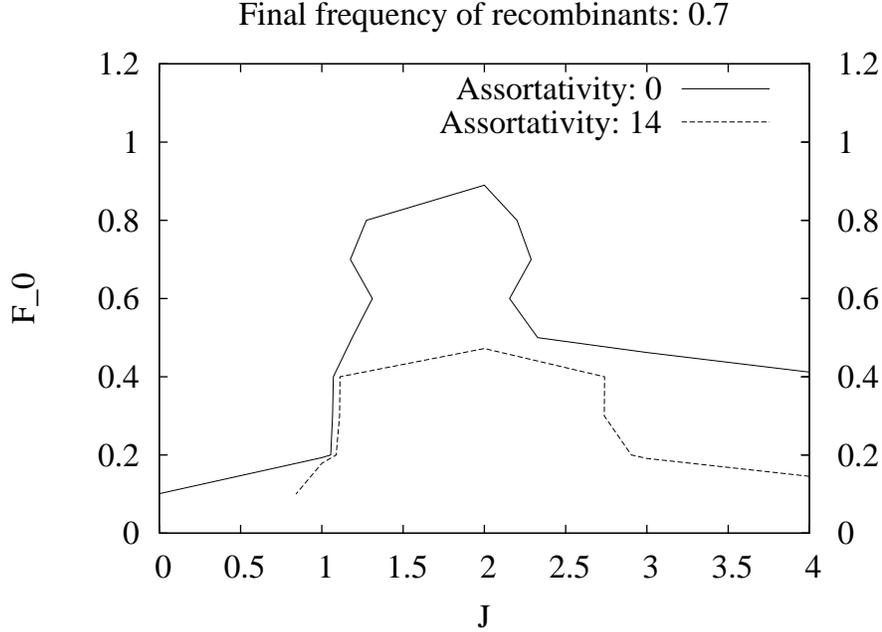}
\end{center}

\caption{Contour plots of the final frequency of recombinants as a function of competition and initial frequency in a steep static fitness landscape ($\beta = 1$, $\Gamma = 14$). The plots refer to a contour level $F_{rec} = 0.7$ for minimal ($\mathcal{A} = 0$) and maximal ($\mathcal{A} = 14$) assortativity.  Initial frequencies of recombinants: from 0.1 to 0.9 with steps of 0.1. Initial distribution parameters: $p = q= 0.5$, initial population size $N_0 = 1000$, carrying capacity $K = 10000$. Annealing parameters: $\mu_0 = 10^{-5}$, $ \mu_{\infty} = 10^{-6}$, $\tau = 500$, $\delta = 50$. Total evolution time: 2500 generations. For each pair of parameters $(J,\mathcal{A})$ the final frequency of recombinants war averaged over 10 independent runs.  }
\label{phase_steep_JF}

\end{figure}

The role of assortativity in a steep static fitness landscape, can be identified if we consider populations
with different initial frequencies of recombinants. One possible approach is to plot the final frequency of
recombinants as a function of their initial frequency and competition for various levels of assortativity.
If we now study the contour plots relative to a given final frequency, for instance 70\%, we notice that the
plots become lower and flatter as assortativity is increased. In other words, as assortativity is increased,
lower and lower initial frequencies of recombinants are needed in order to obtain a final frequency of 70\%.
It can be therefore concluded that high levels of assortativity play a significant role in ensuring the
survival of initially very small recombinant populations. The contour plots are shown in Figure~\ref{phase_steep_JF}.

The contour plots shown in Figure~\ref{phase_steep_JF} show that for any given value of assortativity the initial frequency of recombinants needed to attain a final frequency of 0.7 is low when competition is weak or absent, becomes maximal for intermediate competition values and decreases again for high competition values. In absence of competition ($J=0$) the distributions of both recombinants and non recombinants move towards the fitness maximum in $x=0$; the non recombinants, whose mobility is limited by the low mutation rate, typically form a peak in $x=1$ but they do not force the recombinants to extinction as competition is absent and the recombinants are penalized only by their lower fertility. As a consequence, the recombinants are able to survive until they reach the $x=0$ position where they establish a large peak that causes the disappearance of non recombinants due to random sampling effects. This explains why in the absence of competition a very low frequency of recombinants is sufficient to attain a final level of 0.7. With intermediate values of competition however, the recombinants are penalized not only by their low fertility but also by the non recombinants competition so that small groups of recombinants are bound to extinction and larger initial frequencies are needed  to survive until the $x=0$ and $x=14$ positions are colonized, which leads to a final frequency of $70 \%$. Finally, when competition is high ($J = 4$), the recombinants move so quickly in the phenotype space that even small groups can reach $x=0$ and/or $x=14$ before they are driven to extinction by the non recombinants. As a consequence, the initial frequency required to attain the desired final level of $70 \%$ is again very low.

\section{Conclusions}

A microscopic model has been developed for the study of the evolution of recombination in an environment whose features depend on the population itself (the frequency distribution of phenotypes determines competition and hence fitness).

The simulations showed that in a flat static fitness landscape the twofold price of recombination can be counteracted if competition is introduced. Under those conditions, in fact, recombinants reach the lowest competition positions of the phenotypic space earlier than non recombinants.

A similar pattern can be observed for a steep static fitness landscape: in the absence of competition and with low mutation rates  recombinants colonize the highest fitness position while non recombinants are led to extinction. If competition is introduced it acts as a stabilizing force enabling the survival of non recombinants in regions of the phenotypic space where the low static fitness is counteracted by the low competition pressure.

The simulations also show that recombinants can be successful if the mutation rate is very low, which is what happens in real populations. If the mutation rate is low, this limits the mobility of non recombinants in the phenotype space and as a result, the most favourable regions of the phenotype space will be colonized by recombinants. An unnaturally high mutation rate, on the other hand provides non recombinants with a mobility comparable to that of recombinants that are thus driven to extinction because of their low fertility. As a matter of fact, however, a too high mutation rate is not compatible with life. In fact, while a mutation affecting a gene related to a quantitative trait only causes a small phenotypic effect, a mutation on a gene controlling a qualitative trait, can lead to a serious impairment or even to the death of the bearer. It can be therefore concluded that one of the reasons for the evolution of recombination is that it provides as large a genetic variability of offsprings as an extremely high mutation rate, without involving all the drawbacks of mutation. The fact that recombination increases the velocity of movement in the phenotype space can be regarded as a reformulation of Weismann's statement according to which recombination creates genetic variability upon which selection acts. In particular, in agreement with the most recent studies~\cite{SPO-Ref-7,SPO-Ref-37}, our work confirms that the high mobility in the phenotype space provided by recombination is particularly useful when the environment (phenotype distribution) rapidly changes affecting competition levels and fitness. The ability of recombinants to colonize the ends of the phenotype space also shows that recombination is effective in bringing together several favourable mutations that would be otherwise segregated in different cell lines which is consistent with the results reported by~\cite{Rice-Ref-3,Rice-Ref-24,Rice-Ref-25}.       

We also checked that an increase in genome length leads to an increase in the size of the phenotype space so that higher mutation rates are required to provide both recombinants and non recombinants with a sufficient mobility and attain the same patterns observed with a shorter genome. As the non recombinants are particularly sensitive to this problem, this pattern seems to explain why non recombinant organisms such as the prokaryotes are usually characterized by a shorter genome

Finally, the analysis of phase diagrams showed that contour plots are almost parallel to the assortativity axis implying that competition is the most important factor for the evolution of recombination. Assortativity, on the other hand, turned out to be important for the survival and evolutionary success of small populations of recombinants: a high assortativity in fact, prevents dispersion of offsprings so that small populations of recombinants can survive long enough to reach the most favourable regions of the phenotypic space. 
 
\bibliographystyle{unsrt}

\bibliography{bibliography_ricombinazione}

\end{document}